\documentclass{emulateapj}
\usepackage{natbib}
\usepackage{graphicx}
\usepackage{longtable}
\usepackage{layout}
\usepackage[latin1]{inputenc}
\usepackage{amssymb}
\usepackage{natbib}

\def\Msun{M$_\odot$}

\def\Mbh{$M_{\rm BH}$}

\def\Civ{C\,{\sc iv}}
\def\Mgii{Mg\,{\sc ii}}
\def\Feii{Fe\,{\sc ii}}
\def\Feiii{Fe\,{\sc iii}}

\def\Ha{H$\alpha$}
\def\Hb{H$\beta$}
\def\lsim{\mathrel{\rlap{\lower 3pt \hbox{$\sim$}} \raise 2.0pt \hbox{$<$}}}
\def\gsim{\mathrel{\rlap{\lower 3pt \hbox{$\sim$}} \raise 2.0pt \hbox{$>$}}}

\slugcomment{ApJ accepted 2011 June 23}

\shortauthors{De Rosa et al. 2011}

\begin{document}


\title{Evidence for non-evolving \Feii/\Mgii \ ratios in rapidly accreting $z\sim6$ QSOs}


\author{G. De Rosa\altaffilmark{1}, R. Decarli\altaffilmark{1}, F. Walter\altaffilmark{1}, 
 X. Fan\altaffilmark{2}, L. Jiang \altaffilmark{2},  J. Kurk \altaffilmark{3}, A. Pasquali\altaffilmark{4},
 H.--W. Rix\altaffilmark{1}}


\altaffiltext{1}{Max-Planck-Institut f\"{u}r Astronomie K\"{o}nigstuhl 17, D-69117, Heidelberg}
\altaffiltext{2}{Steward Observatory, University of Arizona, 933 North Cherry Avenue, Tucson, AZ 85721}
\altaffiltext{3}{Max-Planck-Institut f\"ur Extraterrestrische Physik, Gie\ss enbachstra\ss e 1,
85748, Garching}
\altaffiltext{4}{Astronomisches Rechen-Institut, M\"{o}nchhofstr. 12-14, D-69120, Heidelberg}


\begin{abstract}
Quasars (QSOs) at the highest known redshift ($z\sim6$) are unique probes of the early growth 
of supermassive black holes (BHs). 
Until now, only the most luminous QSOs have been studied, often one object at a time. Here 
we present the most extensive consistent analysis to date of $z>4$ QSOs with observed NIR spectra,
combining three new $z\sim6$ objects from our ongoing VLT-ISAAC program with nineteen $4<z<6.5$ sources from the literature. 
The new sources extend the existing SDSS sample towards the faint end of the QSO luminosity function. 
Using a maximum likelihood fitting routine optimized for our spectral decomposition,
we estimate the black hole mass (\Mbh), the Eddington ratio (defined as $L_{bol}/L_{Edd}$) and the \Feii/\Mgii \ 
line ratio, a proxy for the chemical abundance, to characterize both the central 
object and the broad line region gas.  
The QSOs in our sample host BHs with masses of $\sim 10^9$ \Msun \ that are 
accreting close to the Eddington luminosity, consistent with earlier results. 
We find that the distribution of observed Eddington ratios is 
significantly different than that of a luminosity-matched comparison sample of SDSS QSOs at lower redshift 
($0.35<z<2.25$): 
the average $\langle log(L_{bol}/L_{Edd})\rangle=-0.37$ ($L_{bol}/L_{Edd}\sim$0.43) with a scatter of 0.20 dex 
for the $z>4$ sample and the $\langle log(L_{bol}/L_{Edd})\rangle=-0.80$ ($L_{bol}/L_{Edd}\sim$0.16) 
with a scatter of 0.24 dex for the $0.35<z<2.25$ sample. This implies that, at a given luminosity, 
the \Mbh \ at high-z is typically lower than the average \Mbh \ of the lower-redshift population, i.e. 
the $z>4$ sources are accreting significantly faster than the lower-redshift ones. 
We show that the derived \Feii/\Mgii \ ratios depend sensitively on the performed analysis:
 our self-consistent, homogeneous analysis significantly reduces the \Feii/\Mgii \ scatter found in previous studies.
The measured \Feii/\Mgii \ line ratios show no sign of evolution with cosmic time in the redshift range $4<z<6.5$.
If the \Feii/\Mgii \ line ratio is used as a secondary proxy of the Fe/Mg abundance ratio, this implies 
that the QSOs in our sample have undergone a major episode of Fe enrichment
in the few 100 Myr preceding the cosmic age at which they are observed.  
\subjectheadings{cosmology: observations -- quasars: general, emission lines -- galaxies: active, high-redshift, formation}

\end{abstract}

\section{Introduction}
\label{intro}

In the past 10 years, more than 50 QSOs at $z>5.7$ have been discovered \citep{Fan2000, Fan2001, Fan2003, Fan2004, 
Fan2006, Jiang2008, Jiang2009, Willott2007, Willott2010} thanks to the Sloan Digital Sky Survey \citep[SDSS,][]{York2000}
 and other large multi-wavelength surveys such as 
the Canada France High-z Quasar Survey \citep[CFHQS,][]{Willott2007}. These high-redshift QSOs are among the most luminous sources known to date and 
are direct probes of the universe less than 1 Gyr after the Big Bang. They are fundamental force studying 
the early growth of supermassive BHs, galaxy formation and interstellar medium chemical 
evolution \citep[e.g.][]{Kauffmann2000, Wyithe2003, Hopkins2005}.
Strong emission features excited by the central engine and its immediate surroundings can be used 
to infer properties of the powering BH and of the circumnuclear gas. For example, measuring the emission line 
width one can obtain information about the bulk motion of 
the broad-line region (BLR) gas, that can then be used to estimate the BH mass (\Mbh). From the line 
flux ratios we can instead derive chemical abundances of the BLR gas that are fundamental 
to set constraints on the star formation history of the QSO host galaxy. 

Under the assumption that the dynamics of the BLR is dominated by the central gravitational field, it is possible to 
estimate \Mbh \ by using the velocity and the distance of the line-emitting gas from the central 
BH : $M_{BH}\propto R \ v^2$. From reverberation mapping studies of local active galactic nuclei (AGNs) 
it has been found that $R$ is related to the continuum luminosity $L$ \citep[$R-L$ relation, e.g.][]{Kaspi2000}: 
$R\propto L^{0.5}$. 
Measurements of the luminosity and of the gas velocity, via Doppler-broadened line-width, can thus be used 
to determine the \Mbh \  from single epoch spectra. Various strong emission lines can be used as \Mbh \ estimators: 
\Ha, \Hb, \Mgii, \Civ \ \citep[for a review on the single-epoch spectrum method see][]{Peterson2010}. For objects in 
the local universe, the $R-L$ relation and its intrinsic scatter have been investigated in detail using the \Hb \ line 
and the relative continuum. These studies have shown that one can obtain accurate \Mbh \ estimates via the \Hb \ line. 
For sources at high-z a complication is that the \Hb \ emission line is 
redshifted out of the visible window already at modest redshifts. 
Whether or not a particular UV line can be used to estimate \Mbh \ depends on how well the respective 
UV line widths in AGN spectra are correlated. In the
case of \Mgii \ \citet{Shen2008}, using a large collection of
SDSS spectra, found that log ([FWHM(\Hb)] / [FWHM(\Mgii)]) = 0.0062, with a scatter of
only 0.11 dex, suggesting that the \Mgii \ can be used as a proxy for the \Hb \ line, and thus for the BH mass. 
On the other hand, the accuracy achievable in the determination of the BH mass from the \Civ \ line remains somewhat 
controversial, since \Civ \ is a resonance line and absorptions are often detected in its blue wing due to outflows, 
which in turn affects the line width. With the single-epoch spectrum method it has been possible 
to estimate the \Mbh \  for $z\sim6$ QSOs \citep{Barth2003, Willott2003, Jiang2007, Kurk2007, Kurk2009, Willott2010}. 
These studies have shown that these high-z sources from SDSS host BHs with \Mbh \ $\sim 10^9$ \Msun \ 
and are accreting close to the Eddington limit ($L_{bol}/L_{Edd} \sim$1). However, the $z\sim6$ QSOs from SDSS only account for the 
brightest end of the QSO luminosity function.
Only few faint QSOs ($z'_{AB}>20$) have measured \Mbh \  to date: two from the SDSS deep Stripe 
82 \citep{Jiang2008, Jiang2009, Kurk2007, Kurk2009}, a deep imaging survey obtained by repeatedly scanning a stripe 
(260 deg$^2$) along the celestial equator, and nine from CFHQS \citep{Willott2010}. These sources are powered by less massive BHs 
(down to $M_{BH} \sim 10^8$ \Msun) that also appear to accrete close to the Eddington limit. For QSOs at lower 
redshifts \citet{Shen2008} found that the most 
luminous QSOs ($L_{bol}>10^{47}$ erg s$^{-1}$) at redshift $2<z<3$ are characterized by an Eddington ratio
of only $L_{bol}/L_{Edd} \sim$0.25 with a dispersion of 0.23 dex.
It seems therefore that high-redshift QSOs have fundamentally different properties than the lower-redshift ones.    

Regarding the chemical enrichment, photoionization models show that various emission line ratios are good 
estimators for the metallicity of the BLR gas \citep[e.g.][]{Hamann2002, Nagao2006}. 
It is possible to estimate the BLR gas metallicity and to set constraints on the BLR enrichment history 
through measurements of the relative abundances of nitrogen (N) respect to carbon (C) and helium (He), 
since these three elements are formed by different astrophysical processes and 
on different timescales \citep{Hamann1993}: 
C is rapidly produced in the explosion of massive stars; N is a second generation element, i.e. 
slowly produced in stars 
from previously synthesized C and O; He is a primordial element and its abundance does not change significantly 
with cosmic time. 
Using these element ratios \citet{Jiang2007} estimated the BLR metallicity for a sample of six 
luminous QSOs with $5.8<z<6.3$, finding super solar metallicities with a typical value of $\sim4 \ Z_\odot$ and 
no strong evolution up to $z\sim6$.   

The abundance of Fe and $\alpha$ elements (e.g. O, Mg, Ne that are produced in the stars via $\alpha$ processes) 
is of particular interest for understanding the chemical evolution 
of galaxies at high-z. 
Most of Fe in the solar neighborhoods has been produced via the explosion of type Ia supernovae (SNe Ia) while $\alpha$ elements such 
as Mg and O are mainly produced by core collapse supernovae (SNe of types II, Ib and Ic). 
SNeIa are thought to originate from intermediate-mass stars in close binary systems,    
characterized by long life-times, while core 
collapse SNe come from more massive stars which explode very soon after the initial starburst. 
A time delay between $\alpha$ elements and the Fe enrichment is thus expected. This delay depends on the 
Initial Mass Function (IMF) and on the galactic star formation history, and can vary from 0.3 Gyr, for massive 
elliptical galaxies, to 1-3 Gyr, for Milky-Way type galaxies \citep{Matteucci2003}. In this picture the 
Fe/$\alpha$ ratio is expected to be a strong function of age in young systems. 
The observational proxy that is usually adopted to trace the Fe/Mg abundance ratio for the BLR gas 
is the \Feii/\Mgii \ line ratio. 
Unfortunately this line ratio is only a second order proxy, since it depends not only on 
the actual abundance of Fe but also significantly on the excitation conditions 
that determine how strongly \Feii \ lines are emitted (Baldwin et al. 2004). However, even though    
to date no calibration is available to convert the observed relative line strengths into actual abundance 
ratios, the study of the \Feii/\Mgii \ line ratio as a function of look-back
time does in itself carry significant information about the BLR chemical enrichment history.\\
For $z>5.7$ QSOs the \Feii \ and the \Mgii \ lines are redshifted into the near-infrared (NIR). 
Numerous NIR-spectroscopy 
studies of QSO samples including $z\sim6$ sources have been carried out in the past \citep[e.g.][]{Maiolino2001, Maiolino2003,
Pentericci2002, Iwamuro2002, Iwamuro2004, Barth2003, Dietrich2003, Freudling2003, Willott2003, 
Jiang2007, Kurk2007, Kurk2009}. Their results show an increase in the scatter of the measured \Feii/\Mgii \ 
line ratios as a function of the redshift.
One possible way to interpret the increase in the scatter is that some objects are observed such
a short time after the initial starburst that the BLR is not yet fully enriched with Fe \citep{Iwamuro2004}.
Nonetheless, several possible systematics related to the adopted fitting procedure could potentially be the 
cause for the observed discrepancies between different authors: 
the iron template employed, the wavelength limits over which
the template is integrated, the wavelength range over which the template
is actually fitted and the wavelength coverage and S/N of the spectra used.

The goal of this paper is to study BH masses, Eddington ratios and to characterize the \Feii \ and \Mgii \ 
emission in a large sample of high-z QSOs, by fitting their respective NIR spectra in a coherent and homogeneous way. 
With the aim of extending the existing SDSS sample of $z\sim6$ QSOs towards the faint end of the QSO luminosity function, 
we observed three additional $z\sim6$ sources \citep[discovery papers:][De Rosa et al. in prep]{Jiang2008} with ISAAC 
(Infrared Spectrometer And Array Camera) mounted on the Very Large 
Telescope (VLT). The three targets have $19.6< z'_{AB} < 20.7 $ and have been selected from the SDSS and
the SDSS Stripe 82. The observed NIR spectra include the \Mgii \ and \Feii \ emission lines. 
We have also collected literature spectra of $z>4$ QSOs covering the restframe wavelengths 
between $2700$ \AA $<\lambda<3200$ \AA, characterized by the presence of the \Mgii \ line doublet ($\lambda=2796,2803$ \AA) and 
of the \Feii \ line forest.  
    
The paper is structured as follows. In $\S$ 2 we describe the sample, the new observations and the data 
reduction we have performed. In $\S$ 3 the spectral decomposition and the fitting procedure adopted are explained 
together with their results. In $\S$ 4 we estimate the \Mbh \  and the \Feii/\Mgii \ 
line ratio and we discuss their implications. Finally, we give a brief summary in $\S$ 5. We assume the following $\Lambda$CDM 
cosmology throughout the paper: $H_0=70 \ $km$ \ $s$^{-1} \ $Mpc$^{-1}$, $\Omega_M=0.3$, and $\Omega_\Lambda
=0.7$ \citep{Spergel2007}. 

\section{Sample, observations and data reduction}

Our sample is composed of 22 targets: 3 new spectra observed with VLT-ISAAC (see Tab.~\ref{new_source}) and 19
sources from the literature (see Tab.~\ref{lit_spec}), kindly provided by the respective authors. 
Ten of the literature sources have redshift $4.50 <z<5.70$
\citep{Iwamuro2002}, while the remaining 9 have
 $5.70 < z < 6.43$ and z-band magnitudes $z'_{AB}<20.9$ (magnitudes are taken from the discovery papers). 

\subsection{New data}
We have observed 3 SDSS QSOs with magnitudes $19.6<z'_{AB}<20.7$ and redshifts 6.05$<$z$<$6.08 
(see Tab.~\ref{new_source}). The two faintest ones have been selected from the SDSS Stripe 82 and extend the existing sample 
towards the faint end of the QSO luminosity function. The observations were carried out with ISAAC 
on Antu (VLT-UT1) in low resolution mode (LR), using the 1024x1024 Hawaii Rockwell array of the Short Wavelength arm.
For each QSO the \Mgii \ line and the \Feii \ complex were observed: given the redshift of the sources, these features fall 
 in the K band. The selected slit had a width of 1" and combined with the order selection filter it gives a spectral 
resolution $\lambda/\Delta\lambda \sim$450. Table \ref{new_source} summarizes the exposure time for each 
object. For each observation block (OB), sixteen frames of 148 seconds were taken following an ABBA dithering 
pattern (with large offsets among the dithered positions: from 20'' to 30''). Further, small random offsets within a box 
of 4'' to 18'' were applied at each dithered position in order to avoid pixel-related artifacts (jittering). Given the faintness of the sources, the observation setup was chosen so that a bright 
star was always in the slit, in order to allow a correct centering of the target.

The QSO SDSS J2054-0005 has been discovered in the SDSS deep Stripe 82 by \citet{Jiang2008}. 
Our new ISAAC spectrum confirms the weak-line nature of this source \citep[see also the the optical 
discovery spectrum by][]{Jiang2008}. Given the intrinsic weakness of the \Mgii \ emission, we 
do not include this QSO in the following analysis.

\begin{deluxetable*}{lccccccl}
\tablecaption{\label{new_source}
Summary of the newly observed sources. (1) QSO name; (2), (3) QSO right ascension and declination in J2000.0 (from 
discovery papers, col $\left[8\right]$); 
 (4) optical redshift (from discovery papers); (5) $z^*_{AB}$ (from discovery papers); 
 (6) K band nominal magnitude, obtained from the redshifted SDSS QSO template scaled to match the observed J and H-band 
 magnitudes (discovery papers); (7) total on source exposure time;
 (8) discovery paper: DRip, De Rosa et al. (in prep); J08, \citet{Jiang2008}.}
\tablecolumns{8}
\tablewidth{0pc}
\tablehead{
\colhead{QSO name} &   \colhead{RA} & \colhead{DEC} & \colhead{z} & \colhead{z$^*_{AB}$} & \colhead{K} &
\colhead{$t_{exp}$} & \colhead{Reference} \\ \colhead{(1)} & \colhead{(2)} & 
\colhead{(3)} & \colhead{(4)} & \colhead{(5)} & \colhead{(6)} & \colhead{(7)} & \colhead{(8)}  }
\startdata
SDSS J$0353+0104$ & $03 \ 53 \ 49.72$ & $+01 \ 04 \ 04.4$ & 6.05 & 20.5 & 17.81 & 4.60 h & J08 \\
SDSS J$2054-0005$ & $20 \ 54 \ 06.49$ & $-00 \ 05 \ 41.8$ & 6.06 & 20.7 & 17.87 & 5.92 h & J08 \\
SDSS J$0842+1218$ & $08 \ 42 \ 29$ & $+12 \ 18 \ 50.5$ & 6.08 & 19.6 & 17.67 & 4.60 h & DRip \\
\enddata
\end{deluxetable*}

\subsubsection{Data reduction} 

The ESO ISAAC pipeline produces wavelength 
calibrated co-added 2-D spectra from the individual frames that were acquired during each OB. Subsequent reduction
was carried out within IRAF. One-dimensional spectra were extracted using the \texttt{apall} task.
The tracing of the 1-D spectra was performed first on the bright stars in the slit. 
The resulting tracing functions were then used for the extraction of the 
QSO spectra. Individual 1D spectra were corrected for the telluric absorptions using the \texttt{telluric} task.
Usually the telluric correction is performed by dividing the observed spectrum by the one of a telluric standard star observed shortly 
after the science target. This ratio is subsequently multiplied by the model atmosphere corresponding to the spectral 
type of the telluric standard and scaled to its observed K magnitude, in order to recover the correct slope of the QSO spectrum.
This operation allows us also to flux-calibrate the QSO spectrum.
Instead of using the observed telluric standard stars, 
we employed the ESO sky absorption spectrum measured on 
the Paranal site at a nominal airmass of 1. This choice was driven by two reasons: 
i) the spectral regions in the QSO spectrum where telluric absorptions are most severe are characterized by a very 
low signal-to-noise, i.e. insufficient to properly correct for the actual detailed shape of the night sky features; 
ii) the observed telluric standard stars were often characterized by a spectral and 
luminosity class for which accurate model atmosphere could not be computed; the derived QSO continuum slope would have hence been 
distorted by the stellar spectral shape. We thus decided to focus on the spectral regions with higher signal, 
where an accurate correction was possible: we assumed the template sky absorption spectrum, scaled to the 
airmass of our QSO spectra and ignored the stability of the sky transparency. This way, we 
could preserve the intrinsic shape of the QSO continua.
The \texttt{telluric} task corrects for the difference in airmass between the science and calibration spectra via 
the Beer-Lambert-Bouguer law. The telluric absorptions are in any case well removed from the spectra of bright QSOs, 
while significant residuals are left in the spectra of faint QSOs at lower S/N.
The relative flux calibration was obtained with the \texttt{sensfunction}, 
\texttt{standard} and \texttt{calib} tasks. 
The instrument sensitivity function was obtained from the observed telluric stars of luminosity class V (giants) and spectral class B. 
For these stars it is possible to compute reliable model atmospheres since effective temperature and surface 
gravity are well estimated.    
The model atmospheres were computed by interpolating the NIR spectral energy distributions available from \citet{Pickles1998} 
in temperature and surface-gravity in order to match the observed spectral type and magnitudes. 
After the relative flux calibration, the individual 1D QSO spectra were averaged to form a single spectrum.
The absolute flux calibration was performed scaling the observed spectra to match the QSOs K-band magnitudes. 
Since the K-band magnitude was not measured for the QSOs in Tab.~\ref{new_source}, 
we derived it from the SDSS QSO template scaled to the observed J and H-band magnitude. 
The whole reduction procedure was extensively tested on some of
the observed telluric standard stars used as reference targets. The recovered spectra match the theoretical model atmospheres typically 
within 10\%, even in the spectral regions more affected by telluric absorptions. The reduced spectra are shown in Fig.~\ref{fig_new}.

\subsection{Literature Data}

The QSOs taken from the literature are summarized in Tab.~\ref{lit_spec}. 
We have collected a total of 22 spectra (19 different sources) which cover all the features of interest 
at sufficiently high S/N to perform the spectral
decomposition (see Sec.~\ref{sec_SN_mass} and Sec.~\ref{sec_SN_met}). The literature sample is composed 
of 10 sources with $z<5.70$ and 9 sources with $5.70 < z < 6.43$.  The 10 QSOs with $z<5.70$ are selected amongst the 13 sources published
by \citet{Iwamuro2002}. These QSOs have redshifts $4.4<z<5.3$ and were 
observed in the J and H bands with an OH-airglow suppressor spectrograph (OH-S), mounted on the Subaru Telescope. 
The spectral resolution ($\lambda/\Delta\lambda$) is equal to 210 (J band) and 420 (H band). 
We also selected 2 of the 4 SDSS QSOs at $z\sim6$ presented by \citet{Iwamuro2004}: 
H and K observations were carried out with the
 Cooled Infrared Spectrograph and Camera (CISCO)  
mounted on the Subaru telescope, with a spectral resolution of 210 and 330 respectively. 
We used the K-band spectrum of the $z=6.43$ QSO J1148+5251 published by \citet{Barth2003}.
 Data were obtained with the NIRSPEC spectrograph on KeckII (spectral resolution
of 1500). We selected 4 of the 5 sources presented by \citet{Kurk2007} at redshifts $z>5.8$ .
 The K-band observations were carried out with VLT-ISAAC with a  spectral resolution of 450. 
We also included the K-band spectrum of J0303-0019 presented by \citet{Kurk2009}.
This faint $z\sim6$ QSO was selected in SDSS Stripe 82 and its spectrum was taken with VLT-ISAAC.
Finally, we added 4 SDSS QSOs at $z>5.8$ published by \citet{Jiang2007}, who observed them with GNIRS on Gemini South with a spectral 
resolution of 800. 

\begin{deluxetable*}{lcccc}
\tablecolumns{5}
\tablewidth{0pc}
\tablecaption{\label{lit_spec}Literature sample sorted by redshift. (1) QSO name; (2), (3) QSO right ascension and declination in J2000.0; 
(4) optical redshift from discovery
paper; (5) NIR spectrum
  reference paper for the NIR spectra: A \citet{Iwamuro2002}, B \citet{Iwamuro2004}, C \citet{Barth2003}, D \citet{Kurk2007}, E \citet{Kurk2009}, 
F \citet{Jiang2007}.}
\tablehead{
\colhead{QSO name} &   \colhead{RA} & \colhead{DEC} & \colhead{z}  & \colhead{Reference for NIR spectra} \\ \colhead{(1)} & \colhead{(2)} & 
\colhead{(3)} & \colhead{(4)} & \colhead{(5)} }
\startdata
BR $1033-0327$ & $10 \ 33 \ 51.47$ & $-03 \ 27 \ 45.5$ & 4.51 & A \\
BR $0019-1522$ & $00 \ 19 \ 35.90$ & $-15 \ 22 \ 16.0$ & 4.53 & A \\
BR $2237-0607$ & $22 \ 37 \ 17.44$ & $-06 \ 07 \ 59.7$ & 4.56 & A \\   
SDSS J$0310-0014$ & $03 \ 10 \ 36.97$ & $-00 \ 14 \ 57.0$ & 4.63 & A \\
SDSS J$1021-0309$ & $10 \ 21 \ 19.16$ & $-03 \ 09 \ 37.2$ & 4.70 & A \\
SDSS J$0210-0018$ & $02 \ 10 \ 43.17$ & $-00 \ 18 \ 18.4$ & 4.77 & A \\
SDSS J$0211-0009$ & $02 \ 11 \ 02.72$ & $-00 \ 09 \ 10.3$ & 4.90 & A \\
PC $1247+3406$ & $12 \ 47 \ 17.79$ & $+34 \ 06 \ 12.7$ & 4.90 & A \\
SDSS J$0338+0021$ & $03 \ 38 \ 29.31$ & $+00 \ 21 \ 56.3$ & 5.00 & A \\
SDSS J$1204-0021$ & $12 \ 04 \ 41.73$ & $-00 \ 21 \ 49.6$ & 5.03 & A \\
SDSS J$0005-0006$ & $00 \ 05 \ 52.30$ & $-00 \ 06 \ 56.0$ & 5.85 & D \\
SDSS J$1411+1217$ & $14 \ 11 \ 11.30$ & $+12 \ 17 \ 37.0$ & 5.93 & D, F \\
SDSS J$1306+0356$ & $13 \ 06 \ 08.20$ & $+03 \ 56 \ 26.0$ & 5.99 & D, F \\
SDSS J$1630+4012$ & $16 \ 30 \ 33.90$ & $+40 \ 12 \ 10.0$ & 6.05 & B \\
SDSS J$0303-0019$ & $03 \ 03 \ 31.40$ & $-00 \ 19 \ 12.9$ & 6.07 & E \\
SDSS J$1623+3112$ & $16 \ 23 \ 31.80$ & $+31 \ 12 \ 01.0$ & 6.22 & F \\
SDSS J$1048+4637$ & $10 \ 48 \ 45.05$ & $+46 \ 37 \ 18.3$ & 6.23 & B \\
SDSS J$1030+0524$ & $10 \ 30 \ 27.10$ & $+05 \ 24 \ 55.0$ & 6.28 & D, F \\
SDSS J$1148+5251$ & $11 \ 48 \ 16.64$ & $+52 \ 51 \ 50.3$ & 6.43 & C \\
\enddata
\end{deluxetable*}

\section{Data Analysis}
We focus on the spectral region with restframe wavelengths $2000$ \AA \ $< \lambda_{rest} < 3500$ \AA. 
This region is characterized by the presence of the \Mgii \ emission line, the underlying non-stellar continuum, 
the Balmer pseudo continuum and the \Feii \ emission line forest. The last three emission features, that are overlapped  
in the spectral range of interest, are described in the following sections.  

\subsection{Power-law}

The dominant component of a QSO spectrum is the non-stellar continuum, modeled as a power-law: 
\begin{equation}
 F_{\lambda} = F_0 \left( \frac{\lambda}{2500 \ \AA}\right)^\alpha 
\end{equation}

Typically, in our data, 
the determination of the slope coefficient $\alpha$ depends on the adopted 
fitting procedure and on the observed spectral range. 
In case of a wide wavelength coverage it is possible to choose fitting windows free of 
contributions by other emission components. For spectra with restricted 
wavelength coverage, the power-law continuum has to be fitted simultaneously with the other components, resulting in a
local estimate of the slope that might not be fully representative 
of the overall continuum shape of the QSO. \citet{Decarli2010} analyzed a sample of 96 QSOs at $z<3$ 
by fitting the power-law continuum in 8 different windows free of strong features, and obtained a mean value 
for the slope of $-1.3$ with a 1-$\sigma$ dispersion of 1.6. \citet{Shen2010} estimated the local slope for a sample of $\sim$ 100.000 
SDSS QSOs at $z<4.95$, fitting the the power-law plus an iron template to the wavelength range around four broad
 emission lines (\Ha, \Hb, \Mgii \ and \Civ). 
In particular, from the analysis of the wavelength range adjacent the \Mgii \ emission line, 
they obtained a mean value for the local slope of $-1.3$ with a 1-$\sigma$ dispersion of 0.8. From our spectral
 decomposition we find consistent values (see Sec.~\ref{sec_fit}).

\subsection{Balmer pseudo-continuum}

We model the Balmer pseudo-continuum following \citet{Dietrich2003}. We assume partially optically thick gas clouds 
with uniform temperature $T_e=15000$ K. 
For wavelengths below the Balmer edge ($\lambda_{BE}=3646$ \AA), the Balmer spectrum can be parametrized as: 
\begin{equation}
 F_{\lambda}=F_{norm} B_{\lambda}(T_{e})(1-e^{-\tau_{BE}(\frac{\lambda}{\lambda_{BE}})^3}), \qquad \lambda<\lambda_{BE}
\end{equation}
where $B_{\lambda}(T_e)$ is the Planck function at the electron temperature $T_{e}$, $\tau_{BE}$ is the optical 
depth at the Balmer edge \citep[we assume $\tau_{BE}=1$ following][]{Kurk2007}, and $F_{norm}$ is the normalized flux 
density at the Balmer edge \citep{Grandi1982}. The normalization should be determined at 
$\lambda_{rest} \backsimeq 3675$ \AA, where no \Feii \ emission is present. Since this wavelength is 
either not covered or has a very low S/N in our sample, we fix the normalization to a fraction of the continuum strength 
extrapolated at 3675 \AA: $F_{norm}=f_B\cdot F_{power-law}(3675$ \AA $)$.    
To define the relative strength of the two components and monitor the effects on the \Feii \ estimate,
 we have run various tests with $f_B=0.1, 0.3, 0.5, 0.8, 1$. Since differences in the \Feii \ estimates resulting from this test 
were less than measured errors (with $f_B$ only partially affecting the power-law normalization), 
we have fixed $f_B=0.3$ based on the results by \citet{Dietrich2003}.      

\subsection{\Feii \ template}

The \Feii \ ion emits a forest of lines, many of which are blended. We fit the \Feii \ forest using 
a modified version of the emission line template by \citet{Vestergaard2001}. 
This template is based on the high resolution 
spectrum of the narrow line Seyfert 1 galaxy PG0050+124 (z=0.061), observed with the Hubble Space Telescope.
Vestergaard \& Wilkes obtained the emission line template by first  fitting and subtracting  
the power-law continuum and all the absorption/emission features from all the elements but Fe.
Afterward an \Feiii \ model was subtracted from the residual to obtain a pure \Feii \ template. 
As \citet{Kurk2007} pointed out, no \Feii \ emission is left in the 2770-2820 \AA \ range, 
due to the \Mgii \ line subtraction. We modified the template 
following \citet{Kurk2007} by adding a constant flux density between 2770 and 2820 \AA \ equal to the 20\% of the 
mean flux density of the template between 2930 and 2970 \AA. The justification for this operation comes from the 
theoretical \Feii \ emission line strength by \citet{Sigut2003}. Since the \Feii \ and the \Mgii \ are not distributed in the same way
 within the BLR, we decided not to fix the Doppler broadening of the \Feii \ template to the one measured for the \Mgii \ emission line. 
We have instead run many tests broadening the \Feii \ template by convolving it with Gaussian profiles with 
constant FWHM $=7.5, \ 15, \ 22.5$ \AA, corresponding to FWHM $\sim 930, \ 1860, \ 2800$ km$/$s at 2400 \AA \ respectively. 
A Gaussian broadening with $\sigma$ constant in wavelength leads to slightly different velocities over the fitting range, 
but we have found no significant differences in the measured \Feii \  normalization in function of the three broadening.
Given the typical S/N of our spectra the \Feii \ normalization is in fact mainly determined by the fit of 
the template broad bumps rather than individual \Feii \ features. We finally chose to fix FWHM$=15$ \AA \ 
for the \Feii \ template.

\subsection{Fitting procedure}
\label{sec_fit}
\begin{center}
\begin{figure}[h]
\label{fig_fit_errors}
\resizebox{0.45\textwidth}{!}{\includegraphics{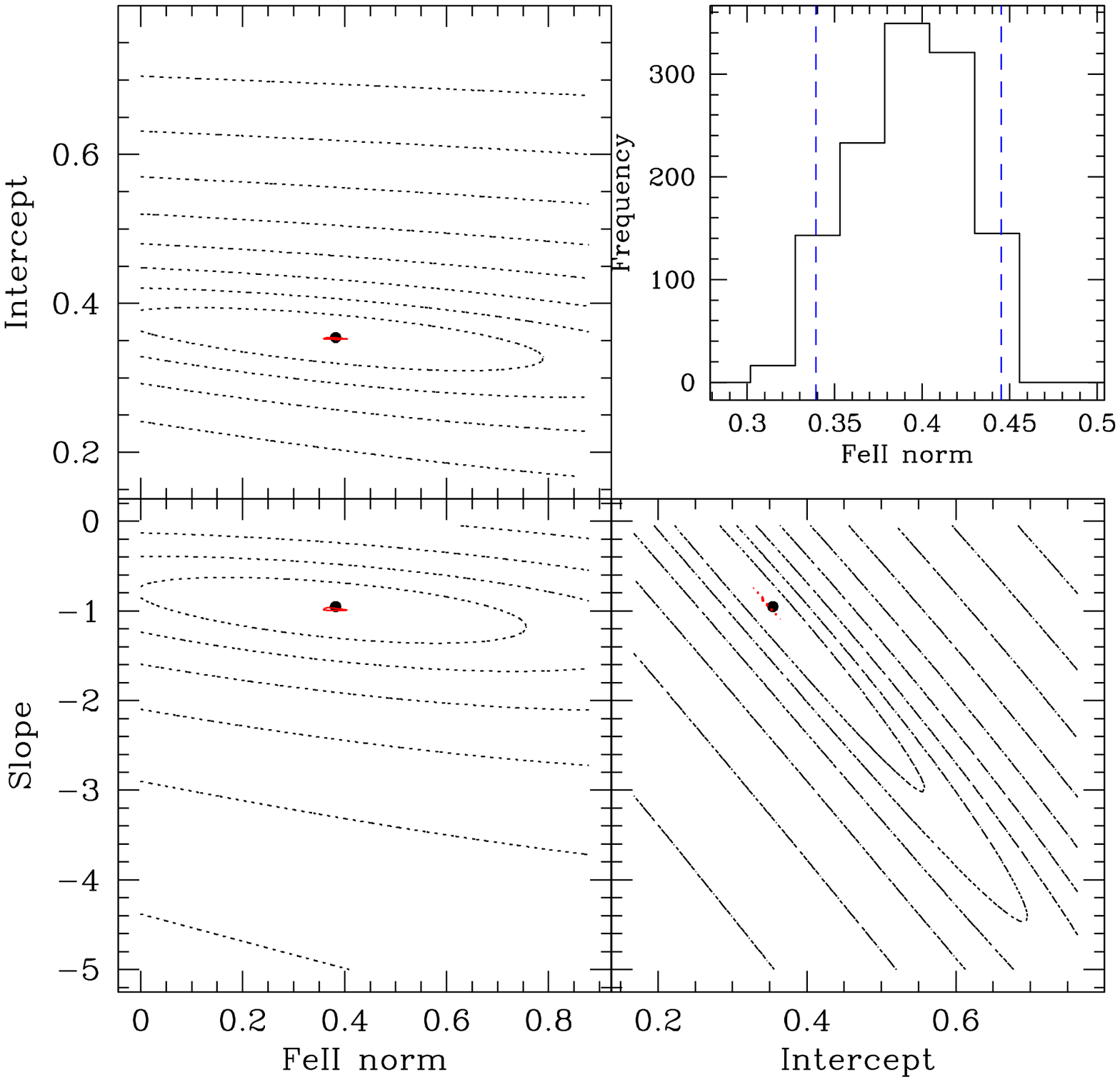}}
\resizebox{0.45\textwidth}{!}{\includegraphics{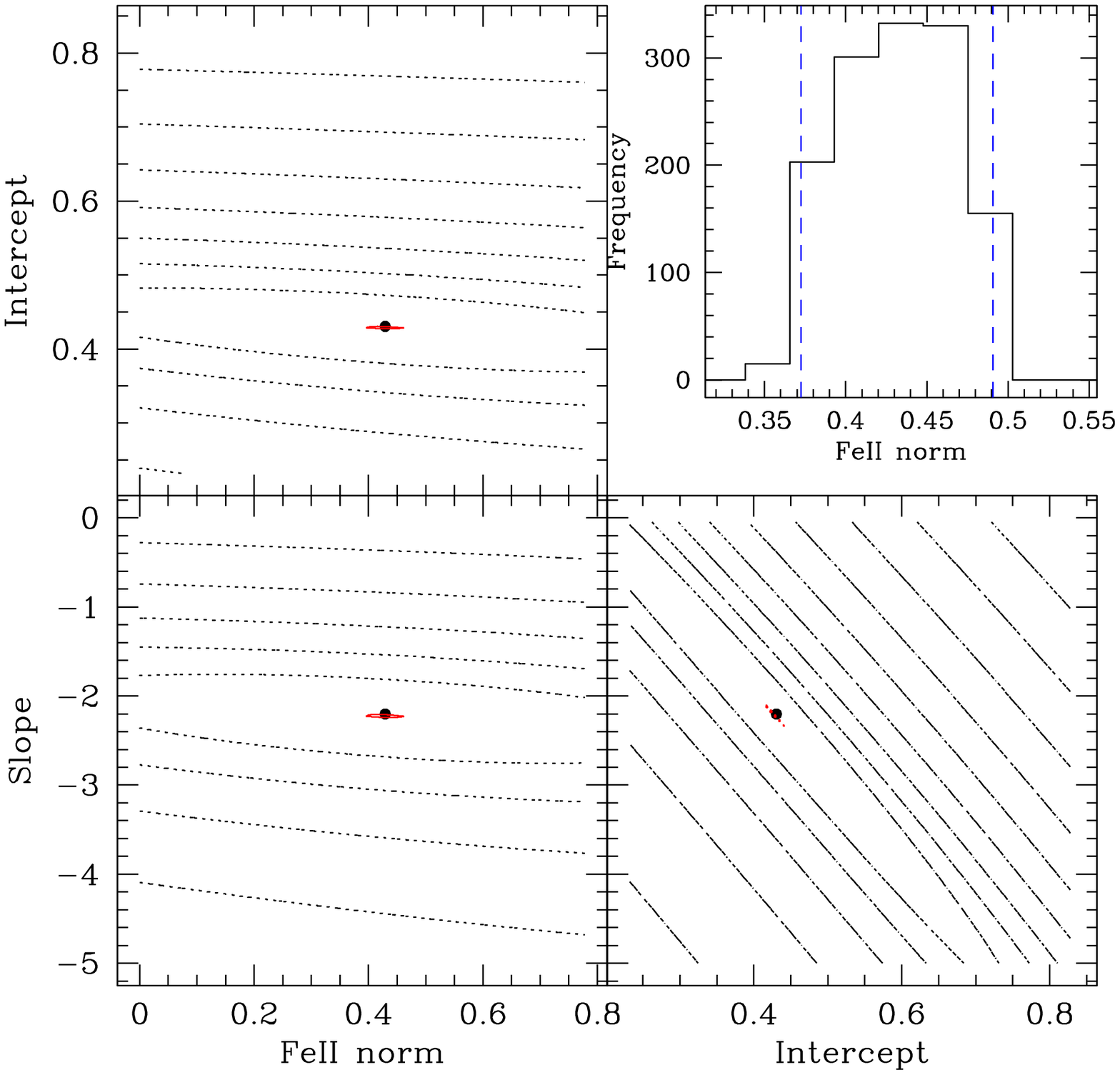}}
\caption{The \Feii \ normalization, obtained from the fit of the first set of components, depends on the power-law 
slope and its normalization (intercept). SDSS J0842+1218 (top panel) and SDSS J0353+0104 (bottom panel): $\chi^2$ domain analysis for \Feii \ error computation. 
 In each panel we show: a) two dimensional projections of the 3D $\chi^2$-surfaces (\Feii \ normalization vs Intercept, upper-left plot;
\Feii \ normalization vs Slope, bottom-left plot; Intercept vs slope, bottom-right plot): 
contours represent iso-$\chi^2$ 
 levels spaced by a factor of 2 while the best fit case is marked with a dot; b) probability distribution for the \Feii \ 
template normalization (upper-right plot): the distribution has been  
 obtained by marginalizing the 3-D probability distribution considering only the triplets  
for which $\chi^2 -\chi_{min}^2< 1$, the dashed vertical lines mark our estimate of the 1-$\sigma$ confidence 
level.}
\end{figure}
\end{center}

All the spectra have been shifted to the rest-frame system of reference using optical redshifts. 
Even if the optical redshifts are slightly different from the NIR ones, this will not affect our results 
since the \Mgii \ peak wavelength is a free parameter in our fitting procedure. Moreover,  
the \Feii \ template is constituted by blended multiplets (broadened by the convolution with Gaussian profiles 
with FWHM $=15$ \AA), for which the definition of peak wavelength is not straightforward. 
We tested a posteriori the difference between the optical redshifts (see Tab.~\ref{new_source} and Tab.~\ref{lit_spec}) 
and the ones measured from the \Mgii \ peak wavelength (see Tab.~\ref{tab_mass}). The average difference is equal to 
$\Delta$z$=$0.02 with a maximum of $\Delta$z$=$0.08 for J$1411+1217$ \citep[][online Appendix Fig.~A.12]{Jiang2007}
: in this case the \Mgii \ line profile is 
severely affected by the atmospheric absorptions.
Our spectra have different wavelength coverage because of the different redshifts of the sources and of the various  
instruments used to collect the data. Similarly
 the sky contamination varies across the sample with redshift. For these reasons it is 
impossible to define fixed fitting windows for the entire sample. We focus on the rest-frame region 
within 2000 and 3500 \AA, choosing the fitting windows in a way as homogeneous as possible, as a function of the 
spectral coverage and of the sky contamination.     
The fit is performed in two steps. A first set of spectral components is given by the sum of the power-law continuum, the Balmer 
pseudo-continuum and the \Feii \ template. The free parameters are the power-law slope ($\alpha$) and 
its normalization ($\beta=log(F_0)$)  
and the \Feii \ template normalization ($\gamma$). Since the three components are overlapped in this wavelength range, 
the fit is performed with a $\chi^2$ minimization on a suitable grid in the parameter space.  
In this way, we minimize the possibility that our solution represents only a local minimum of the $\chi^2$ domain. 
The errors on the \Feii \ normalization are computed by marginalizing the probability distributions 
in 3-D parameter space, selecting 
all the cases for which $\chi^2 -\chi_{min}^2< 1$ (1$\sigma$ confidence level). To ensure a reliable estimate of these 
errors we need enough triplets satisfying the $\chi^2$ condition. After many tests we decided to sample the parameter space with 16 
million points: 100 ($\alpha$) $\times$ 100 ($\beta$) $\times$ 1600 ($\gamma$). This way, for each fit, 
we sample the $\gamma$ parameter 
space with an average bin of $0.001\times 10^{-17}$ erg s$^{-1}$ cm$^{-2}$ and we obtain 
$\sim 100$ triplets satisfying the $\chi^2$, with a minimum of 15 (for the QSO BR $1033-0327$). 
This condition is not satisfied for J$0303-0019$ \citep[][online Appendix Fig.~A.17]{Kurk2009}: in this case the \Feii \ major 
features are severely affected by the telluric absorptions, resulting in a weaker constraint on the \Feii \ 
template normalization which is in any case consistent with 0 (see Tab.~\ref{tab_fit}).
In Fig.~\ref{fig_fit_errors} we show two examples of 
the 3D $\chi^2$-cube projections and of the relative probability distribution for the \Feii \ 
template normalization (similar plots for all sources discussed here are shown in the online Appendix). 
The $\chi^2$-maps are overall regular (non-patchy), implying the absence of secondary 
local minina. The degeneracy between power-law slope and intercept is evident from the bottom-right plots in the 
two panels.  

The fitted components are then subtracted and we proceed to fit the \Mgii \ emission line. 
The \Mgii \ line fit is performed with a least-squares procedure.
Since the \Mgii \ doublet is not resolved in the majority of our spectra, we model the emission 
line as a simple Gaussian (three free parameters: central wavelength, width and normalization). 
If there was a significant narrow second component that we do not resolve, this would lead to a slight underestimate 
of the black-hole mass.  

\begin{center}
\begin{figure*}[h]
\label{fig_new}
\begin{center}
\resizebox{0.45\textwidth}{!}{\includegraphics{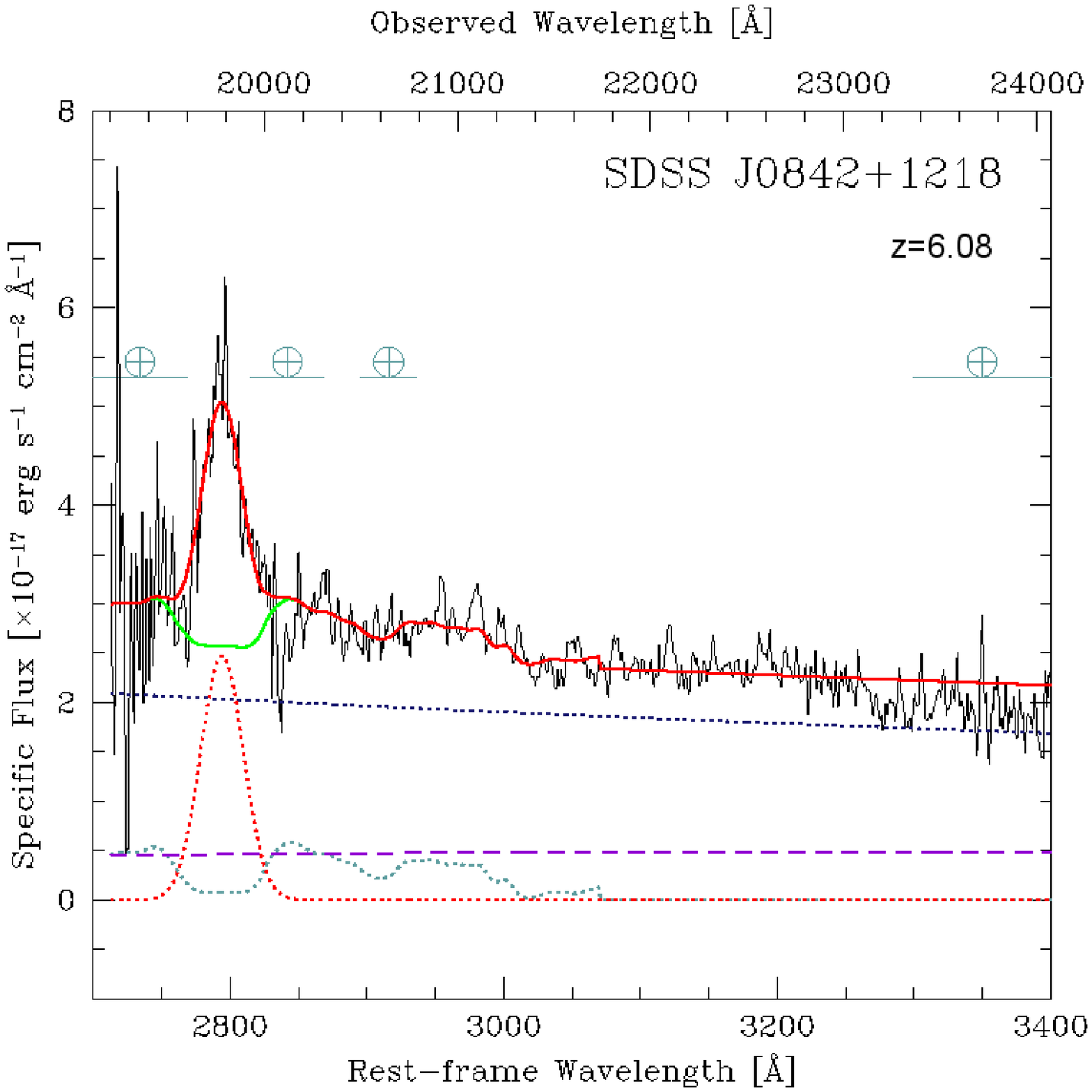}}
\resizebox{0.45\textwidth}{!}{\includegraphics{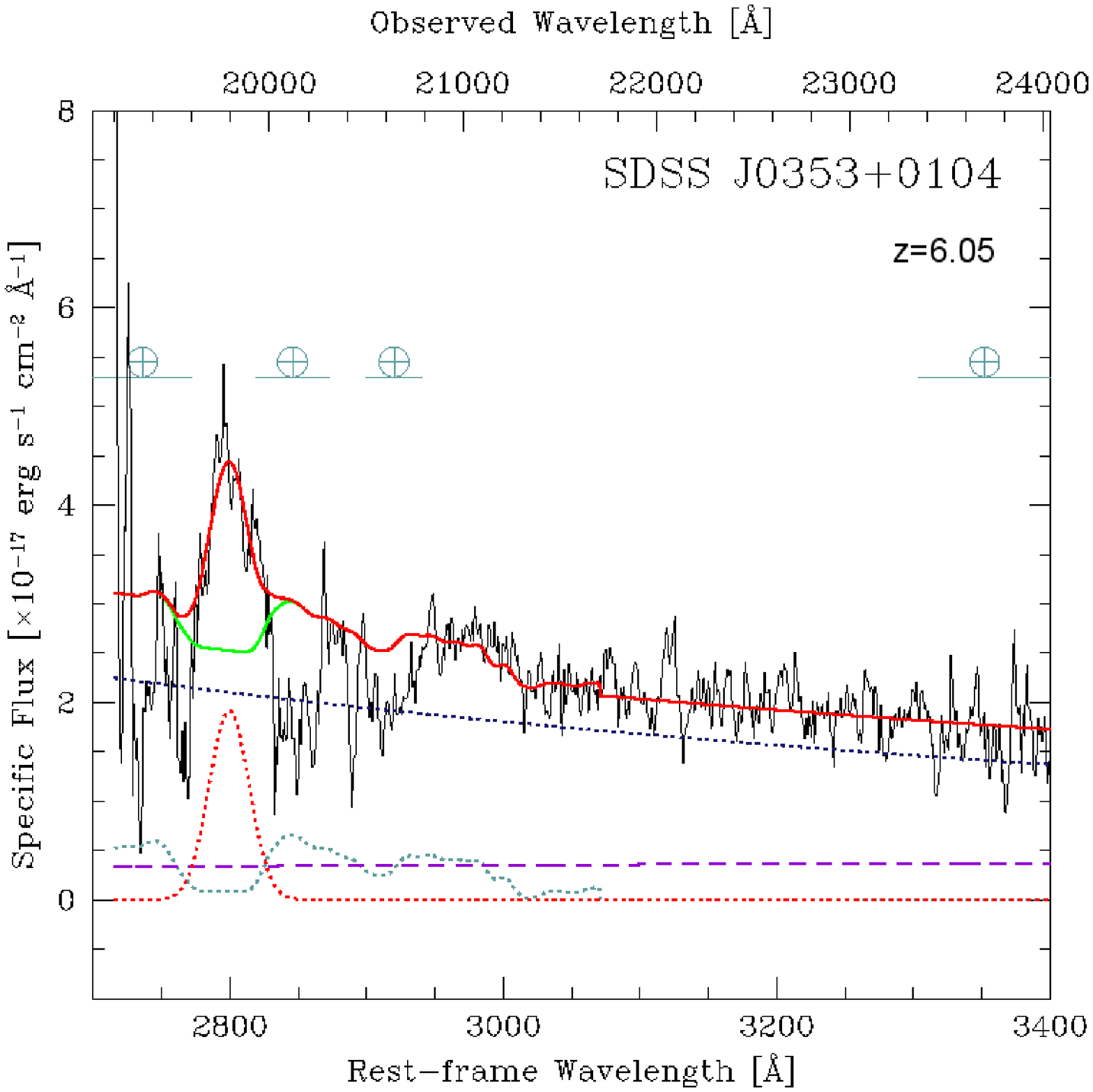}}

\begin{center}
\resizebox{0.45\textwidth}{!}{\includegraphics{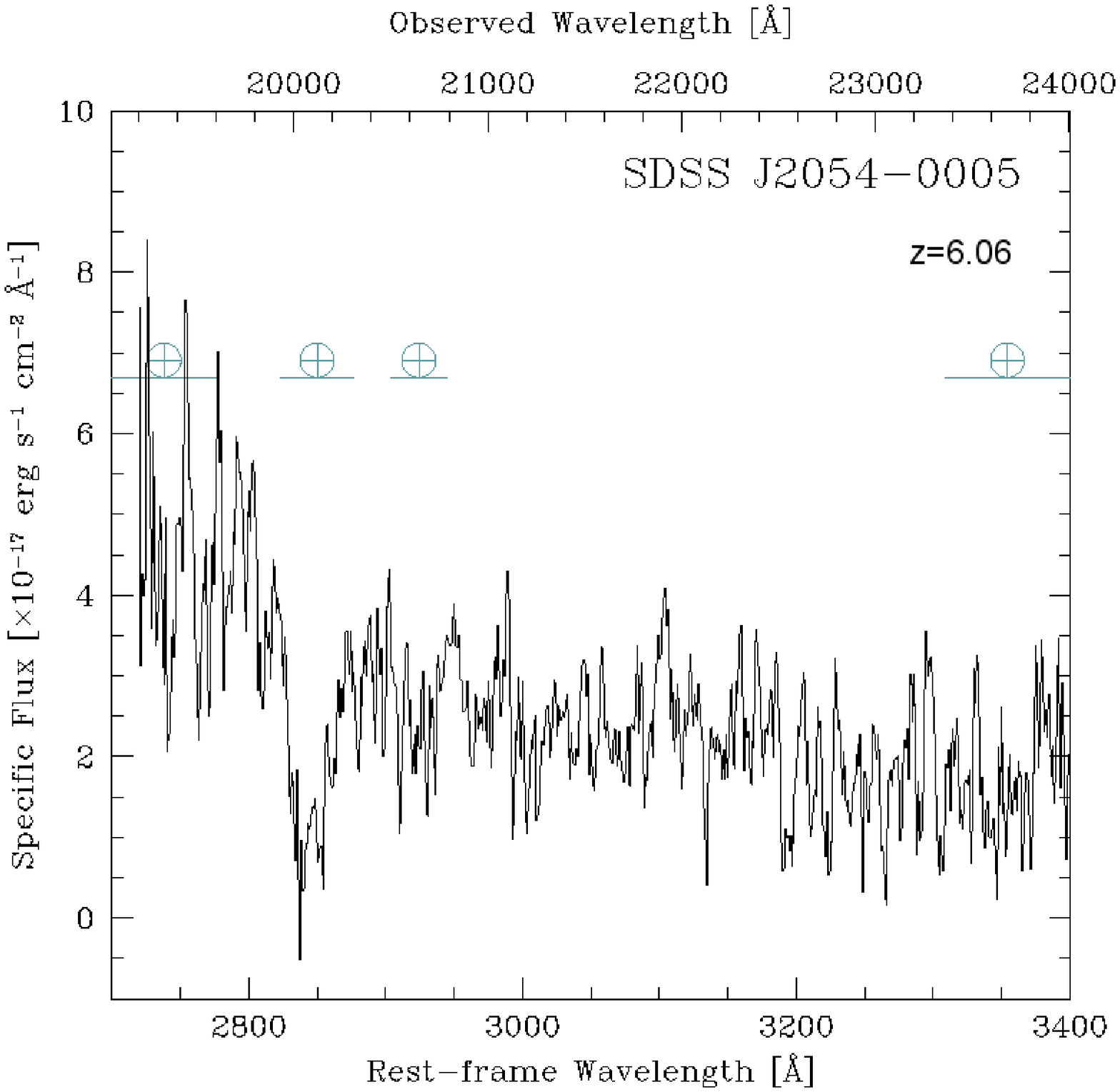}}
\end{center}
\caption{Reduced spectra of the QSOs SDSS J0842+1218 (upper-left panel), SDSS J0353+0104 (upper-right panel) and SDSS J2054-0005 (lower panel). 
The observed spectra are shown in black continuous line. The modeled components are: power-law continuum (blue dotted 
line), Balmer pseudo continuum (purple dashed line), \Feii \ normalized template (light blue dotted line), 
\Mgii \ emission line (red dotted line). The sum of the first set of components (power-law continuum $+$ Balmer pseudo continuum $+$ 
\Feii \ normalized template) is overplotted to the spectrum as a green solid line, while the sum of all the components is over plotted 
as a red solid line. 
The spectral decomposition was not performed for the weak line QSO SDSS J2054-0005. 
Telluric absorption bands are indicated over the spectra with the symbol  $\Earth$: they are extracted from the ESO sky absorption spectrum measured on 
the Paranal site at a nominal airmass of 1.
}
\end{center}
\end{figure*}
\end{center}

Examples of the spectral decomposition are shown in Fig.~\ref{fig_new}. The results of the fit and relative 
$\chi^2$ maps for the literature sample are shown in the online Appendix. 
The fitted parameters are listed in Tab.~\ref{tab_fit}. Even if we do not overcome the degeneracy between the power-law 
slope and intercept, we can compare the distribution of our local slope estimates 
with those in the literature. We obtain a mean value $\alpha=-1.5\pm1.2$ which is in agreement within the uncertainties with both the 
local slope estimate by \citet{Shen2008} and with the global one by \citet{Decarli2010}.       

\section{Results}

\subsection{Black hole mass estimate}
\label{sec_MBH}
We estimate the \Mbh \  using scaling 
relations, calibrated on local AGNs, that are based on broad emission line widths and continuum luminosities. 
Under the assumption that the dynamics of the broad line region is dominated by the gravity of the central BH, 
the virial theorem states:
\begin{equation}
 M_{BH}=G^{-1} \ R_{BLR} \ v^2_{BLR}
\end{equation}
where \Mbh \ is the black hole mass, $R_{BLR}$ is the characteristic radius of the BLR and $v_{BLR}$ is the 
orbital velocity of the clouds emitting at $R_{BLR}$. The cloud velocity can be obtained from the width of the broad emission lines: $v_{BLR} = f \cdot 
FWHM$, where $f$ is a geometrical factor that accounts for the de-projection of $v_{BLR}$ from the line of sight 
\citep[see][]{McGill2008, Decarli2008}, and FWHM is the full width at half maximum of the line profile. 
Even if the BLR size cannot be directly measured by single epoch spectra, it can still be evaluated since it is strongly 
correlated with the continuum luminosity of the AGN \citep[e.g.][]{Kaspi2000}. It is then possible to estimate 
\Mbh \  for high-redshift QSOs using a single spectrum covering the blended \Mgii \ line doublet ($\lambda=2795,2803$ \AA) 
and the redward continuum ($\lambda=3000$ \AA). In particular for $R_{BLR}$ we use the relation provided by \citet{McLure2002}: 
\begin{equation}
\label{eq_mass_1}
 \frac{R_{BLR}(\rm{Mg \ II})}{10 \ \rm{light-days}} = (2.52 \pm 0.03) \left[ \frac{\lambda L_{\lambda}(3000  \ \rm{\AA})}{10^{44} \ \rm{erg \ s}^{-1}} \right] ^{0.47}
\end{equation}
and for the geometrical factor the value provided by \citet{Decarli2008}: $f($\Mgii$)=1.6$, obtained assuming that  
the \Mgii \ and \Hb-emitting regions have a similar geometry. 
All these relations are based on low redshift 
objects ($z<0.3$), and the underlying assumption here is that they are still valid at high-redshift. 
Using relation (\ref{eq_mass_1}) 
 \citet{McLure2002} reproduced \Mbh \  obtained with the full reverberation mapping method with an accuracy
of 0.4 dex. This intrinsic scatter of the estimator dominates the measurement uncertainties. 
To compare our results with the ones published in previous works on $z\sim6$ sources \citep{Barth2003, Jiang2007, Kurk2007, Kurk2009}, 
we also estimate \Mbh \ using the relation obtained by \citet{McLure2004}:
\begin{equation}
\label{eq_mass_2}
 \frac{M_{BH}}{M_\odot}= 3.2 \left[ \frac{\lambda L_\lambda (3000 \ \rm{\AA})}{10^{37} \ \rm{W}} \right] ^{0.62} \left[\frac{FWHM(\rm{Mg \ II})}{\rm{km \ s}^{-1}}\right]^2
\end{equation}
from a sample of 17 low redshift AGNs ($z<0.7$), 
with luminosities comparable to those of high-z 
QSOs ($\lambda L_\lambda> 10^{44} $ erg s$^{-1}$). 

From the estimated \Mbh \  we compute the QSO Eddington ratios defined as the ratio between the measured  
bolometric luminosity $L_{bol}$ and the theoretical Eddington luminosity $L_{Edd}$, as computed from the \Mbh. 
We obtain the observed monochromatic luminosity 
$\lambda L_\lambda(3000 \rm{\AA})$ from the continuum component of the fit $f_\lambda(3000 \rm{\AA})$ 
and from the luminosity distance $D_L$ as:
\begin{equation}
 \lambda L_\lambda(3000 \rm{\AA})=\lambda \ F_\lambda(3000 \rm{\AA}) \ 4\pi D_L^2
\end{equation}
\noindent
Since $L(3000 \rm{\AA})$ is only a fraction of the total electromagnetic luminosity coming from the QSO, 
we apply the bolometric correction by \citet{Shen2008} to obtain $L_{bol}$: 
\begin{equation}
\label{eq_bol}
L_{bol}=5.15 \ \lambda L_\lambda(3000 \rm{\AA})
\end{equation}
The Eddington luminosity is defined as the maximum luminosity attainable, at which the radiation 
pressure acting on the gas counterbalances the gravitational attraction of the BH:
\begin{equation}
 L_{Edd}=1.3 \cdot 10^{38} \left(\frac{M_{BH}}{M_\odot}\right) \rm{erg \ s^{-1}}
\end{equation}

The obtained \Mbh \  and the relative QSO Eddington ratios are listed in Tab.~\ref{tab_mass}. 
The two relations, Eq.~\ref{eq_mass_1} and Eq.~\ref{eq_mass_2}, lead to a difference in the 
mass estimates of a factor $\sim 1.7$. The results obtained for the sample of published sources via 
Eq.~\ref{eq_mass_2} are in good agreement 
with previous estimates, indicating that different fitting procedures imply variations 
smaller than the errors due to the intrinsic scatter of the estimator. 
The average difference between the BH masses estimated by us and the ones previously published is equal to 0.1 dex.
Only in the case of the QSO SDSS J0005-0006 \citep[][online Appendix Fig.~A.11]{Kurk2007} this difference is higher (0.7 dex): 
in this case the \Mgii \ line is severely affected by the atmospheric absorption and the line fit becomes 
significantly procedure-dependent. 

\begin{deluxetable*}{lccccc}
\tablecolumns{6}
\tablewidth{0pc}
\tablecaption{\label{tab_fit} Fitted spectral properties. 
The spectral decomposition has been performed on the specific fluxes $F_{\lambda}$
 in units of $10^{-17}$ erg s$^{-1}$ cm$^{-2}$ \AA$^{-1}$.
(1) QSO name. For sources with multiple observations: $^*$, \citet{Jiang2007}; $^+$, \citet{Kurk2007}.; 
(2) power-law slope; (3) \Feii \ normalization; (4) \Mgii \ central wavelength 
$\left[ \rm \AA \right]$; 
(5) FWHM $\left[ \rm {km \ s}^{-1}\right]$; 
(6) measured specific flux of the continuum at 3000 \AA, in units of  $10^{-17}$ erg s$^{-1}$ cm$^{-2}$ \AA$^{-1}$. 
J$0303-0019$: the \Feii \ major features are severely affected by the telluric absorptions, 
resulting in a weaker constraint on the \Feii \ 
template normalization which is in any case consistent with 0.}     
\tablehead{
\colhead{QSO name} &   \colhead{slope} & \colhead{\Feii \ norm} & \colhead{\Mgii \ $\lambda_{peak}$}  
 & \colhead{\Mgii \ FWHM} & \colhead{$F_{\lambda}\left(3000 \rm \AA\right)$} \\ \colhead{(1)} 
& \colhead{(2)} & \colhead{(3)} & \colhead{(4)} & \colhead{(5)} & \colhead{(6)}} 
\startdata
BR $1033-0327$    & -2.4 & 0.84$\pm$0.02 & 2804.6$\pm$2.3 & 4248$\pm$120 & 8.20$\pm$1.59 \\ 
BR $0019-1522$ 	  & -2.3 & 1.01$\pm$0.03 & 2802.6$\pm$0.6 & 4317$\pm$120 & 7.78$\pm$1.50 \\ 
BR $2237-0607$ 	  & -1.3 & 1.57$\pm$0.36 & 2796.3$\pm$0.6 & 3811$\pm$163 & 19.98$\pm$0.12 \\
SDSS J$0310-0014$ & -1.5 & 0.24$\pm$0.05 & 2797.3$\pm$1.1 & 4087$\pm$260 & 2.83$\pm$0.05 \\ 
SDSS J$1021-0309$ & -1.7 & 0.26$\pm$0.09 & 2797.4$\pm$0.6 & 3100$\pm$162 & 2.76$\pm$0.09 \\ 
SDSS J$0210-0018$ & -2.7 & 0.11$\pm$0.06 & 2798.4$\pm$2.1 & 6543$\pm$803 & 3.42$\pm$0.05 \\ 
SDSS J$0211-0009$ & -2.4 & 0.38$\pm$0.05 & 2795.7$\pm$1.5 & 5975$\pm$468 & 1.44$\pm$0.28 \\ 
PC $1247+3406$	  & -2.2 & 0.30$\pm$0.10 & 2797.7$\pm$0.7 & 4094$\pm$197 & 3.86$\pm$0.08 \\
SDSS J$0338+0021$ & -1.8 & 0.51$\pm$0.05 & 2798.4$\pm$0.4 & 2969$\pm$128 & 3.20$\pm$0.05 \\ 
SDSS J$1204-0021$ & -2.3 & 0.83$\pm$0.03 & 2796.2$\pm$3.0 & 5753$\pm$208 & 4.36$\pm$0.04 \\ 
SDSS J$0005-0006$ & 0.0  & 0.18$\pm$0.04 & 2796.2$\pm$0.1 & 1036$\pm$65 & 0.94$\pm$0.02 \\
SDSS J$1411+1217^*$ & -1.7 & 0.20$\pm$0.07 & 2800.4$\pm$1.2 & 2208$\pm$317 & 2.39$\pm$0.06 \\ 
SDSS J$1411+1217^+$ & 1.2  & 0.36$\pm$0.04 & 2787.8$\pm$0.4 & 2824$\pm$168 & 3.03$\pm$0.05 \\ 
SDSS J$1306+0356^*$ & -2.3 & 0.21$\pm$0.12 & 2809.5$\pm$0.5 & 3158$\pm$145 & 1.92$\pm$0.05 \\ 
SDSS J$1306+0356^+$ & -2.5 & 0.13$\pm$0.13 & 2809.8$\pm$1.1 & 4134$\pm$351 & 1.97$\pm$0.04 \\ 
SDSS J$1630+4012$ & -0.7 & 0.03$\pm$0.03 & 2798.0$\pm$2.0 & 3366$\pm$533 & 1.46$\pm$0.30 \\ 
SDSS J$0303-0019$ & 2.0 & 0.00$+$0.15 & 2802.28$\pm$0.4 & 2307$\pm$63 & 0.60$\pm$0.01 \\ 
SDSS J$1623+3112$ & 0.0 & 0.31$\pm$0.02 & 2795.0$\pm$0.3 & 3587$\pm$118 & 1.81$\pm$0.05 \\ 
SDSS J$1048+4637$ & -1.7 & 0.56$\pm$0.03 & 2797.9$\pm$1.4 & 3366$\pm$532 & 4.22$\pm$0.04 \\ 
SDSS J$1030+0524^*$ & -1.9 & 0.16$\pm$0.07 & 2807.2$\pm$0.6 & 3449$\pm$151 & 1.76$\pm$0.05 \\ 
SDSS J$1030+0524^+$ & -1.3 & 0.17$\pm$0.08 & 2805.9$\pm$0.9 &3704$\pm$231 & 1.88$\pm$0.03 \\
SDSS J$1148+5251$ & -2.1 & 0.91$\pm$0.01 & 2793.7$\pm$0.7 & 5352$\pm$289 & 4.22$\pm$0.08 \\ 
SDSS J$0353+0104$ & -2.2 & 0.43$\pm$0.06 & 2799.6$\pm$0.7 & 3682$\pm$281 & 2.16$\pm$0.10 \\ 
SDSS J$0842+1218$ & -0.9 & 0.38$\pm$0.06 & 2794.2$\pm$0.7 & 3931$\pm$257 & 2.38$\pm$0.12 \\ 
\enddata
\end{deluxetable*}

\subsection{Dependence on the S/N and on the systematics in the observations}
\label{sec_SN_mass}
\begin{figure}[b]
\begin{center}
\label{fig_SN_mass}
\begin{center}
 \resizebox{0.3\textwidth}{!}{\includegraphics[bb=18 144 592 718]{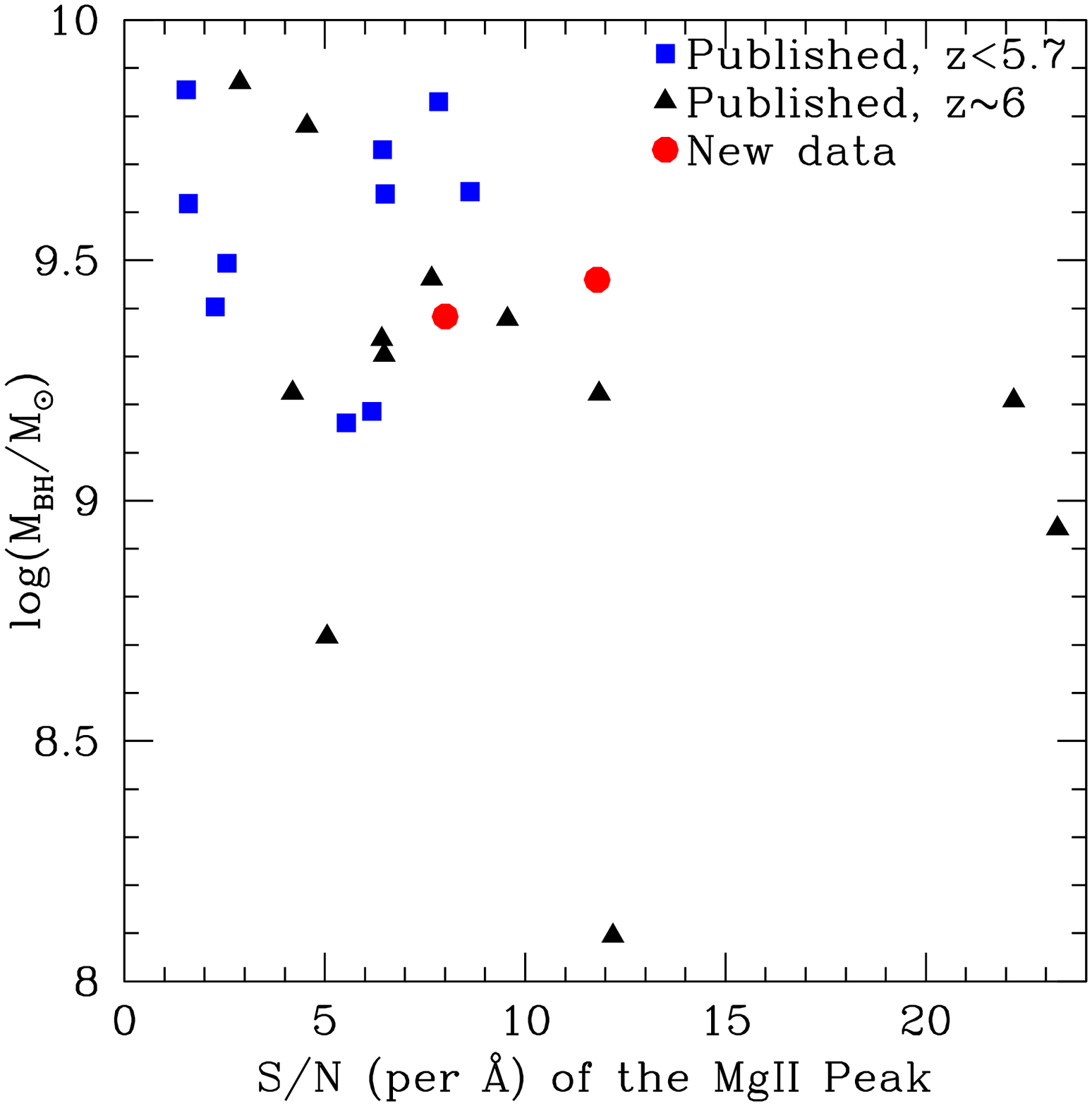}}
\end{center}
\end{center}
 \caption{\Mbh \ estimates as a function of the S/N per \AA \ at the \Mgii \ peak. Blue squares: literature sample,  
at $z<5.7$ \citep{Iwamuro2002}; black triangles: literature sample, $z\sim6$ \citep{Barth2003, Iwamuro2004, Jiang2007,
 Kurk2007, Kurk2009};  red circles: this study. Our \Mbh \ estimates do not depend on the S/N of the QSO 
spectra.}
\end{figure}

In Fig.~\ref{fig_SN_mass} we show the \Mbh \ obtained via Eq.~\ref{eq_mass_1} as a function of the 
S/N per \AA \ of the \Mgii \ peak. The noise was derived from regions of the spectra free of line emission.
There is no evidence of any dependence of the \Mbh \ estimates on the S/N of the \Mgii \ line peak. 
For SDSS J$1411+1217$, SDSS J$1306+0356$ and SDSS J$1030+0524$ we are in the fortunate situation of having 
two independent observations by \citet{Jiang2007}, with Gemini-GNIRS, and by \citet{Kurk2007}, with VLT-ISAAC. 
We can thus use these to analyze the dependence of the \Mbh \ estimates on the systematics in the observations. 
For these 3 sources \citet{Jiang2007} and \citet{Kurk2007} derived (using Eq.~\ref{eq_mass_2}) 0.6, 1.1, 1.0 
$\times10^9 \ M_\odot$ and 1.1, 2.4, 1.4 $\times10^9 \ M_\odot$, respectively. 
Using our fitting results and Eq.~\ref{eq_mass_2}, 
we obtain 0.5, 0.9, 1.1 $\times10^9 \ M_\odot$ and 1.0, 1.7, 1.4 $\times10^9 \ M_\odot$ respectively, 
in good agreement with the published values. 
For those objects whose redshifts make their \Mgii \ line fall on top of atmospheric absorption bands,
the \Mbh \ estimated from the two independent spectra can be significantly different 
(up to 0.3 dex in the case of SDSS J$1306+0356$): therefore we argue that the 
atmospheric contamination is a major contributor to systematics.

\subsection{Comparison with SDSS QSO sample at $0.35<z<2.25$}  
\label{mass_ev}
To study the evolution of the BH population with cosmic time, 
we compare our sources with a sample of SDSS QSOs at lower redshift 
\citep[SDSS Data Release 7,][]{Shen2010}. 
For objects with multiple observations we consider the weighted mean of the individual measurements.
Since the \Mbh \  estimates significantly depend on the adopted estimator 
\citep[see][]{Peterson2010,Shen2008}, we select among the $\sim$100,000 QSOs in the SDSS DR7 sample those  
with measurements of the \Mgii \ FWHM and of $\lambda L_\lambda(3000 \rm \AA)$: i.e., a subset of $\sim$47,000 sources with 
$0.35<z<2.25$. We also consider a sample of nine $z\sim6$ QSOs at lower luminosities ($M_{1450}<-24$) 
from \citet{Willott2010a}, with published 
measurements of the \Mgii \ FWHM and of $\lambda L_\lambda(3000 \rm \AA)$.  
We obtain $L_{bol}$ using the  
bolometric correction by \citet{Shen2008} (see Eq.~\ref{eq_bol}) and we compute \Mbh \ using Eq.~\ref{eq_mass_1}, in analogy 
with the analysis performed in our sample. In Fig.~\ref{fig_mass_shen} we plot the estimated \Mbh \ as a function of the bolometric luminosity for 
the three samples. The black solid lines indicate regions of the parameter space with constant Eddington ratios: 
$L_{bol}/L_{Edd}=0.01, \ 0.1, \ 1.0$. 
The $0.35<z<2.25$ and our high-z QSO populations occupy different regions of the parameter space: the $0.35<z<2.25$
redshift population shows an average $\langle log(L_{bol}/L_{Edd})\rangle\sim-1.2$ dex ($L_{bol}/L_{Edd}\sim$0.06) 
with a dispersion of 0.3 dex, while the typical ratio for the high-z QSOs is $\langle log(L_{bol}/L_{Edd})\rangle
\sim-0.3$ dex ($L_{bol}/L_{Edd}\sim$0.5) with a dispersion of 0.3 dex.
The lower luminosity $z\sim6$ sample \citep{Willott2010a} is characterized by 
lower \Mbh \ but comparable Eddington ratios than the most luminous SDSS QSOs at the same redshift. 
We therefore conclude that the average Eddington ratio for the $z\sim6$ QSO sample studied 
to date is significantly higher than the typical SDSS QSOs at $0.35<z<2.25$.

We have to specify that \citet{Shen2010} fit the spectra considering the power-law 
non--stellar continuum, the \Feii \ line forest \citep[modeled using the template by][]{Vestergaard2001}, 
and both the narrow (simple Gaussian) and broad (simple/double Gaussian) components of the \Mgii \ doublet. 
Since in our sample we are not subtracting the narrow component of the \Mgii \ doublet (not resolved), 
the mass estimates of the high-redshift QSOs may be systematically biased 
towards slightly lower values with respect to the ones in the $0.35<z<2.25$ redshift sample.
{Nevertheless, we note that our fitting procedure yields results consistent with the ones published by 
\citet{Jiang2007} who, except for the cases in which the line profile 
is severely affected by the atmospheric absorption, fits the \Mgii \ emission 
line with a double Gaussian instead of a single one.
Therefore, we conclude that the \Mbh \ estimates do not strongly depend on the adopted fitting procedure.

\begin{center}
\begin{figure*}[t]
\label{fig_mass_shen}
\begin{center}
\resizebox{0.7\textwidth}{!}{\includegraphics{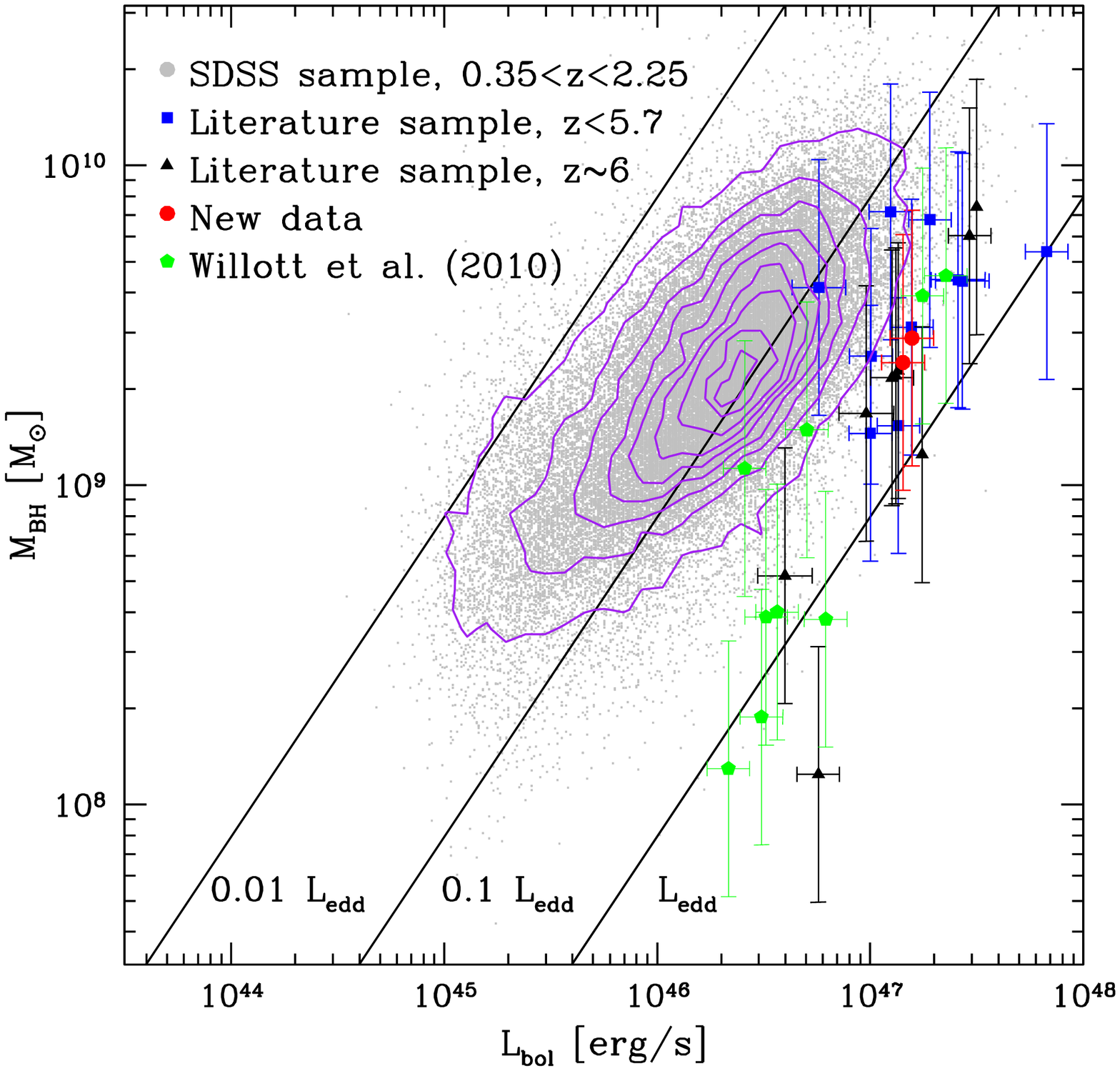}}
\end{center}
\caption{\Mbh \  as a function of the bolometric luminosity: 
comparison between the present dataset and the SDSS QSOs sample at $0.35<z<2.25$. Blue squares: literature 
sample, $z<5.7$ \citep{Iwamuro2002}; black triangles: literature sample, $z\sim6$ \citep{Barth2003, Iwamuro2004, Jiang2007,
 Kurk2007, Kurk2009}; red diamonds: new observations; green pentagons: CFHQS QSO sample \citep{Willott2010}; 
grey dots: SDSS QSOs sample $0.35<z<2.25$ \citep[DR7,][]{Shen2010}; purple lines: isodensity contours. 
The black solid lines indicate regions of the parameter
space with constant Eddington ratios: $L_{bol}/L_{Edd}=0.01, \ 0.1, \ 1.0$. The $0.35<z<2.25$ and our high-z QSO 
populations occupy different regions of the parameter space: the $0.35<z<2.25$
redshift population shows $\langle log(L_{bol}/L_{Edd})\rangle\sim-1.2$ dex with a dispersion of 0.3 dex,
 while the typical ratio for the high-z QSOs is $\langle log(L_{bol}/L_{Edd})\rangle
\sim-0.3$ dex with a dispersion of 0.3 dex.
}
\end{figure*}
\end{center}

\subsection{Comparison with luminosity matched SDSS QSO sample}
In order to perform a statistically robust comparison we now extract only objects with the same (high) bolometric 
luminosity: $10^{47}<L_{bol} \ \left[erg \ s^{-1} \right] <3\times 10^{47}$
 from our sample and from the SDSS comparison 
sample at $0.35<z<2.25$.
In Fig.~\ref{fig_mass_er} we plot the \Mbh \ and the Edddington ratio as a function of redshift. 
The average \Mbh \  at high-z is systematically lower than the 
typical \Mbh \  of the $0.35<z<2.25$ population at similar luminosities, while the Eddington ratio is higher. 
Comparing the Eddington ratio distribution of the two luminosity matched samples (Fig.~\ref{fig_smirnov}, 
$10^{47}<L_{bol} \ \left[erg \ s^{-1} \right] <3\times 10^{47}$) we obtain that 
for the $0.35<z<2.25$ population $\langle log(L_{bol}/L_{Edd})\rangle
\sim-0.80$ dex ($L_{bol}/L_{Edd}\sim$0.16) with a dispersion of 0.24 dex, 
while for the high-z one $\langle log(L_{bol}/L_{Edd})\rangle
\sim-0.37$ dex ($L_{bol}/L_{Edd}\sim$0.43) with a dispersion of 0.20 dex. If we perform a Kolmogorov-Smirnov test 
on the two distributions we obtain that
the probability that they come from the same parent distribution is $7\times 10^{-9}$, i.e. negligible. 
This implies that the high-redshift QSOs in this luminosity bin are building up their mass 
more rapidly than the ones at $z\sim2$ \citep[in agreement with what has been found for subsamples in 
previous studies, e.g.][]{Kurk2009,Willott2010}.
 
Another bias that could possibly affect our comparison between the high-z Eddington 
ratio distribution and the lower-z one is the intrinsic luminosity of the QSO broad lines. 
For instance the equivalent width of high ionization lines in AGN is known to anti--correlate with the continuum luminosity 
\citep{Baldwin1977, Baskin2004, Shang2003, Xu2008}. This effect is a function of the line ionization potential 
and not observed in low ionization lines \citep[e.g. \Mgii \ line,][]{Espey1999}. 
High-z QSOs are usually selected in function of their photometric colors as drop-out objects. 
This criterion is to some extent insensitive to the line brightness. 
For the SDSS QSOs sample with $0.35<z<2.25$ the existence of a broad line feature is instead required in addition 
to the color selection, resulting into a bias against weak line QSOs.  
\citet{Decarli2011} found that for the DR7 SDSS QSO sample \citep{Shen2010}
the \Mgii \ line luminosity positively correlates with the continuum luminosity. 
Selecting objects within a given \Mgii \ line luminosity bin would then result 
in a continuum luminosity cut, like the one previously taken into account.

\begin{center}
\begin{figure}[h]
\label{fig_mass_er}
\begin{center}
\resizebox{0.45\textwidth}{!}{\includegraphics{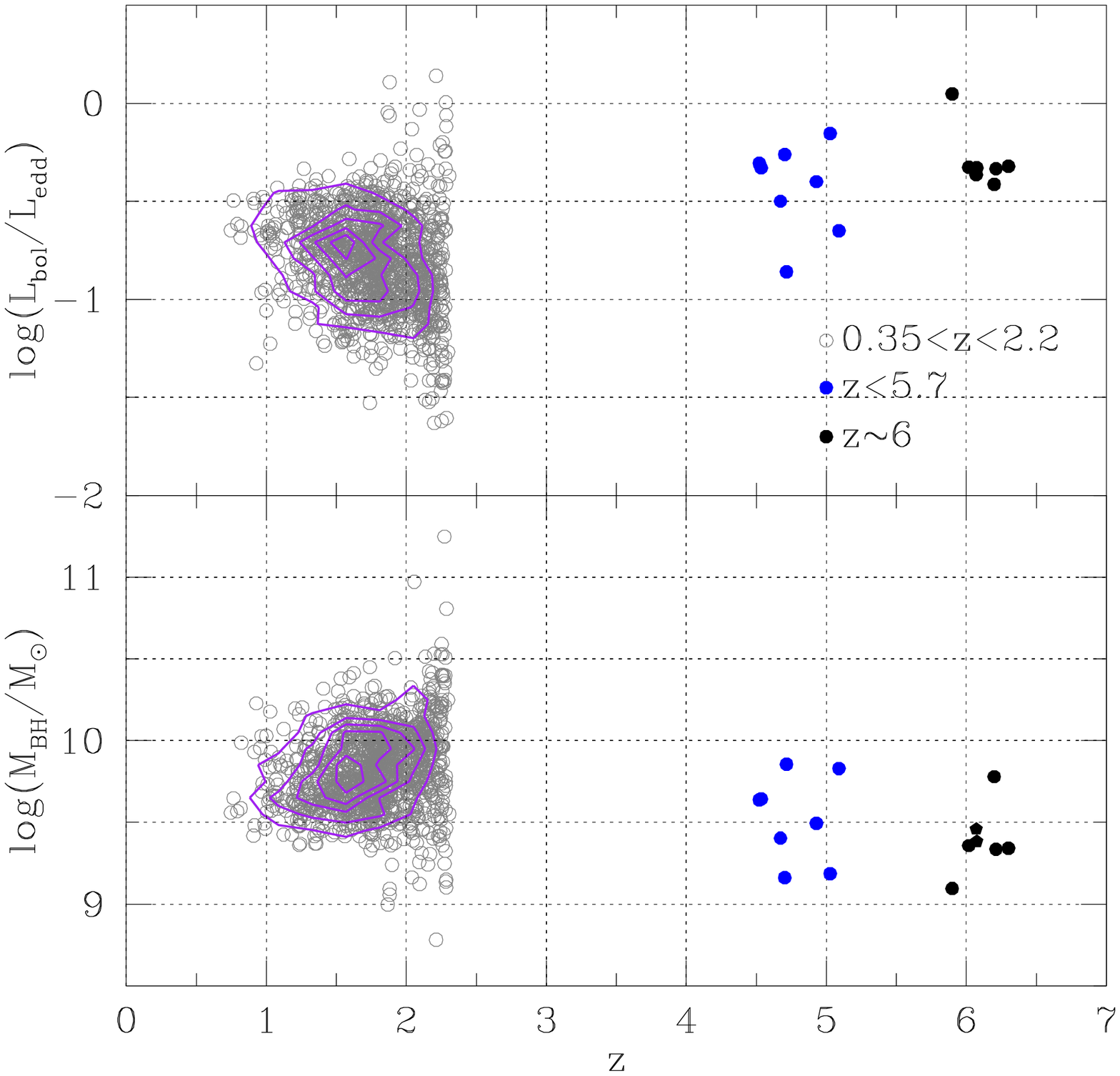}}
\end{center}
\caption{Eddington ratio (upper panel) and \Mbh \ (lower panel) as a function of redshift for subsamples of QSOs
 with bolometric luminosities $10^{47}<L_{bol} \ \left[\rm {erg \ s}^{-1} \right] <3\times10^{47}$ 
extracted from the $0.35<z<2.25$ sample and from our high-z sample. Blue circles: $z<5.7$ \citep{Iwamuro2002}; 
black circles: $z\sim6$ \citep[][this study]{Barth2003, Iwamuro2004, Jiang2007, Kurk2007, Kurk2009}; 
grey circles: SDSS QSOs sample $0.35<z<2.25$ \citep[DR7,][]{Shen2010}; purple lines: isodensity contours.
The average \Mbh \  at high-z is lower than the typical \Mbh \  of the $0.35<z<2.25$ population at similar luminosities, 
while the Eddington ratio is higher.}
\end{figure}
\end{center}

\begin{center}
\begin{figure}[h!]
\label{fig_smirnov}
\begin{center}
\resizebox{0.45\textwidth}{!}{\includegraphics{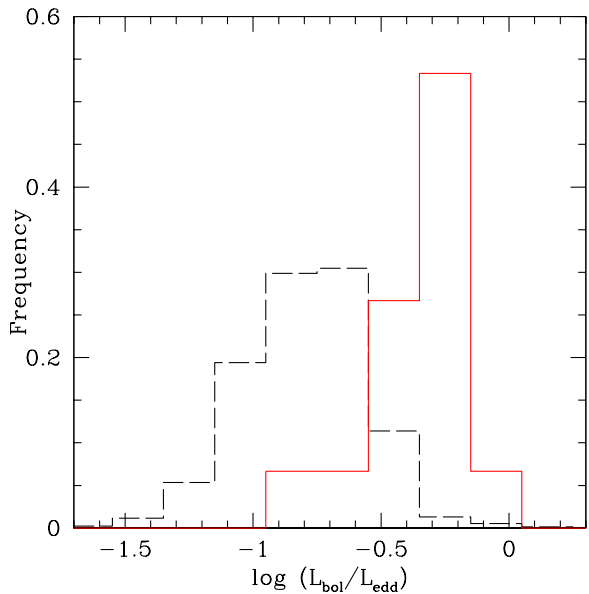}}
\end{center}
\caption{Eddington ratio distributions for sub samples of QSOs
 with bolometric luminosities $10^{47}<L_{bol} \ \left[\rm {erg \ s}^{-1} \right] <3\times10^{47}$ 
extracted from the $0.35<z<2.25$ sample and from our high-z sample: 
a) black dashed line: SDSS QSOs at $0.3<z<2.2$ \citep[SDSS DR7,][]{Shen2010}; b) red solid line: our sample at 
$4.7<z<6.5$. The two distributions are intrinsically different: for the $0.35<z<2.25$ population 
$\langle log(L_{bol}/L_{Edd})\rangle
\sim-0.80$ dex with a dispersion of 0.24 dex, 
while for the high-z one $\langle log(L_{bol}/L_{Edd})\rangle
\sim-0.37$ dex with a dispersion of 0.20 dex. The Kolmogorov-Smirnov probability that 
the two distributions come from the same parent distribution is negligible.}
\end{figure}

\end{center}

\subsection{Impact on BH formation history}

Accurate measurements of BH masses and relative 
Eddington ratios of $z\sim6$ QSOs can be used to give constraints on the formation processes of super-massive BHs 
(SMBHs) in the early universe. 
The three outstanding theories for SMBH seed formation differ 
both in the astrophysical processes considered and in the resulting masses of the seeds: 
(i) BHs seeds of few hundreds of $M_\odot$ can be produced by the first generation of stars (Pop III stars), formed 
out of zero metallicity gas; (ii) BH seeds of $10^2-10^4 \  M_\odot$ can be obtained via stellar-dynamical 
processes amongst Pop II stars; (iii) BH seeds of $10^4-10^5 \  M_\odot$ can be produced via gas-dynamical processes 
(direct collapse of dense gas clouds)  \citep[see][for a review on the SMBH formation]{Volonteri2010}. 

We here consider only the 20 QSOs at $z\sim6$, 
since they provide tighter constraints on the SMBH formation: 
11 sources from our sample ($z>5.7$) and 9 QSOs by 
\citet{Willott2010a}. The resulting sample includes all the 
faint $z\sim6$ QSOs known to date with estimated \Mbh \ from the \Mgii \ line. 
For this $z\sim6$ population we obtain $\langle log(M_{BH}/M_\odot)\rangle
\sim9.06$ dex ($M_{BH}\sim1.1\times 10^9 \ M_\odot$) with a dispersion of 0.5 dex 
and $\langle log(L_{bol}/L_{Edd})\rangle
\sim-0.23$ dex ($L_{bol}/L_{Edd}\sim$0.6) with a dispersion of 0.29 dex. 
The time needed by a BH seed to grow at a constant rate from an initial mass 
$M_0$ to a final mass $M_t$ is equal to \citep{Volonteri2005}:
\begin{equation}
\label{EQ_BH}
t=0.45 \ \rm{Gyr} \times \left[\frac{\epsilon}{1-\epsilon}\right] \times \frac{L_{edd}}{L_{bol}} \times \rm{ln} \left(\frac{M_t}{M_0}\right)
\end{equation}
where 0.45 Gyr is the Salpeter time, $\epsilon$ is the radiative efficiency \citep[$\epsilon\sim0.07$,][]{Volonteri2005}, 
and $L_{bol}/L_{edd}$ is the Eddington ratio. A BH seed with mass $M_0=10^2, \ 10^5 \ M_\odot$ accreting constantly 
at the Eddington ratio characteristic of our population ($L_{bol}/L_{edd}\sim0.6$), would then need a time 
$t\sim0.91, \ 0.53$ Gyr, respectively, to obtain a final mass $M_t$ equal to the mean \Mbh \ of our sample. 

If the characteristic Eddington ratio of the $\sim6$ QSO population is well 
represented by our mean estimate, models with BH seed masses $\sim10^2 \ M_\odot$ become problematic 
since the time needed for the seed to grow is comparable to the age of the universe at $z\sim6$ ($\sim 0.9$ Gyr).
Moreover, since our QSOs are selected from flux-limited surveys the resulting distribution of the Eddington ratios should be 
biased towards higher values: BHs with low accretion rates may not pass the QSO 
selection magnitude limits. \citet{Kelly2010} and \citet{Willott2010a} have shown that the intrinsic 
Eddington ratio distribution for a volume limited QSO sample is indeed shifted towards lower values 
with respect to a luminosity selected population. Our estimate hence represents an upper 
limit of the characteristic Eddington ratio of the $z\sim6$ QSO population:
 lower Eddington ratios, according to Eq.~\ref{EQ_BH}, would yield even longer time 
for the BH seeds to grow. 
In the future, with complete samples of QSOs at $z\sim6$, it will be possible to give better constraints on 
the super massive BHs formation models.    

\begin{deluxetable*}{lcccccc}
\tablecolumns{7}
\tablewidth{0pc}
\tablecaption{\label{tab_mass} Estimated \Mbh,  QSO Eddington ratios and \Feii/\Mgii \ line ratios. 
The mass estimates have an accuracy of 0.4 dex 
that dominates the measurement uncertainties. 
(1) QSO name.  For sources with multiple 
observations: $^*$, \citet{Jiang2007}; $^+$, \citet{Kurk2007}; 
(2) redshift estimate from \Mgii \ line; 
(3) \Mbh \  in units of $10^9$ \Msun \ estimated using Eq.~\ref{eq_mass_1};
(4) \Mbh \  in units of $10^9$ \Msun \ estimated using Eq.~\ref{eq_mass_2}; 
(5) Eddington accretion ratio derived from \Mbh \ col. [2];    
(6) Eddington accretion ratio derived from \Mbh \ col. [3];    
(7) \Feii/\Mgii \ ratio.
}
\tablehead{
\colhead{QSO name} & \colhead{z} & \multicolumn{2}{c}{\Mbh \ $\left[10^9 \ M_\odot\right]$} & \multicolumn{2}{c}{$L_{bol}/L_{edd}$} & 
\colhead{\Feii/\Mgii} \\ \colhead{} & \colhead{} & \colhead{Eq.~\ref{eq_mass_1}} & \colhead{Eq.~\ref{eq_mass_2}}
& \colhead{Eq.~\ref{eq_mass_1}} & \colhead{Eq.~\ref{eq_mass_2}} & \colhead{} \\ \colhead{(1)} & \colhead{(2)} & 
\colhead{(3)} & \colhead{(4)} & \colhead{(5)} & \colhead{(6)}& \colhead{(7)}}
\startdata
BR $1033-0327$      & 4.521$\pm$0.005 & 4.3 & 2.8 & 0.5 & 0.8 & 2.6$\pm$0.1 \\   
BR $0019-1522$ 	    & 4.534$\pm$0.001 & 4.4 & 2.8 & 0.5 & 0.7 & 2.7$\pm$0.1 \\    
BR $2237-0607$ 	    & 4.561$\pm$0.001 & 5.4 & 4.0 & 1.0 & 1.3 & 2.1$\pm$0.5 \\    
SDSS J$0310-0014$   & 4.672$\pm$0.002 & 2.5 & 1.4 & 0.3 & 0.6 & 2.7$\pm$0.6 \\    
SDSS J$1021-0309$   & 4.703$\pm$0.001 & 1.5 & 0.8 & 0.5 & 1.0 & 3.3$\pm$1.1 \\    
SDSS J$0210-0018$   & 4.715$\pm$0.004 & 7.2 & 4.1 & 0.1 & 0.2 & 1.5$\pm$0.9 \\    
SDSS J$0211-0009$   & 4.894$\pm$0.003 & 4.1 & 2.1 & 0.1 & 0.2 & 4.2$\pm$0.6 \\    
PC $1247+3406$	    & 4.929$\pm$0.001 & 3.1 & 1.9 & 0.4 & 0.7 & 2.1$\pm$0.7 \\    
SDSS J$0338+0021$   & 5.027$\pm$0.001 & 1.5 & 0.9 & 0.7 & 1.2 & 5.0$\pm$0.6 \\    
SDSS J$1204-0021$   & 5.090$\pm$0.007 & 6.8 & 4.1 & 0.2 & 0.4 & 3.3$\pm$0.1 \\    
SDSS J$0005-0006$   & 5.844$\pm$0.001 & 0.1 & 0.06 & 3.6 & 7.1 & 4.7$\pm$1.1 \\    
SDSS J$1411+1217^*$ & 5.854$\pm$0.003 & 0.9 & 0.5 & 1.3 & 2.2 & 2.0$\pm$0.7 \\    
SDSS J$1411+1217^+$ & 5.903$\pm$0.001 & 1.6 & 1.0 & 0.9 & 1.5 & 2.9$\pm$0.5 \\    
SDSS J$1306+0356^*$ & 6.017$\pm$0.001 & 1.7 & 0.9 & 0.6 & 1.0 & 4.8$\pm$2.7 \\    
SDSS J$1306+0356^+$ & 6.018$\pm$0.003 & 2.9 & 1.7 & 0.3 & 0.6 & 1.5$^{+1.6}_{-1.5}$ \\    
SDSS J$1630+4012$   & 6.058$\pm$0.005 & 1.7 & 0.9 & 0.5 & 0.8 & 0.5$\pm$0.5 \\    
SDSS J$0303-0019$   & 6.079$\pm$0.001 & 0.5 & 0.3 & 0.6 & 1.3 & 0.0$+$5.8 \\    
SDSS J$1623+3112$   & 6.211$\pm$0.001 & 2.2 & 1.2 & 0.5 & 0.8 & 4.3$\pm$3.3 \\    
SDSS J$1048+4637$   & 6.198$\pm$0.004 & 6.0 & 3.9 & 0.4 & 0.6 & 4.1$\pm$0.3 \\    
SDSS J$1030+0524^*$ & 6.302$\pm$0.001 & 2.0 & 1.1 & 0.5 & 0.9 & 2.9$\pm$1.3 \\    
SDSS J$1030+0524^+$ & 6.299$\pm$0.002 & 2.4 & 1.4 & 0.5 & 0.8 & 2.3$\pm$1.2 \\    
SDSS J$1148+5251$   & 6.407$\pm$0.002 & 7.4 & 4.9 & 0.3 & 0.5 & 4.6$\pm$0.2 \\   
SDSS J$0353+0104$   & 6.072$\pm$0.002 & 2.4 & 1.4 & 0.5 & 0.8 & 4.7$\pm$0.7 \\    
SDSS J$0842+1218$   & 6.069$\pm$0.002 & 2.9 & 1.7 & 0.4 & 0.7 & 3.1$\pm$0.5 \\
\enddata
\end{deluxetable*}

\subsection{\Feii/\Mgii}
\label{sec_SN_met}
An estimate of the Fe/Mg abundance ratio 
at $z\sim6$ may serve as an indication of the onset of star formation in the highest-z QSOs (see Sec.~\ref{intro}). 
The proxy usually used to trace the Fe/Mg abundance ratio at high-z is the \Feii/\Mgii \ line ratio.
In the past, numerous NIR-spectroscopy 
studies have been carried out to analyze the \Feii/\Mgii \ line ratio in high-z QSOs 
\citep[e.g.][]{Maiolino2001, Maiolino2003,
Pentericci2002, Iwamuro2002, Iwamuro2004, Barth2003, Dietrich2003, Freudling2003, Willott2003, 
Jiang2007, Kurk2007}. The combined results revealed an increase in the scatter 
of the measured \Feii/\Mgii \ line ratios as a function of redshift \citep[see][]{Kurk2007}. 
A proposed explanation for the increased scatter is that some young objects have been observed such
a short time after the initial starburst that the BLR has not been fully enriched with Fe yet.

We compute the \Feii \ flux by integrating the normalized \Feii \ template over the rest-frame 
wavelength range $2200 \ \rm \AA <\lambda_{rest}<3090$ \AA. 
The \Mgii \ flux is computed by integrating the fitted single Gaussian over the range 
$\lambda_{peak}\pm5\sigma$. The resulting \Feii/\Mgii \ flux ratios and relative errors are 
listed in Tab.~\ref{tab_mass}. In Fig.~\ref{fig_met} we show the evolution of 
the \Feii/\Mgii \ as a function of z 
(for objects with multiple observations we consider the weighted mean of the individual measurements). 
The reported scatter at $z\sim6$ \citep{Kurk2007} 
is now significantly reduced: our measurements of the line ratios span a range $0<$\Feii/\Mgii$<6$, 
while previous results were distributed up to \Feii/\Mgii$\sim15$.  
The \Feii/\Mgii \ line measurements do not show any clear dependence on the estimated S/N per \AA \ of the 
spectral continuum (see Fig.~\ref{fig_SN_met}). On the other hand, our 
flux ratio estimates are well in agreement with previous results by \citet{Kurk2007} who performed an analysis 
similar to ours. We can conclude that this measurement is significantly 
dependent on the performed analysis, rather than on the S/N of the spectra. 
This result immediately implies that we do not need to invoke
different chemical evolutionary stages of the BLR gas. We do not see evidence for evolution of the estimated \Feii/\Mgii \ 
line ratio as a function of cosmic age for $4<z<6.5$. 
Moreover, if we consider the SDSS QSO sample at $0.35<z<2.25$ \citep[][see Sec.~\ref{mass_ev}]{Shen2010}, 
and we compute the \Feii/\Mgii \ line ratio starting from the published \Feii \ and \Mgii \ flux measurements, 
we see no evolution of the line ratio also for $0.35<z<2.25$. However it is not possible to perform a direct 
comparison between the two samples, since these line ratio measurements are highly dependent on
the adopted fitting procedure. 

To constrain the onset of the star formation in these early QSOs, we would need to derive an estimate of the 
Fe/Mg abundance ratio from the measurements of the \Feii/\Mgii \ line ratio. Unfortunately,  
an accurate conversion cannot be performed since the detailed formation of the \Feii \ bump in QSO 
spectra has not been fully understood yet. 
\citet{Wills1985} discussed the suitability of the \Feii/\Mgii \ line ratio as a proxy of the Fe/Mg abundance ratio, 
as the regions in which the two ions are produced and the radiative transfer of the two lines are both very different.  
Computing the \Feii \ and \Mgii \ line strength for a gas with cosmic abundances and using a wide range of 
hydrogen densities and ionization parameters, they predicted typical \Feii/\Mgii \ 
line ratios between 1.5 and 4.0 (this is indicated in Fig.~\ref{fig_met} as a grey-shaded area), and attributed higher \Feii/\Mgii \ 
values to an overabundance of Fe with respect to Mg.  
However these computations were 
based on a limited 70-level model of the Fe atom. 
\citet{Baldwin2004} used a 371-level Fe$^+$ model to reproduce the observed \Feii \  
emission properties in AGNs considering a large set of models of broad emission line region clouds. 
From their results, all the observed \Feii \ features can be reproduced only if: (i) the BLR is 
characterized by the presence of significant microturbolence; (ii) the \Feii \ emitting gas has different properties 
(density and/or temperature) with respect to the gas emitting other broad lines. 
In any case the strength of the \Feii \ emission relative to the emission 
line of other ions (e.g the \Feii/\Mgii \ line ratio) depends as much on the Fe abundance 
as it does on other physical parameters of the BLR (e.g. turbulence velocity), making it difficult to convert the observed line ratios in 
abundance ratios. Nevertheless, the study of the \Feii/\Mgii \ line ratio as a function of look-back
time can be used to give constraints on the BLR chemical enrichment history, under the assumption that physical 
conditions of the BLR that determine the \Feii \ emission are not evolving.

The observed lack 
of evolution in the measured \Feii/\Mgii \  line ratio can then be explained with 
an early chemical enrichment of the QSO host. I.e. the QSOs in our sample must 
have undergone a major episode of Fe enrichment in a few hundreds Myr before the cosmic age at which they have been 
observed ($\sim0.9$ Gyr). \citet{Matteucci2003} showed that for a massive elliptical galaxy 
characterized by a very intense but short star-formation history, the typical timescale for the maximum SN Ia rate 
can be as short as 0.3 Gyr. This implies that if the Fe in high-z QSO hosts is mainly produced via SN Ia explosions, 
it would be possible to observe fully enriched BLR at $z\sim6$. 
On the other hand, \citet{Venkatesan2004} pointed out that SNe Ia are not 
necessarily the main contributors to the Fe enrichment, and that stars with a present-day initial mass function 
are sufficient to produce the observed \Feii/\Mgii \ line ratios at $z\sim6$. 
Fe could also be generated by Pop III stars: these very metal poor stars with typical masses $M\gtrsim$ 
100 \Msun \ might be able to produce large amounts of Fe within a few Myr \citep{Heger2002}.

\begin{center}
\begin{figure}[h!]
\label{fig_met}
\begin{center}
\resizebox{0.45\textwidth}{!}{\includegraphics{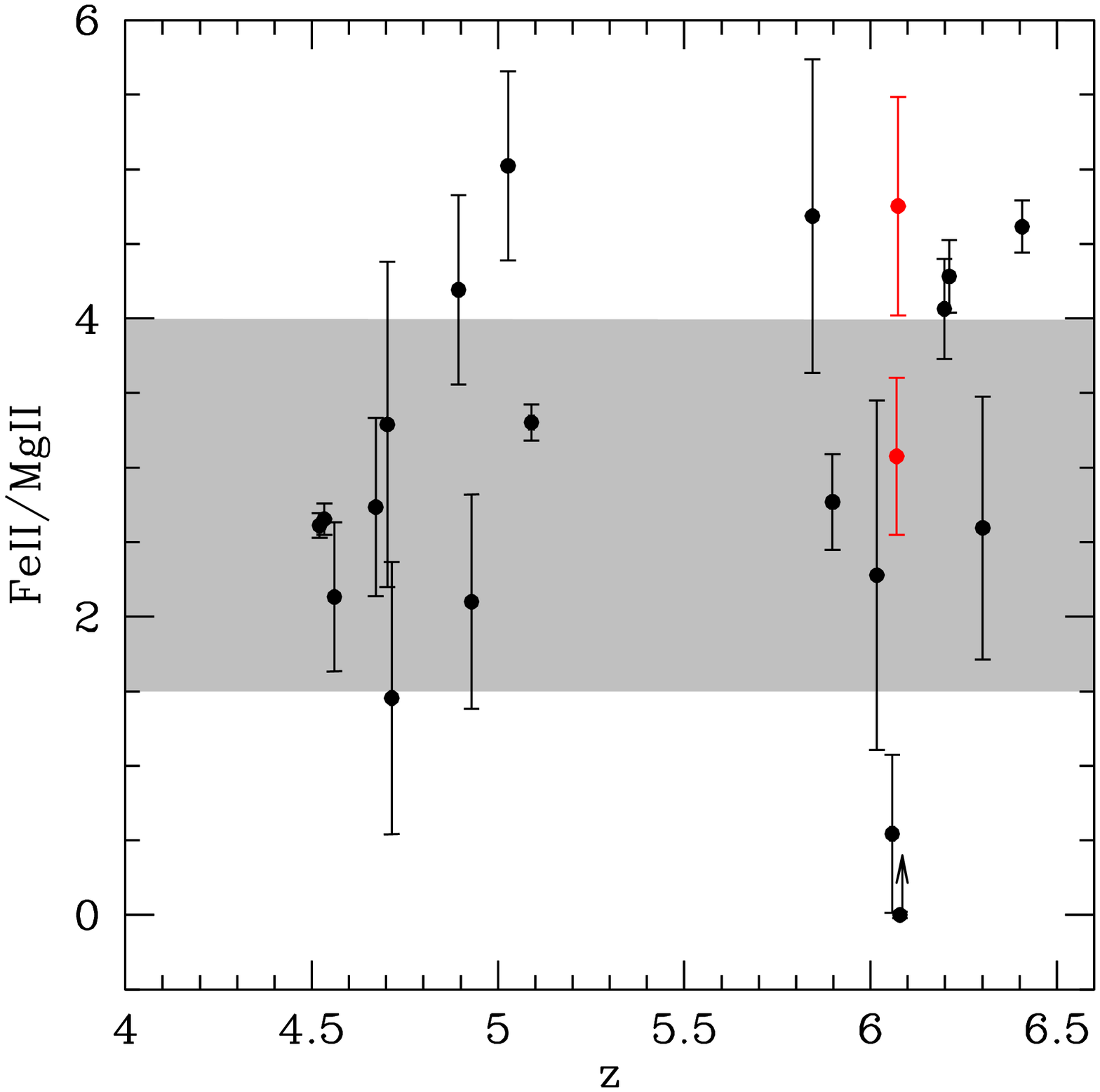}}
\end{center}
\caption{Evolution of the \Feii/\Mgii \ line ratio for $z>4$. 
Black points: literature sample, $z>4.4$; red points: new targets. Grey-shaded area: expected \Feii/\Mgii \ line ratio 
obtained by modeling the \Feii \ and \Mgii \  
line strength considering gas with cosmic abundance and a range of ionizing continua \citep{Wills1985}. 
The reported scatter at $z\sim6$ \citep{Kurk2007} 
is now significantly reduced: our measurements of the line ratios span a range $0<$\Feii/\Mgii$<6$, 
while previous results were distributed up to \Feii/\Mgii$\sim15$. No redshift evolution of the \Feii/\Mgii \ 
ratio is observed for $4<z<6.5$.}
\end{figure}
\end{center}

\begin{center}
\begin{figure}[h!]
\label{fig_SN_met}
\begin{center}
\resizebox{0.3\textwidth}{!}{\includegraphics{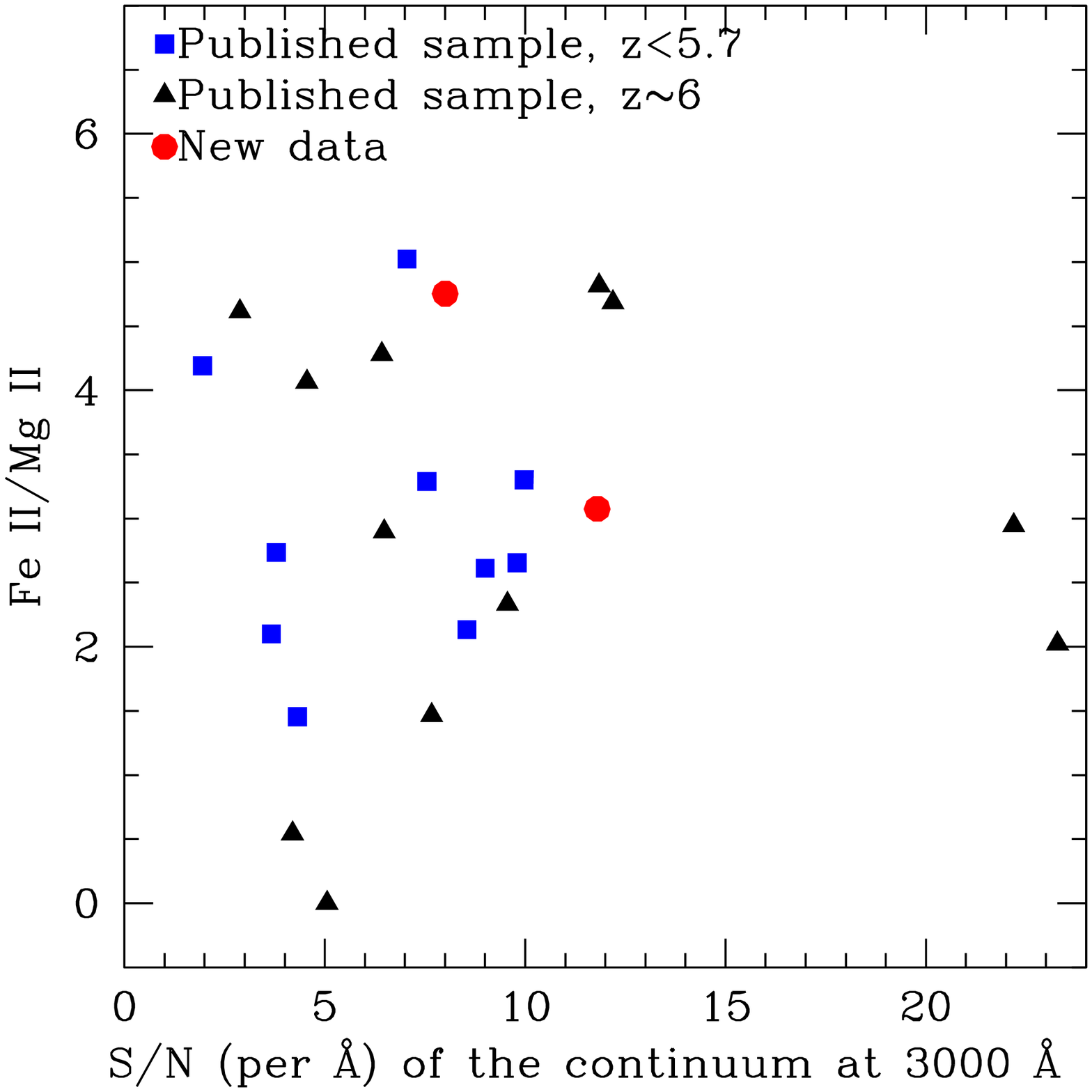}}
\end{center}
\caption{\Feii/\Mgii \ line ratio estimates as a function of the S/N per \AA \ of spectral continuum at 3000 \AA. 
Blue squares: literature sample,  at $z<5.7$ \citep{Iwamuro2002}; black triangles: 
literature sample, $z\sim6$ \citep{Barth2003, Iwamuro2004, Jiang2007,
 Kurk2007, Kurk2009};  red circles: new observations. The \Feii/\Mgii \ line measurements do not show a 
dependence on the S/N per \AA \ of the spectral continuum.}
\end{figure}
\end{center}

\section{Summary and conclusions}

We presented NIR-spectroscopic observations of three $z\sim6$ SDSS QSOs. 
Our NIR spectra cover the \Mgii \ and \Feii \ emission features, which are powerful 
probes of \Mbh \  and of the chemical enrichment of the BLR. The new data extend 
the existing SDSS sample towards the faint end of the QSO luminosity function. We have collected 
22 literature spectra (19 different sources) of high-redshift ($z>4$) QSOs covering the rest-frame wavelength range 
$2700 \ \rm \AA <\lambda<3200$ \AA. 
The final sample is composed of 22 sources: our three new sources at $z\sim6$, 10 spectra from the literature of 
QSOs with $4<z<5.7$ \citep{Iwamuro2002} and 9 spectra of QSOs with $z\sim6$ 
\citep{Barth2003,Iwamuro2004,Jiang2007,Kurk2007,Kurk2009}. 
In order to see how the derivation of physical parameters depend on the 
fitting strategy employed, we performed a new consistent analysis of the sample and we gave 
an estimate of the \Mbh, of the $L_{bol}/L_{Edd}$ and of the \Feii/\Mgii \ line ratio. 

Our results and conclusions can be summarized as follows:

\begin{enumerate}
\item{We have estimated the \Mbh \  from the \Mgii \ emission line using empirical mass scaling relations. The  
QSOs in our sample host BHs with masses of $\sim 10^9$ \Msun. Our results are in good agreement 
with previous estimates, indicating that different fitting procedures imply variations 
smaller than the errors due to the intrinsic scatter of the \Mbh \ estimator. }
 \item{High-redshift QSOs are accreting close to the Eddington luminosity: $\langle log(L_{bol}/L_{Edd})\rangle
\sim-0.3$ dex ($L_{bol}/L_{Edd}\sim$0.5) with a dispersion of 0.3 dex. The distribution of observed 
Eddington ratios is significantly different (average Eddington ratio $\sim8$ times higher) from that
of a comparison sample at $0.35<z<2.25$ (SDSS, DR7). This result is not biased by the luminosity selection 
of the high-z sample: the difference in the two Eddington ratio distributions persists even considering two luminosity 
matched subsamples ($z\sim6$ average Eddington ratio $\sim3$ higher than lower redshift one).  
The high-z sources are accreting faster than the ones at $0.35<z<2.25$.}
\item{We calculated fluxes for the \Mgii \ and \Feii \ lines. The obtained values are 
not always in agreement with the published ones, indicating that these
measurements are dependent on the performed analysis. The previously observed scatter in the \Feii/\Mgii \  
line ratio at $z\sim6$ is significantly reduced in this work: the BLRs in the highest-z QSOs studied 
to date all show comparable level of chemical enrichment.}
\item{No redshift evolution of the \Feii/\Mgii \ ratio is observed for $4<z<6.5$. If we consider the \Feii/\Mgii \ line ratio 
as a second 
order proxy for Fe/Mg, this indicates that the QSOs in our sample must have undergone a major episode of Fe enrichment
in the few hundreds Myr preceding the cosmic time at which the sources are observed.}
\item{Our results do not show any significant dependence on the S/N of the spectra: 
the \Mbh \ estimates do not correlate with the S/N per \AA \ of the \Mgii \ line and 
the \Feii/\Mgii \ line ratios do not show any clear dependence on the S/N per \AA \ of the spectral continuum.}
\end{enumerate}

Even though the currently faintest known QSOs are included in the study we presented here \citep{Willott2010a,Kurk2009}, 
we are still not probing the bulk of QSOs that should be present at that redshift. Future studies of QSOs that populate 
the faint end of the QSO luminosity function at $z\sim6$ are needed to investigate whether or not the results found 
here are applicable to all $z\sim6$ QSOs.

\begin{acknowledgments}
We are indebted to A. J. Barth, W. Freudling, F. Iwamuro,  Y. Ju\`{a}rez and R. Maiolino who 
have generously provided their spectra for this analysis. GDR thanks Marianne Vestergaard, Mauricio Cisternas 
and Mario Gennaro for the useful discussion. GDR is supported by the Deutsche Forschungsgemeinschaft 
priority program 1177. XF and LJ acknowledge support by NSF grant AST-0806861 and Packard Fellowship 
for Science and Engineering. JK thanks the DFG support via German-Israeli Project Cooperation grant 
STE1869/1-1.GE625/15-1. 
\end{acknowledgments}

\appendix

\section{Fit results: literature data}
\label{a_fit}
Hereafter we show the fit results for the literature sample and the relative $\chi^2$ maps for \Feii \ error computation 
(analogous to Fig.~\ref{fig_fit_errors}). 
The sources are sorted in redshift. 

In the right panels we show the spectral decomposition.
The observed spectra are shown as a black continous line. The modeled components are: power-law continuum (blue dotted 
line), Balmer pseudo continuum (purple dashed line), \Feii \ normalized template (light blue dotted line), 
\Mgii \ emission line (red dotted line). The sum of the first set of components (power-law continuum $+$ Balmer pseudo continuum $+$ 
\Feii \ normalized template) is overplotted to the spectrum as green solid line, while the sum of all the components is overplotted 
as a red solid line. 
Telluric absorption bands are indicated over the spectra with the symbol  $\Earth$: they are extracted from the ESO sky absorption spectrum measured on 
the Paranal site at a nominal airmass of 1.

The \Feii \ normalization, obtained from the fit of the first set of components, depends on the power-law 
slope and its normalization (intercept). In the left panel we show the $\chi^2$ domain analysis for \Feii \ 
error computation: a) two dimensional projections of the 3D $\chi^2$-surfaces 
(\Feii \ normalization vs Intercept, upper-left plot;
\Feii \ normalization vs Slope, bottom-left plot; Intercept vs slope, bottom-right plot): 
contours represent iso-$\chi^2$ 
levels spaced by a factor of 2 while the best fit case is marked with a dot; 
b) probability distribution for the \Feii \ 
template normalization (upper-right plot): the distribution has been  
 obtained by marginalizing the 3-D probability distribution considering only the triplets  
for which $\chi^2 -\chi_{min}^2< 1$, the dashed vertical lines mark our estimate of the $1-\sigma$ confidence level.

\begin{center}
\begin{figure*}[h]
\centering
\resizebox{0.3\textwidth}{!}{\includegraphics{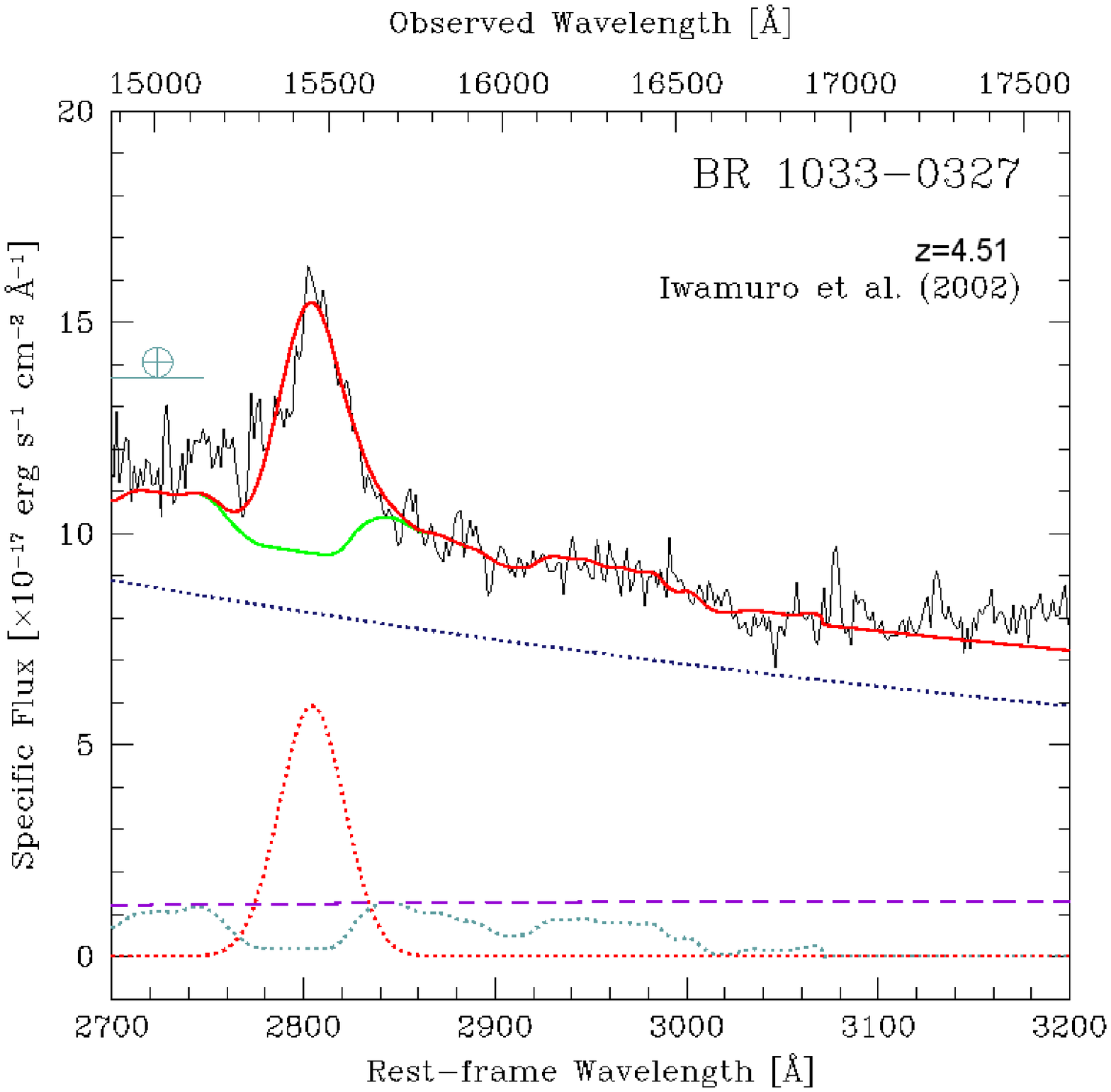}}
\resizebox{0.3\textwidth}{!}{\includegraphics{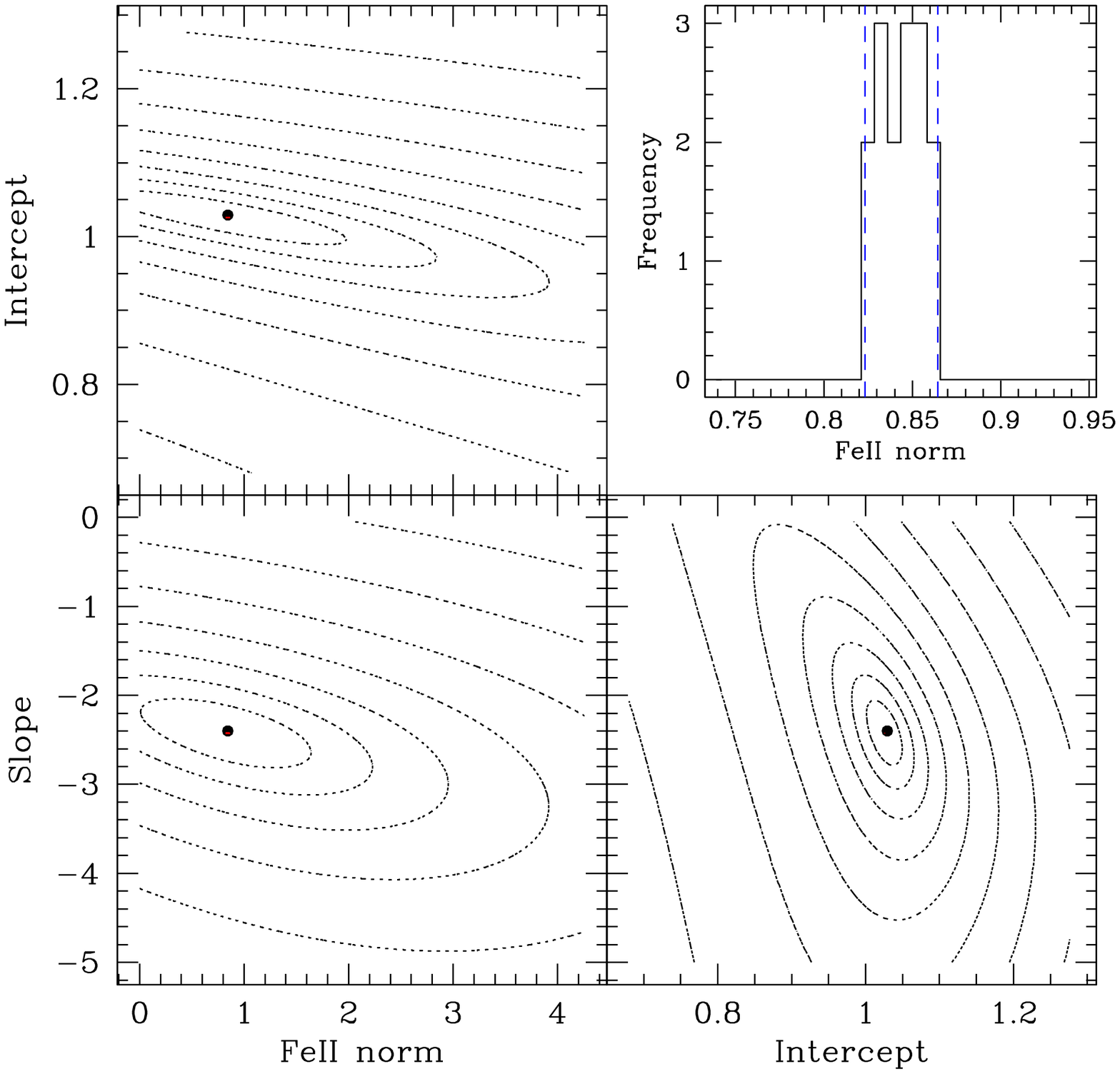}}
\end{figure*}
\end{center}

\begin{center}
\begin{figure*}[h]
\centering
\resizebox{0.3\textwidth}{!}{\includegraphics{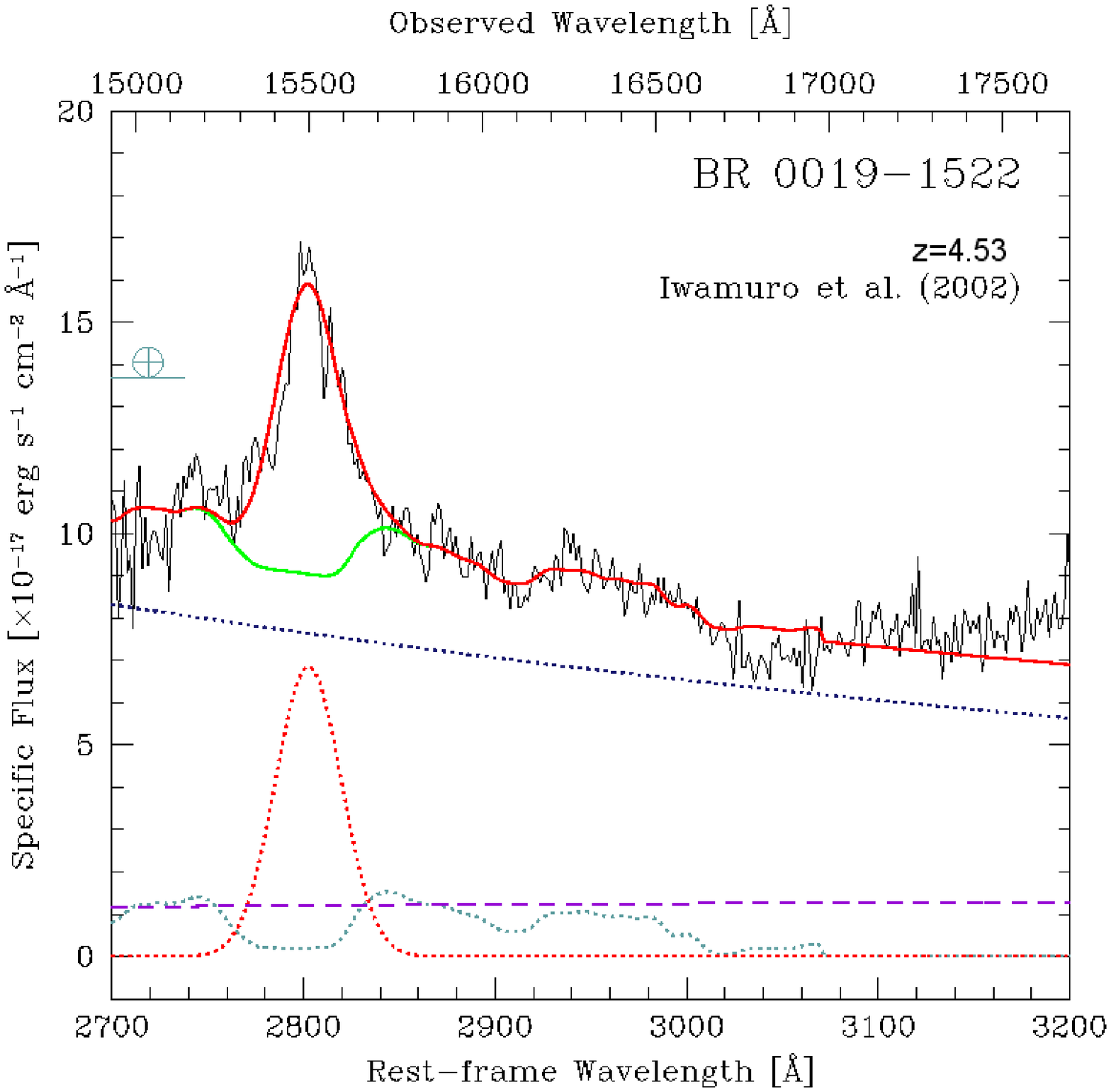}}
\resizebox{0.3\textwidth}{!}{\includegraphics{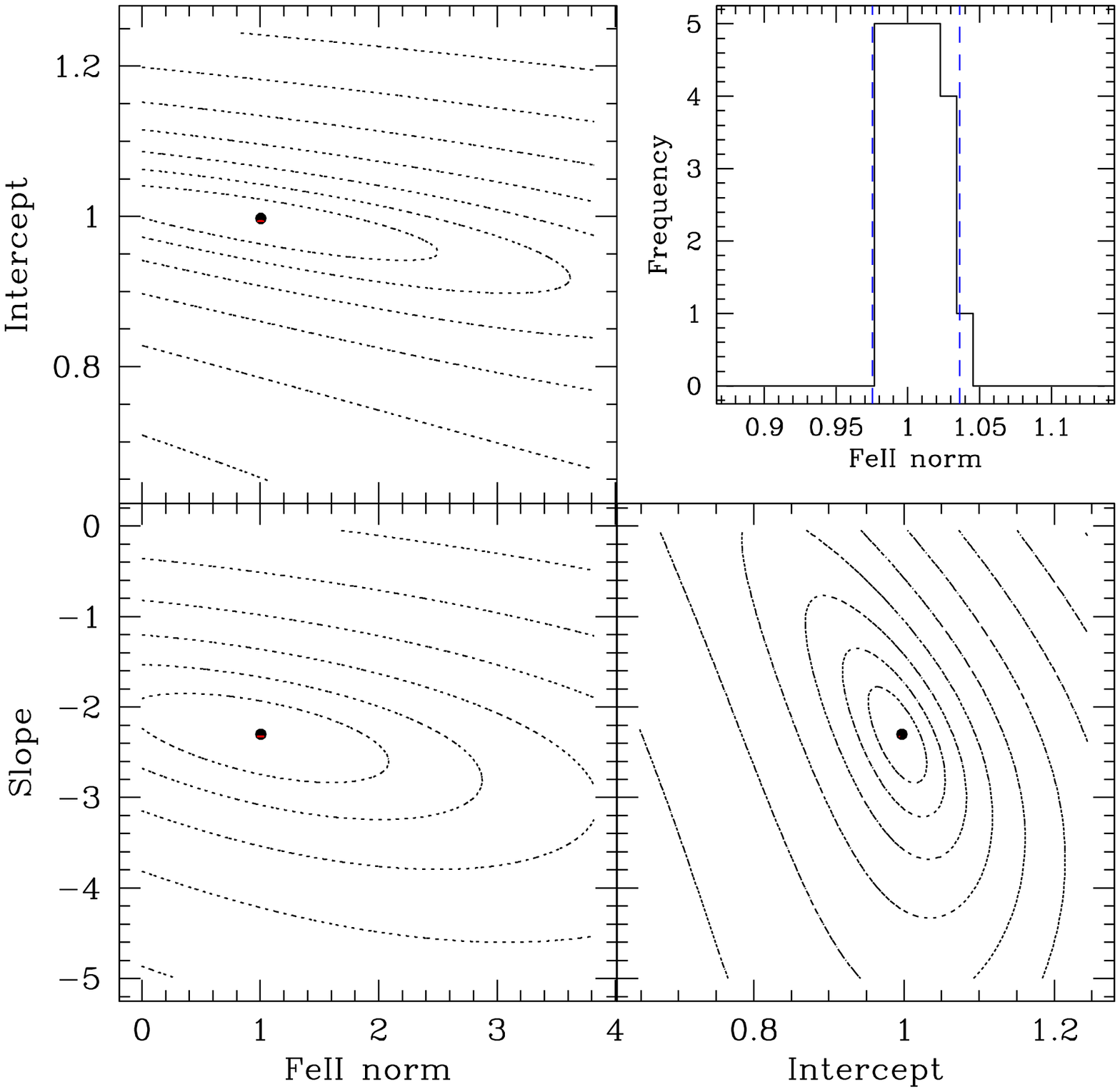}}
\end{figure*}
\end{center}

\begin{center}
\begin{figure*}[h]
\centering
\resizebox{0.3\textwidth}{!}{\includegraphics{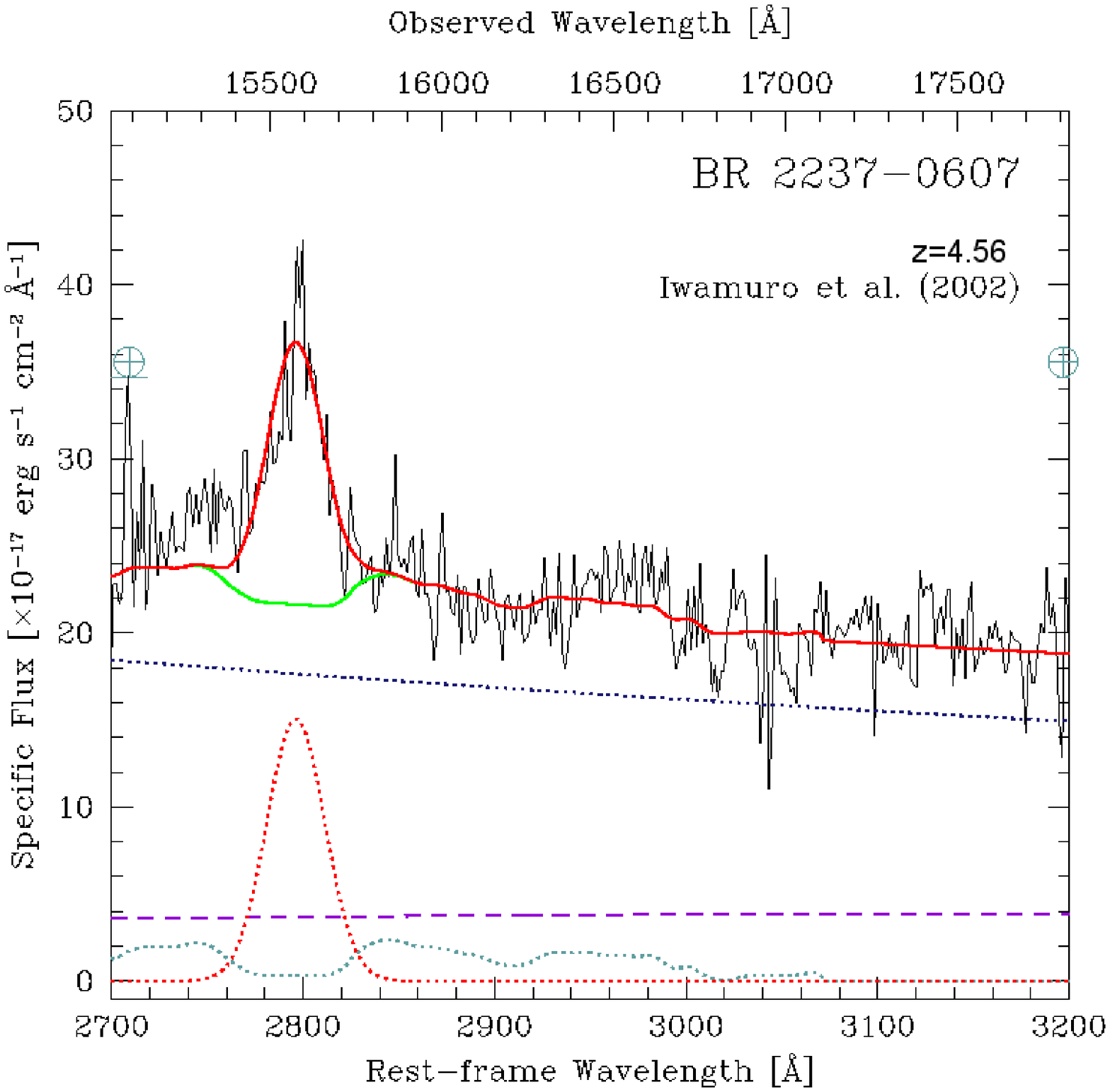}}
\resizebox{0.3\textwidth}{!}{\includegraphics{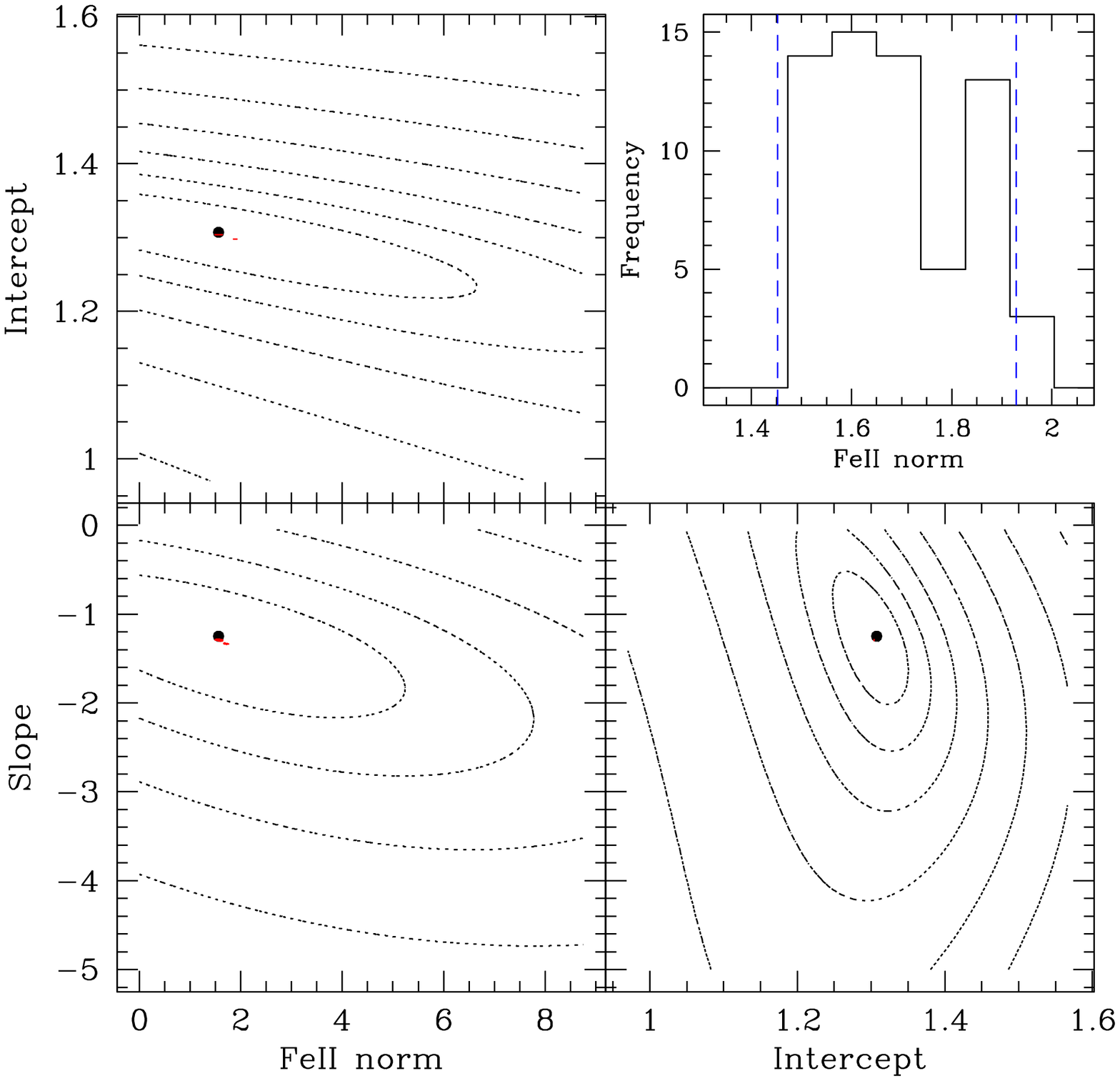}}
\end{figure*}
\end{center}

\begin{center}
\begin{figure*}[h]
\centering
\resizebox{0.3\textwidth}{!}{\includegraphics{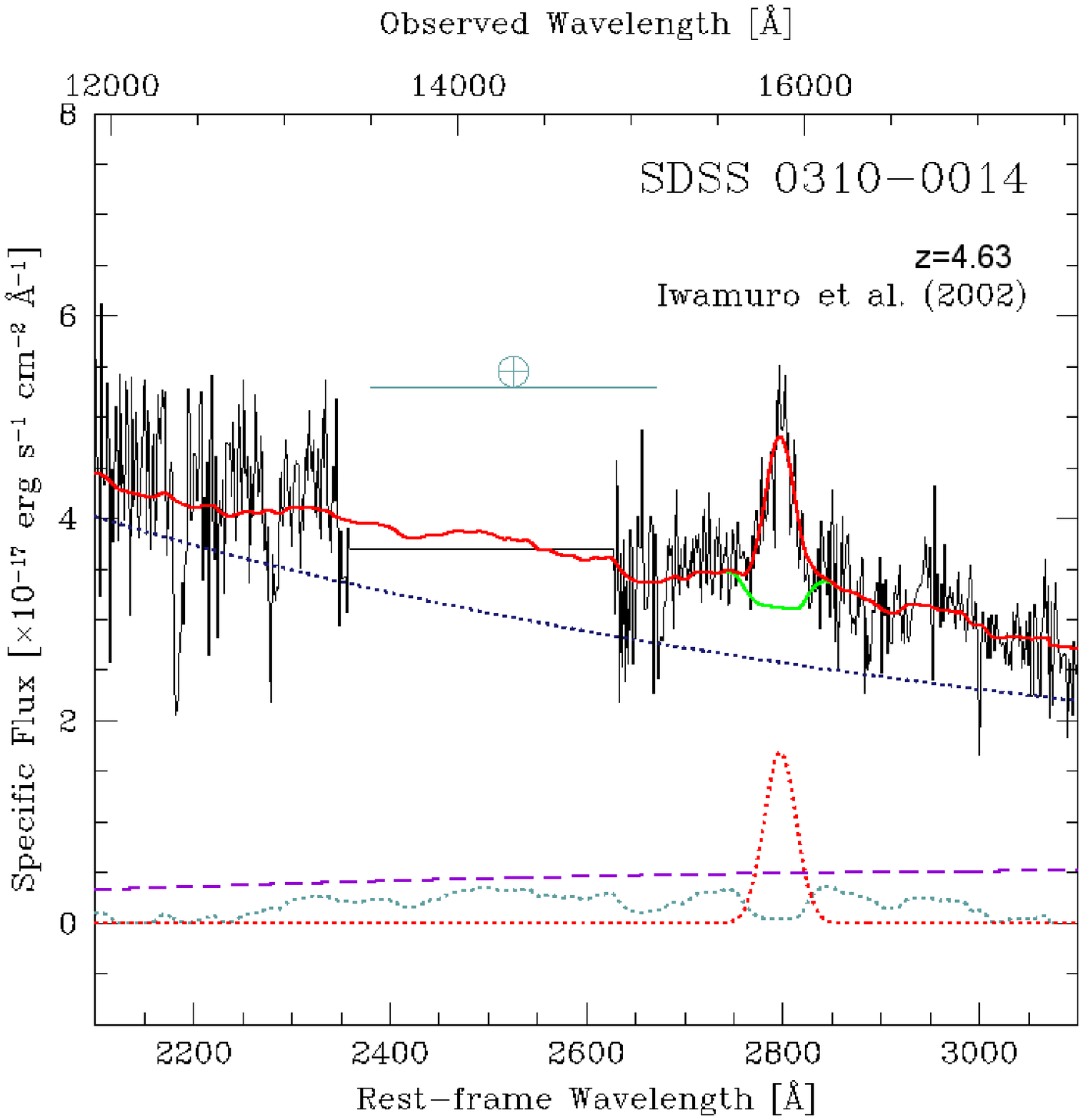}}
\resizebox{0.3\textwidth}{!}{\includegraphics{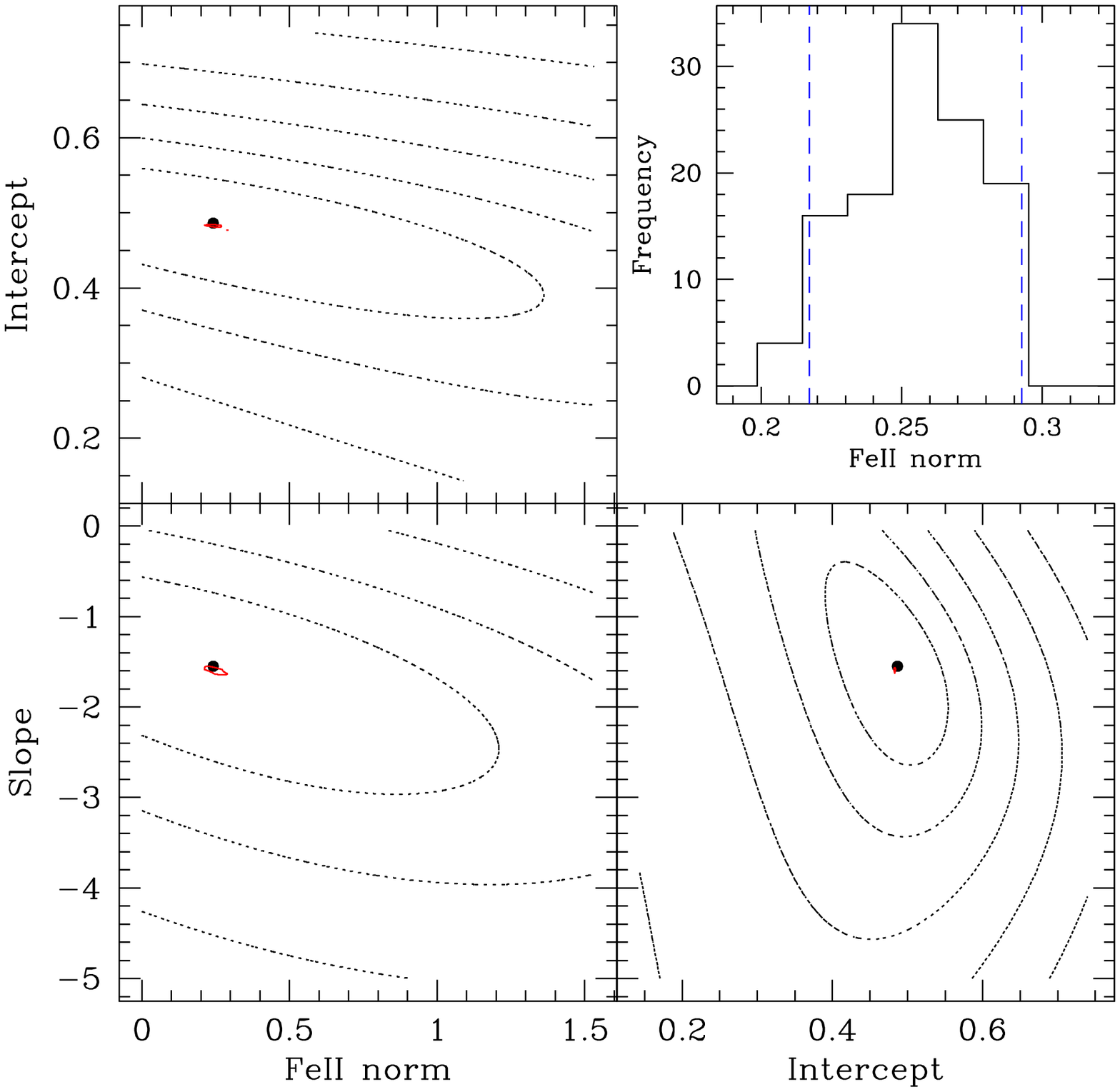}}
\end{figure*}
\end{center}

\begin{center}
\begin{figure*}[h]
\centering
\resizebox{0.3\textwidth}{!}{\includegraphics{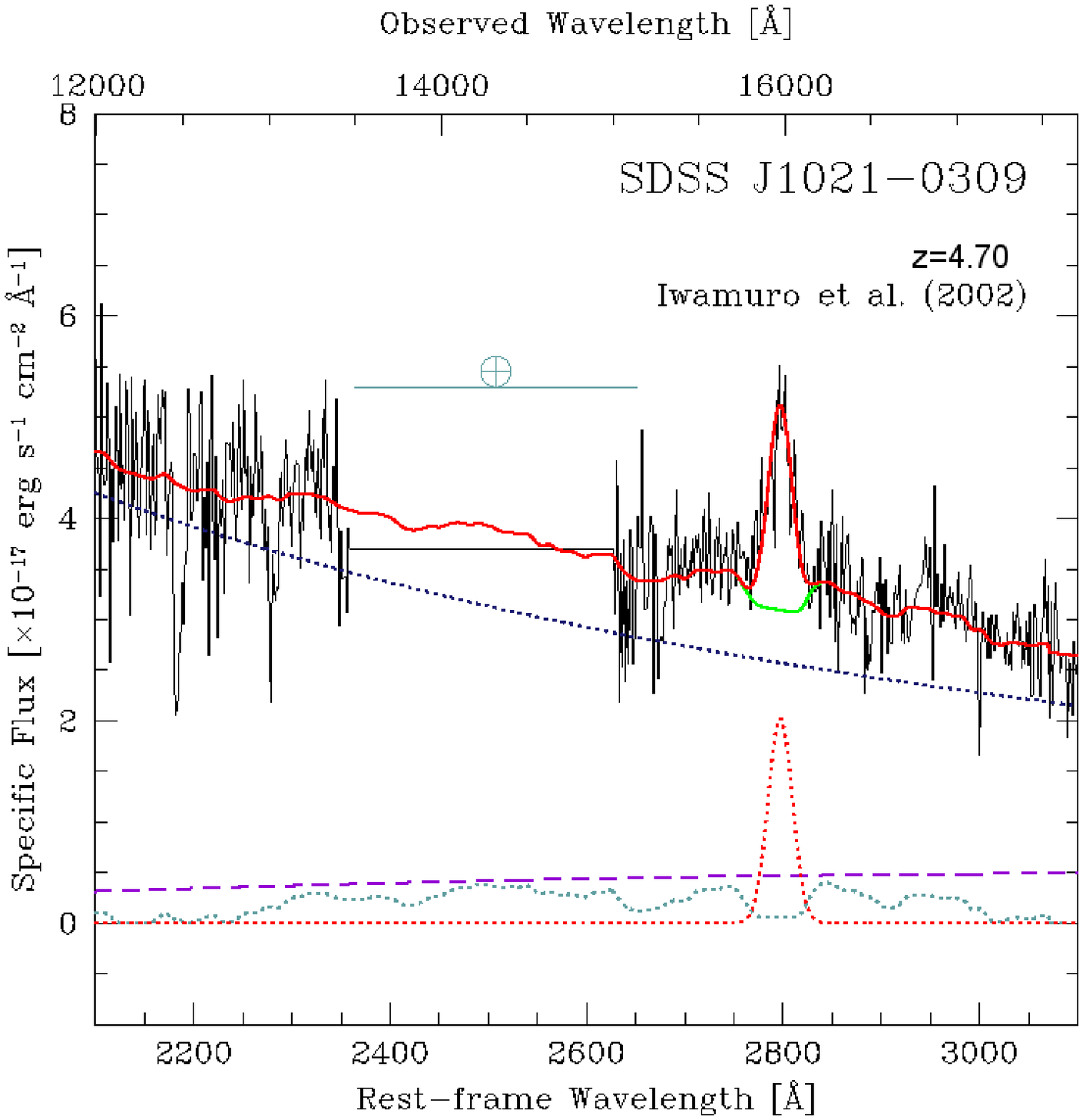}}
\resizebox{0.3\textwidth}{!}{\includegraphics{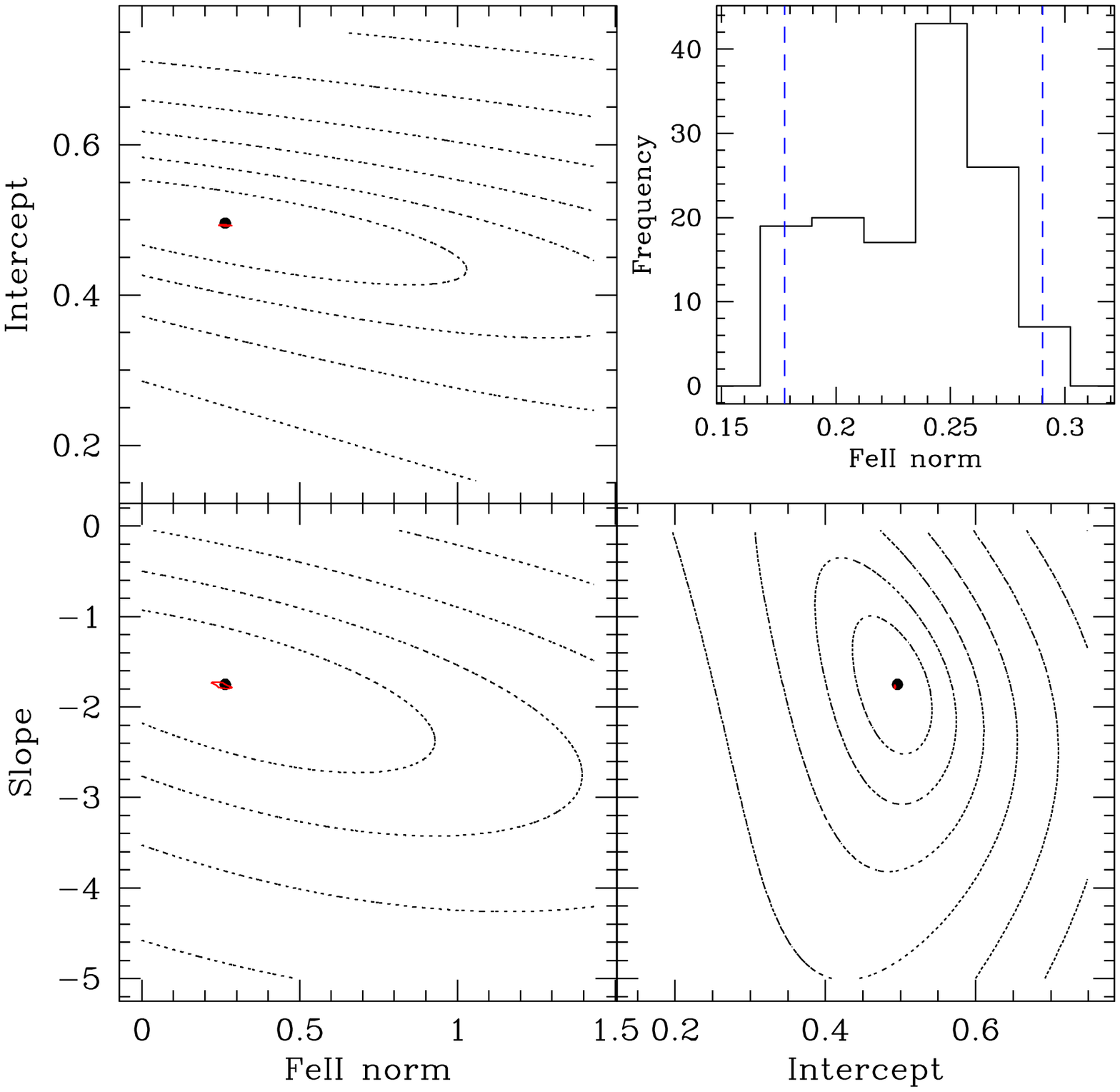}}
\end{figure*}
\end{center}

\begin{center}
\begin{figure*}[h]
\centering
\resizebox{0.3\textwidth}{!}{\includegraphics{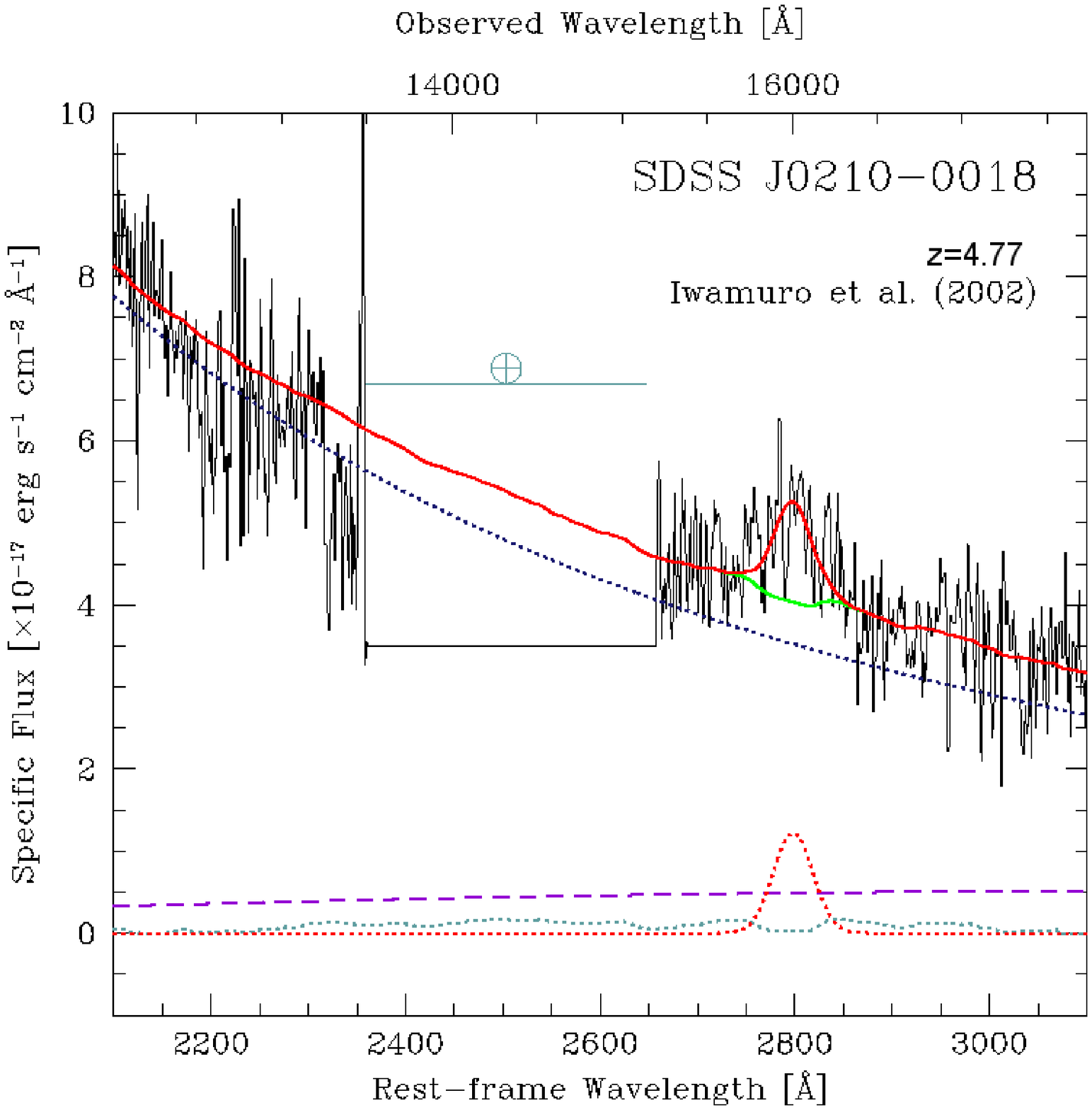}}
\resizebox{0.3\textwidth}{!}{\includegraphics{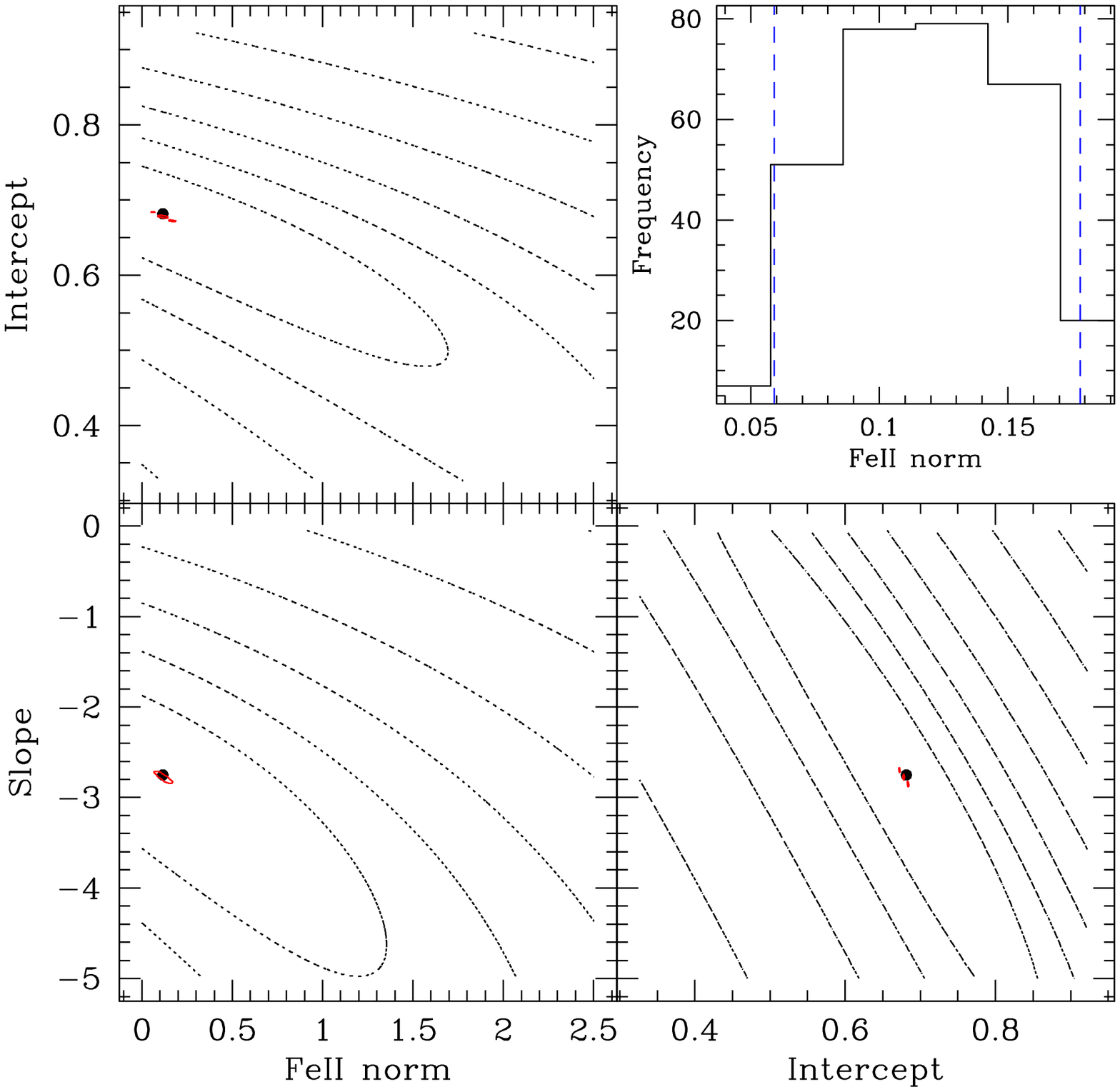}}
\end{figure*}
\end{center}

\begin{center}
\begin{figure*}[h]
\centering
\resizebox{0.3\textwidth}{!}{\includegraphics{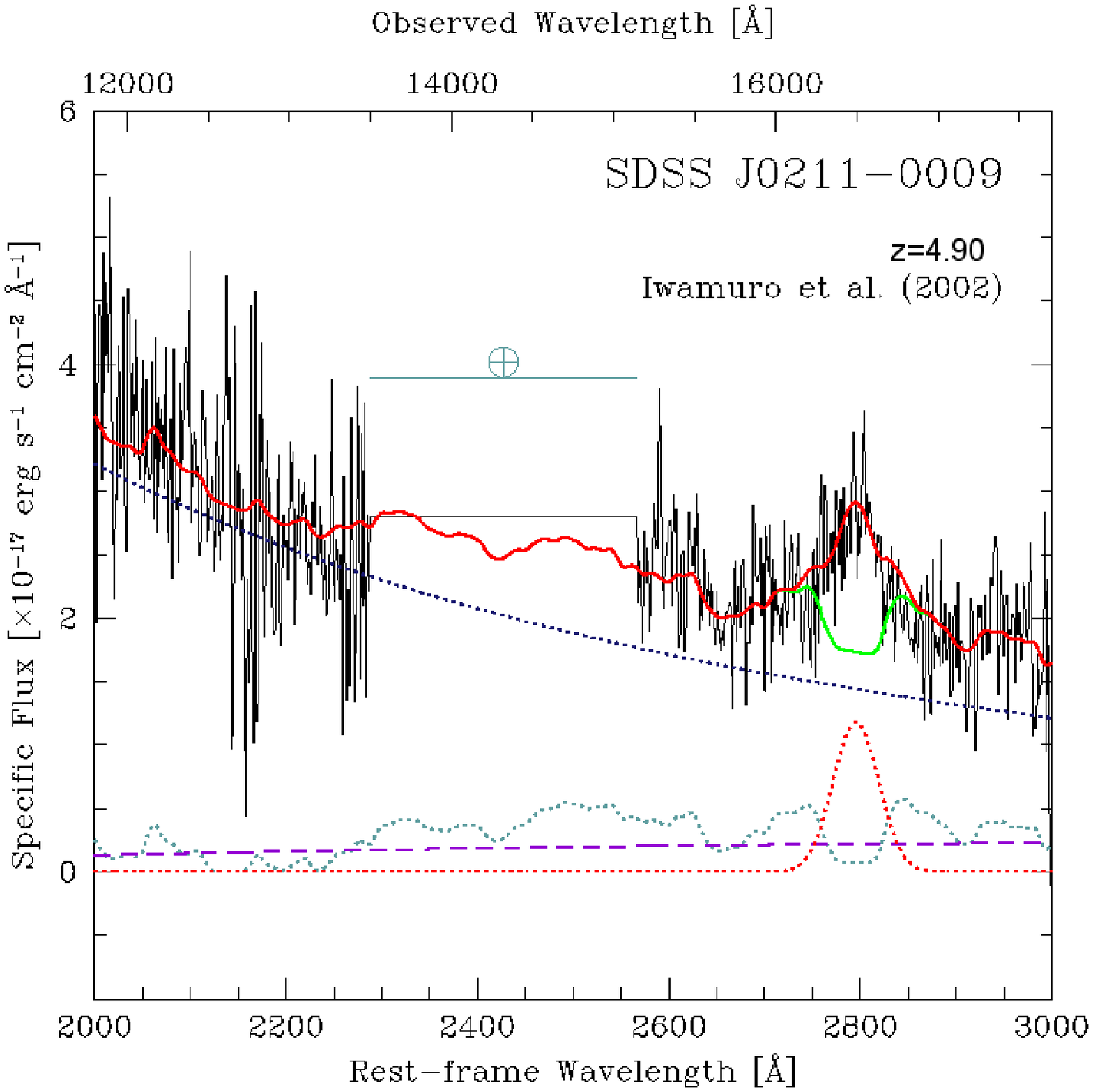}}
\resizebox{0.3\textwidth}{!}{\includegraphics{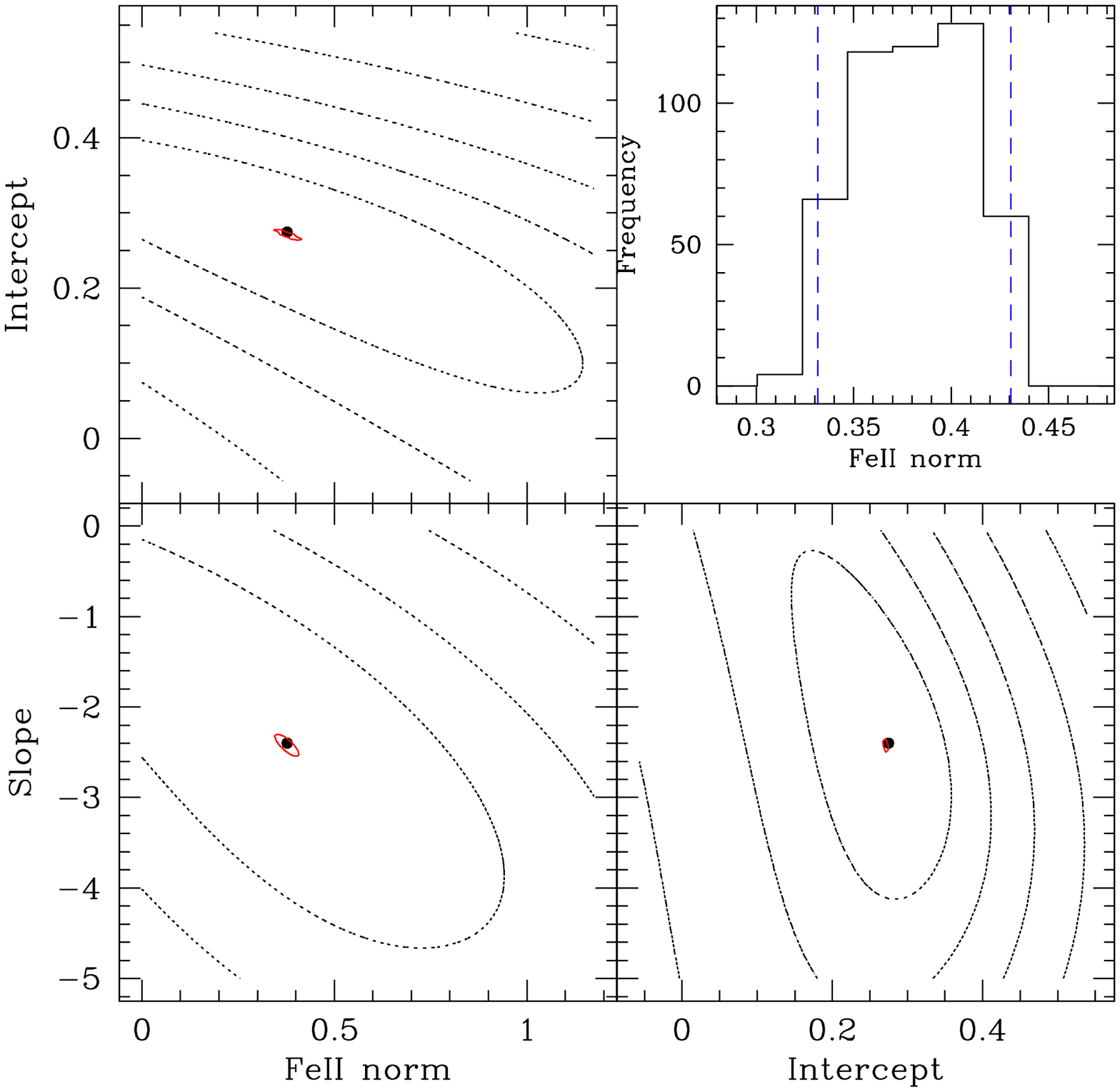}}
\end{figure*}
\end{center}

\begin{center}
\begin{figure*}[h]
\centering
\resizebox{0.3\textwidth}{!}{\includegraphics{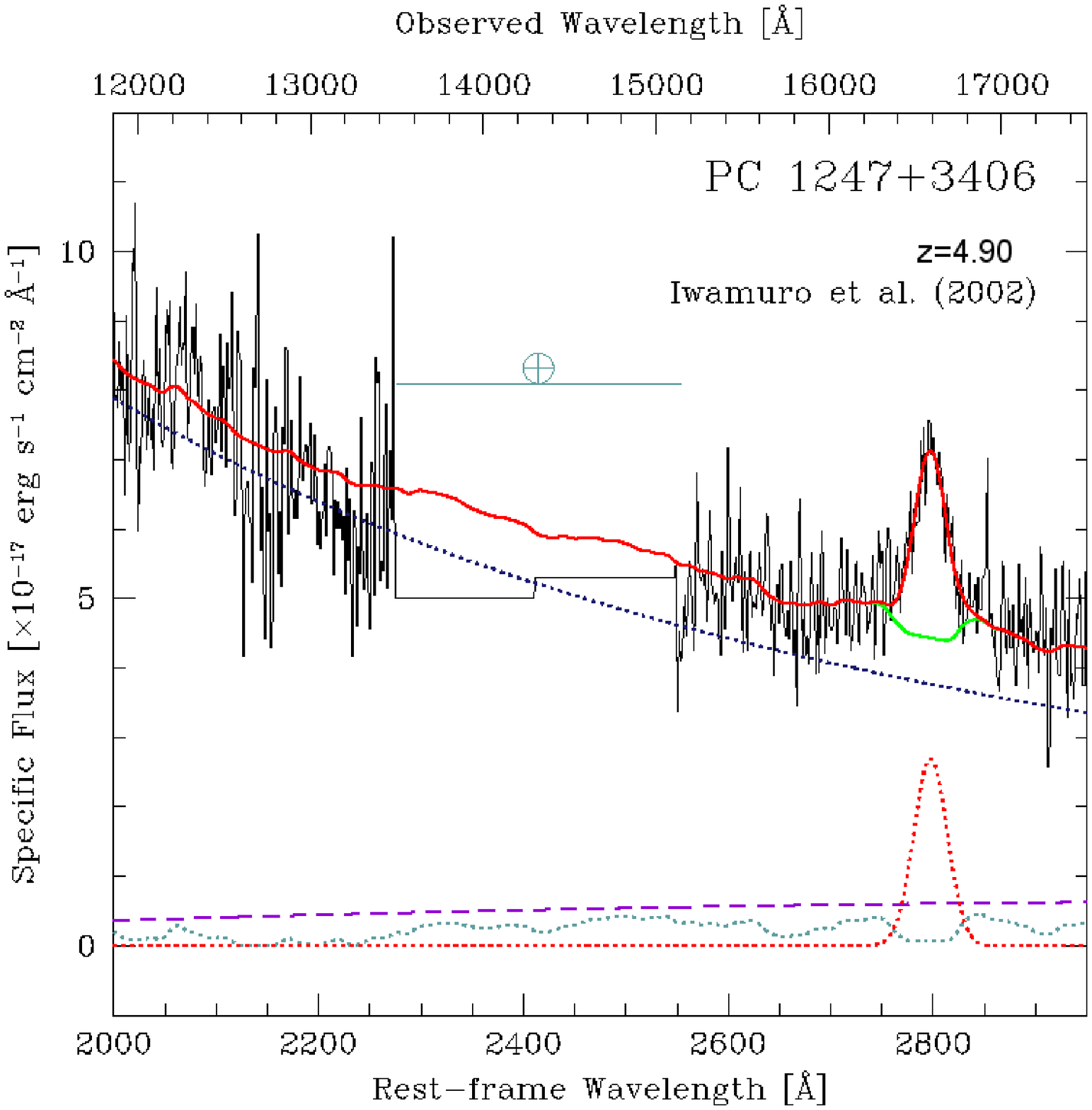}}
\resizebox{0.3\textwidth}{!}{\includegraphics{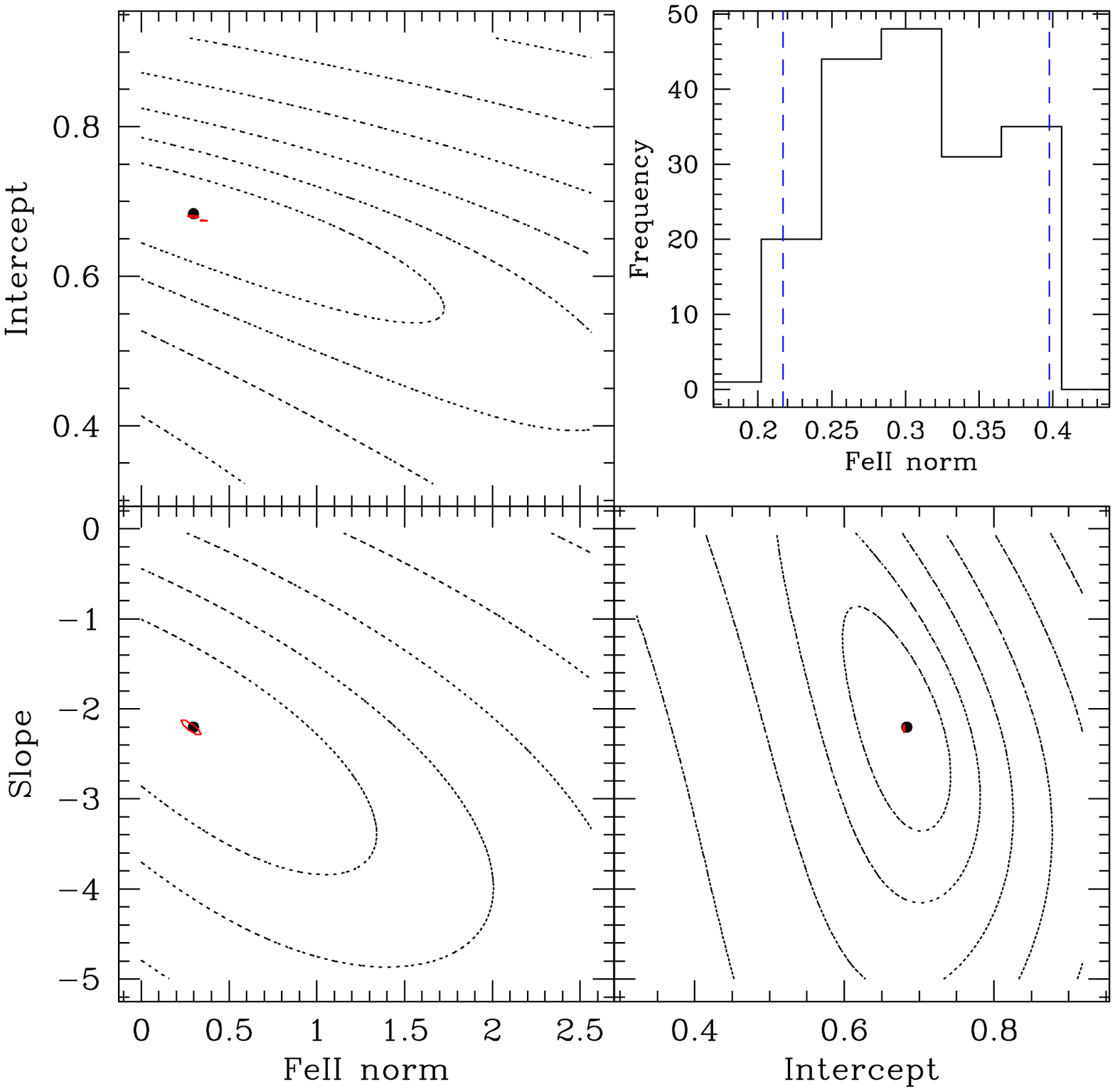}}
\end{figure*}
\end{center}

\begin{center}
\begin{figure*}[h]
\centering
\resizebox{0.3\textwidth}{!}{\includegraphics{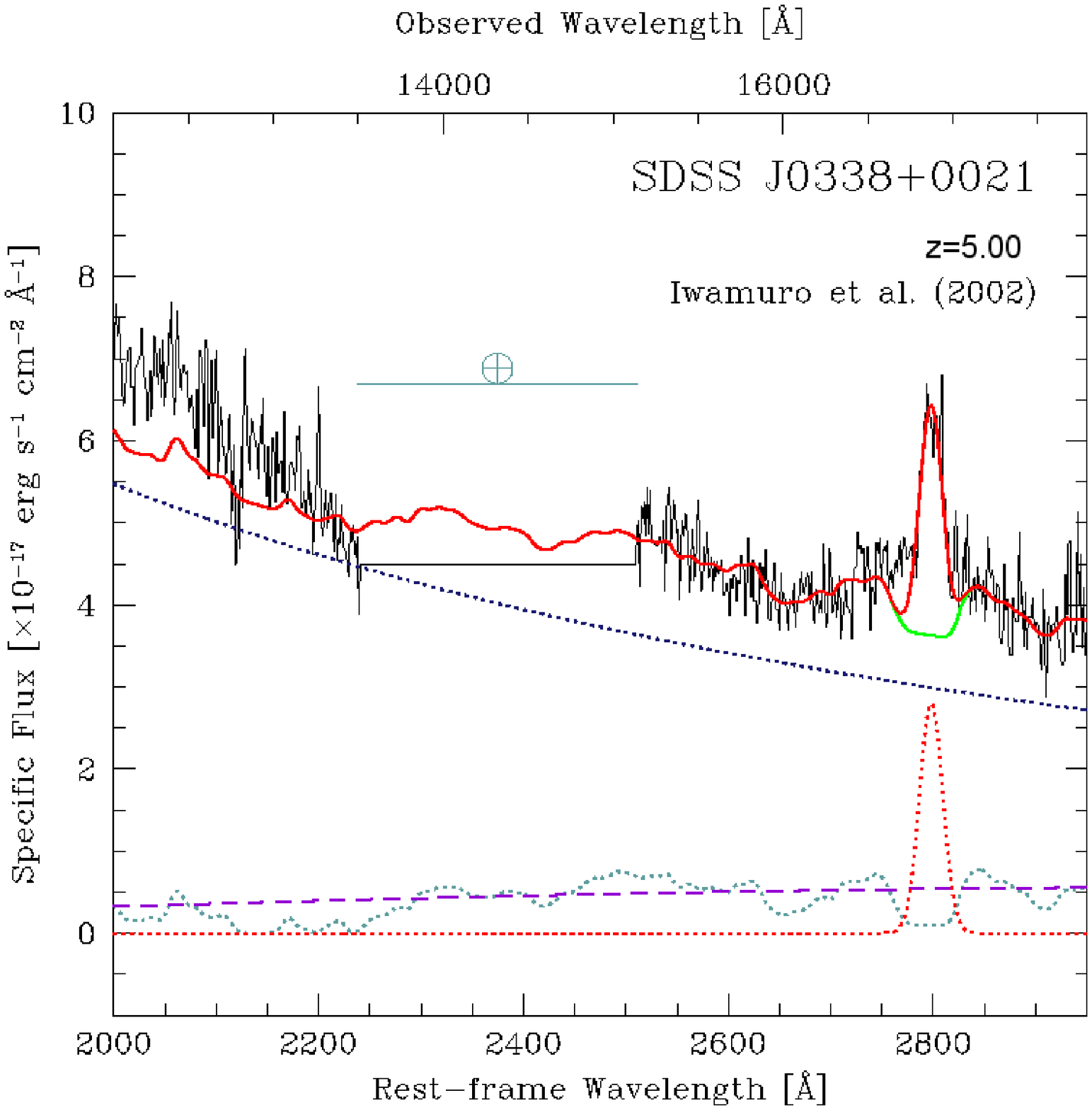}}
\resizebox{0.3\textwidth}{!}{\includegraphics{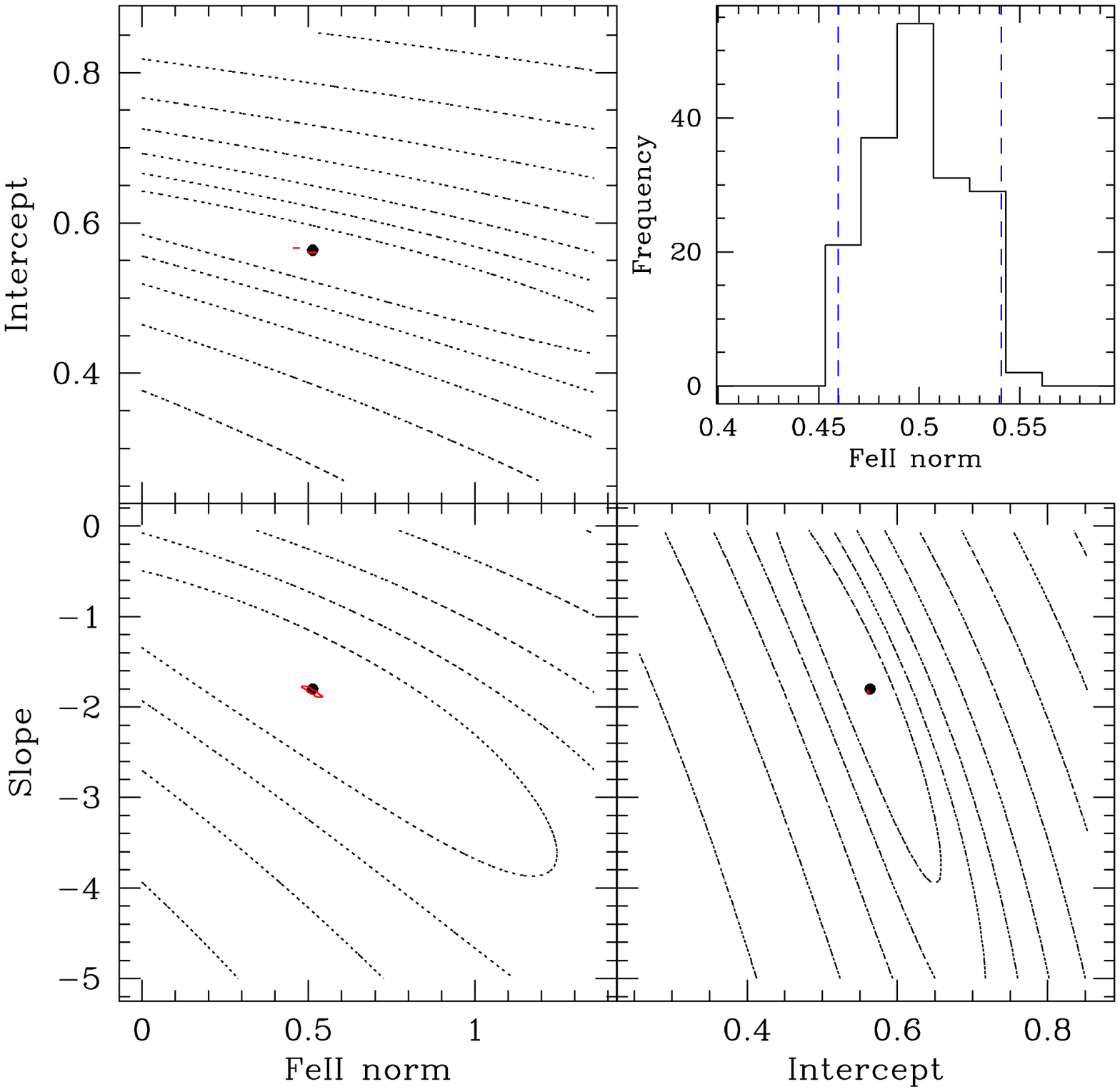}}
\end{figure*}
\end{center}

\begin{center}
\begin{figure*}[h]
\centering
\resizebox{0.3\textwidth}{!}{\includegraphics{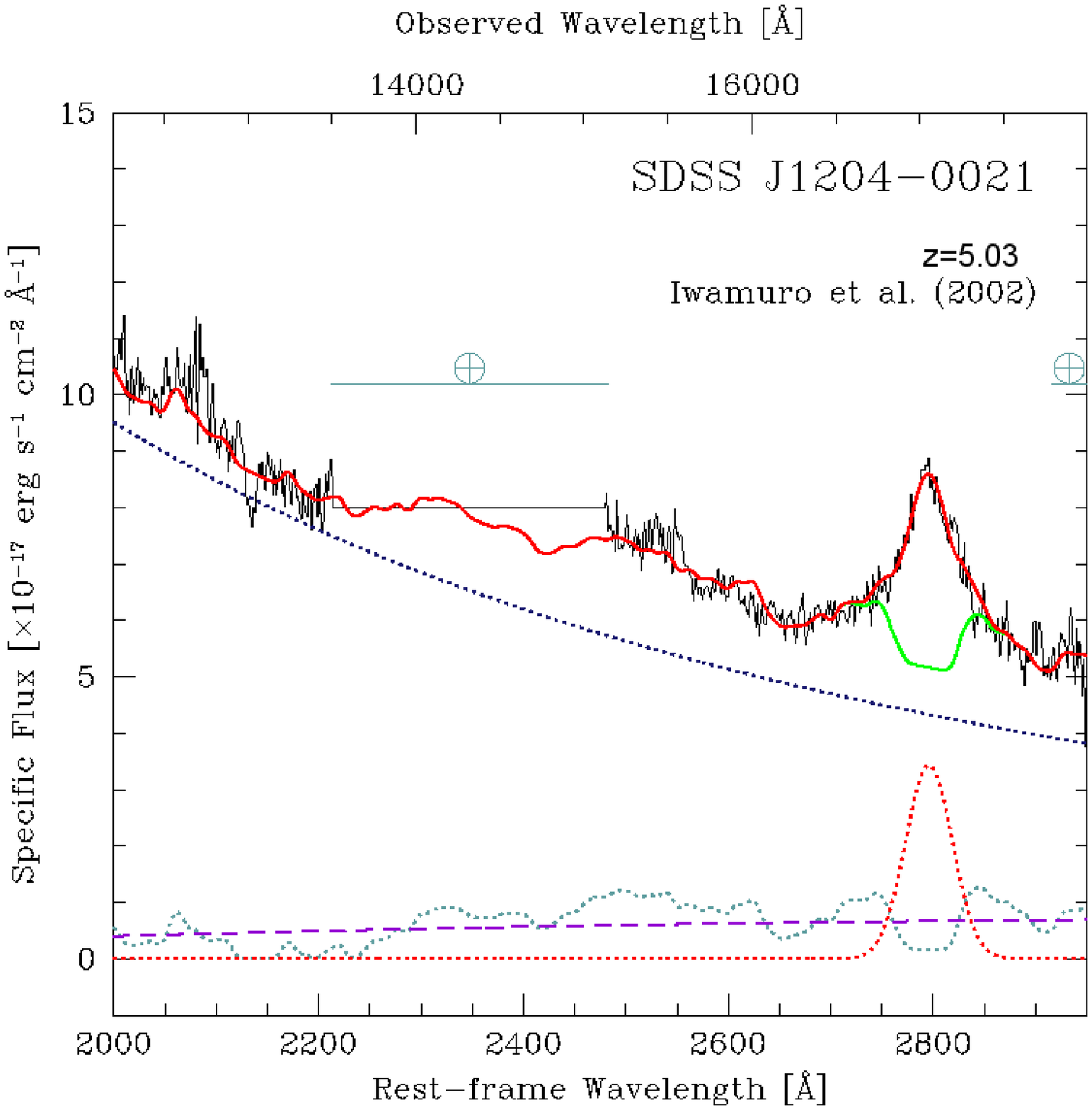}}
\resizebox{0.3\textwidth}{!}{\includegraphics{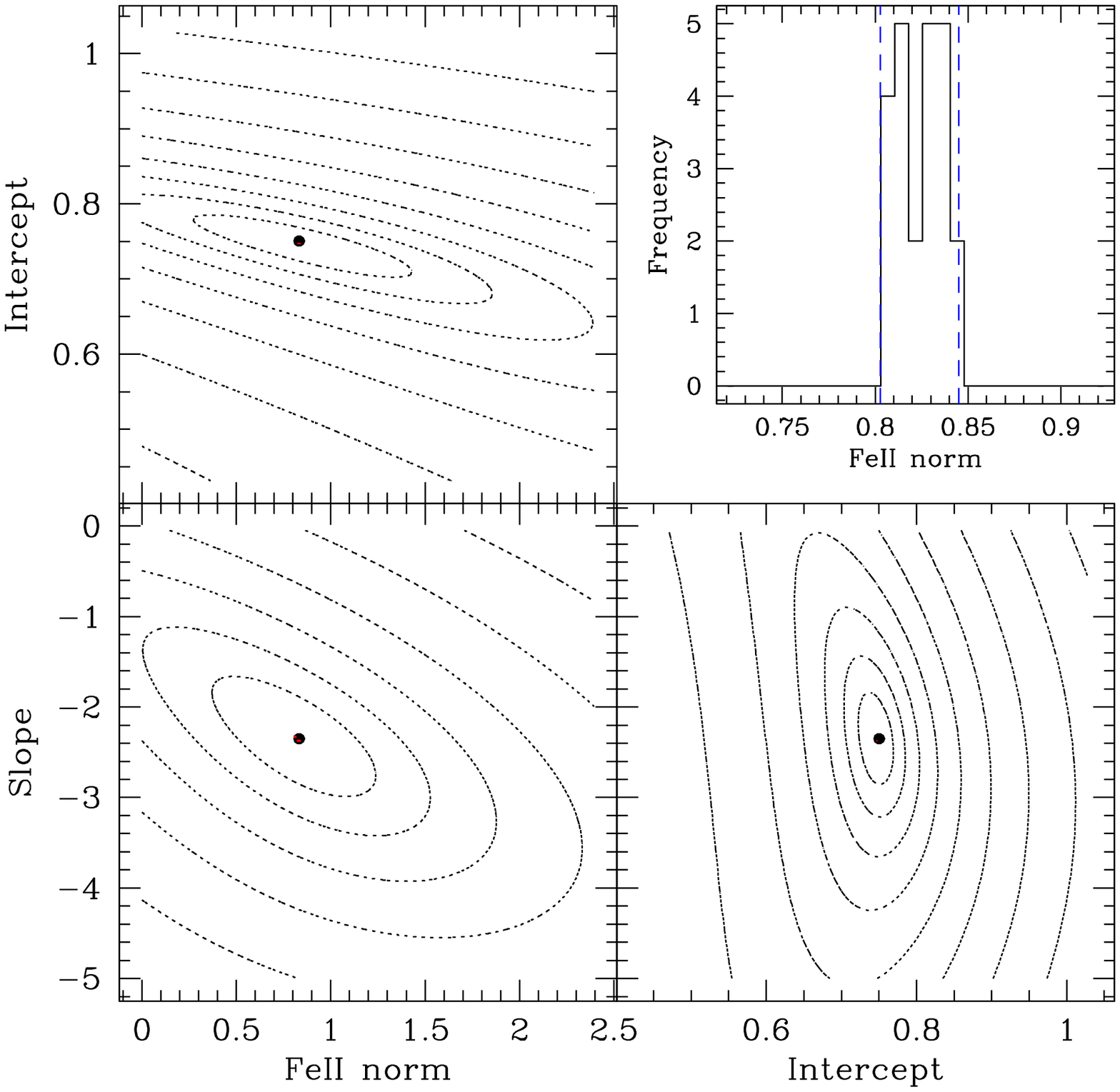}}
\end{figure*}
\end{center}

\begin{center}
\begin{figure*}[h]
\centering
\resizebox{0.3\textwidth}{!}{\includegraphics{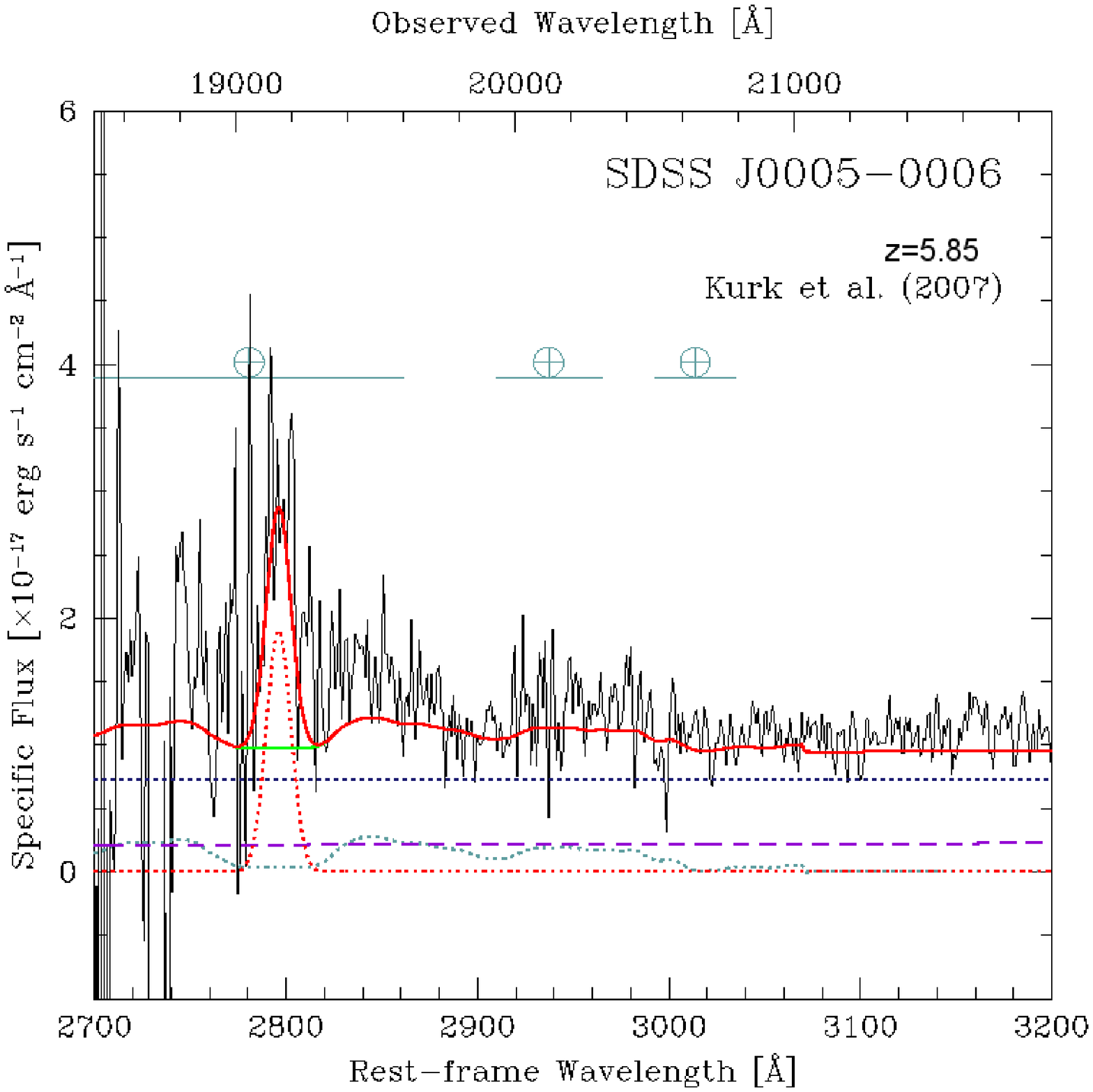}}
\resizebox{0.3\textwidth}{!}{\includegraphics{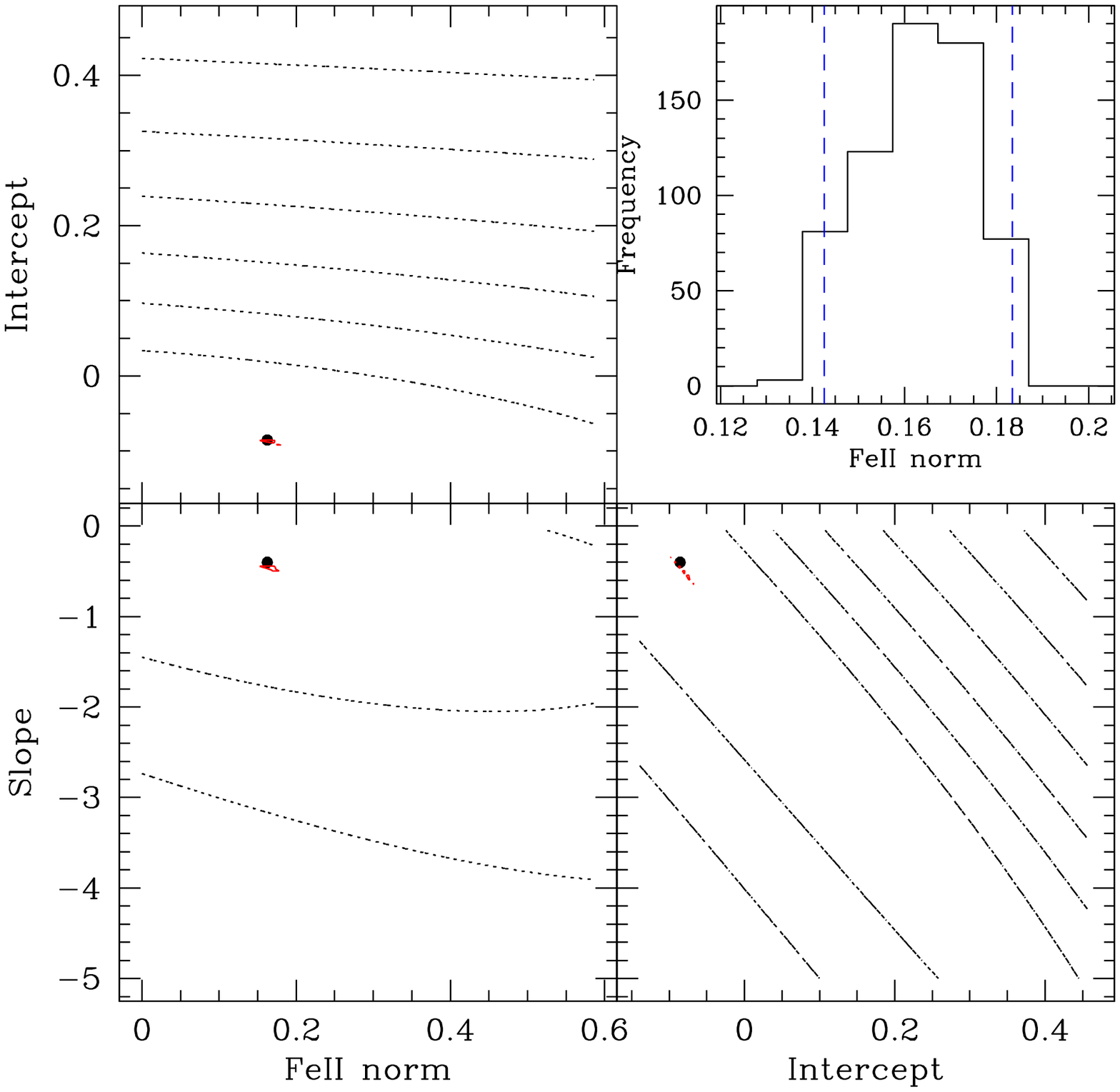}}
\end{figure*}
\end{center}

\begin{center}
\begin{figure*}[h]
\centering
\resizebox{0.3\textwidth}{!}{\includegraphics{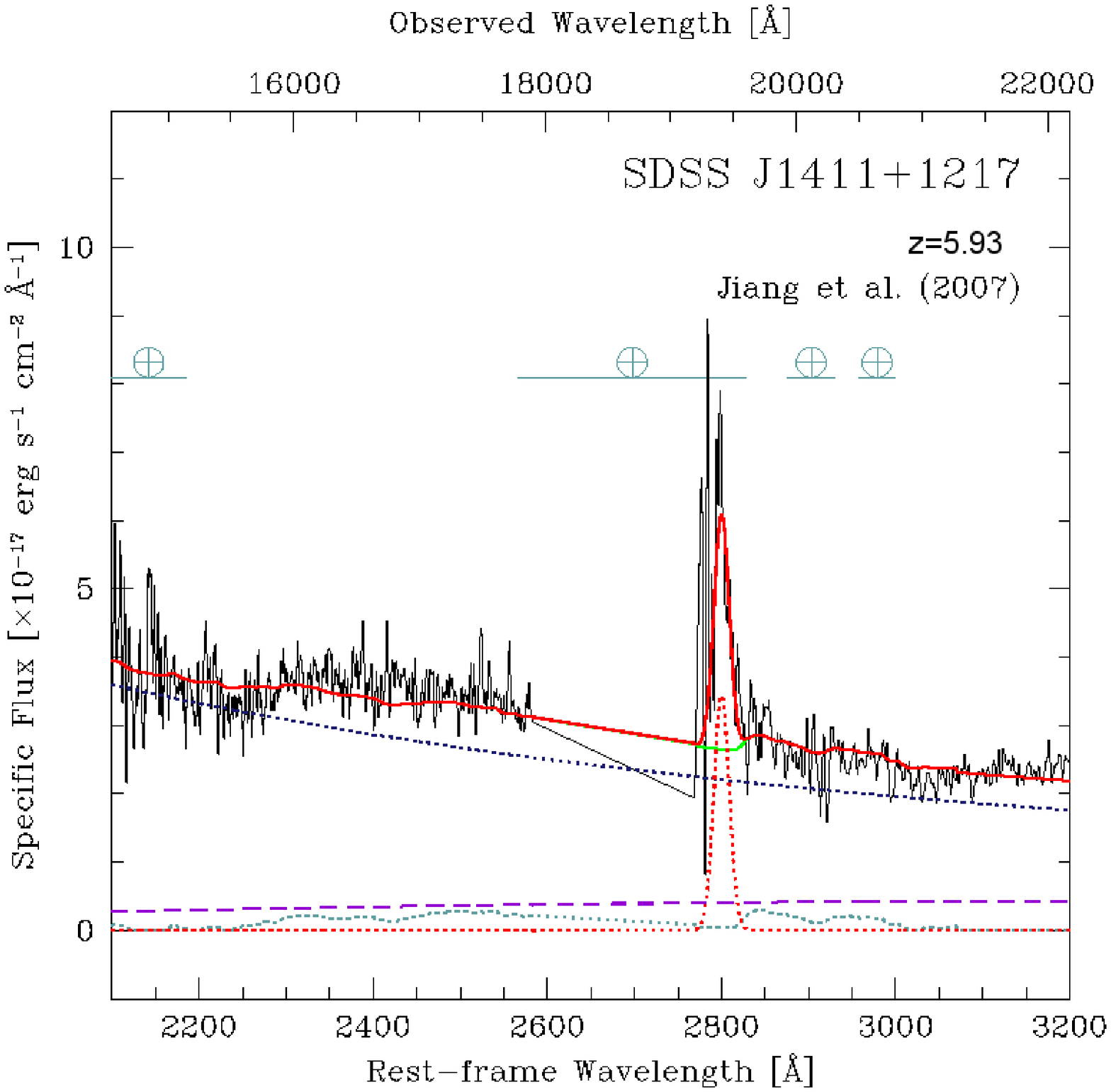}}
\resizebox{0.3\textwidth}{!}{\includegraphics{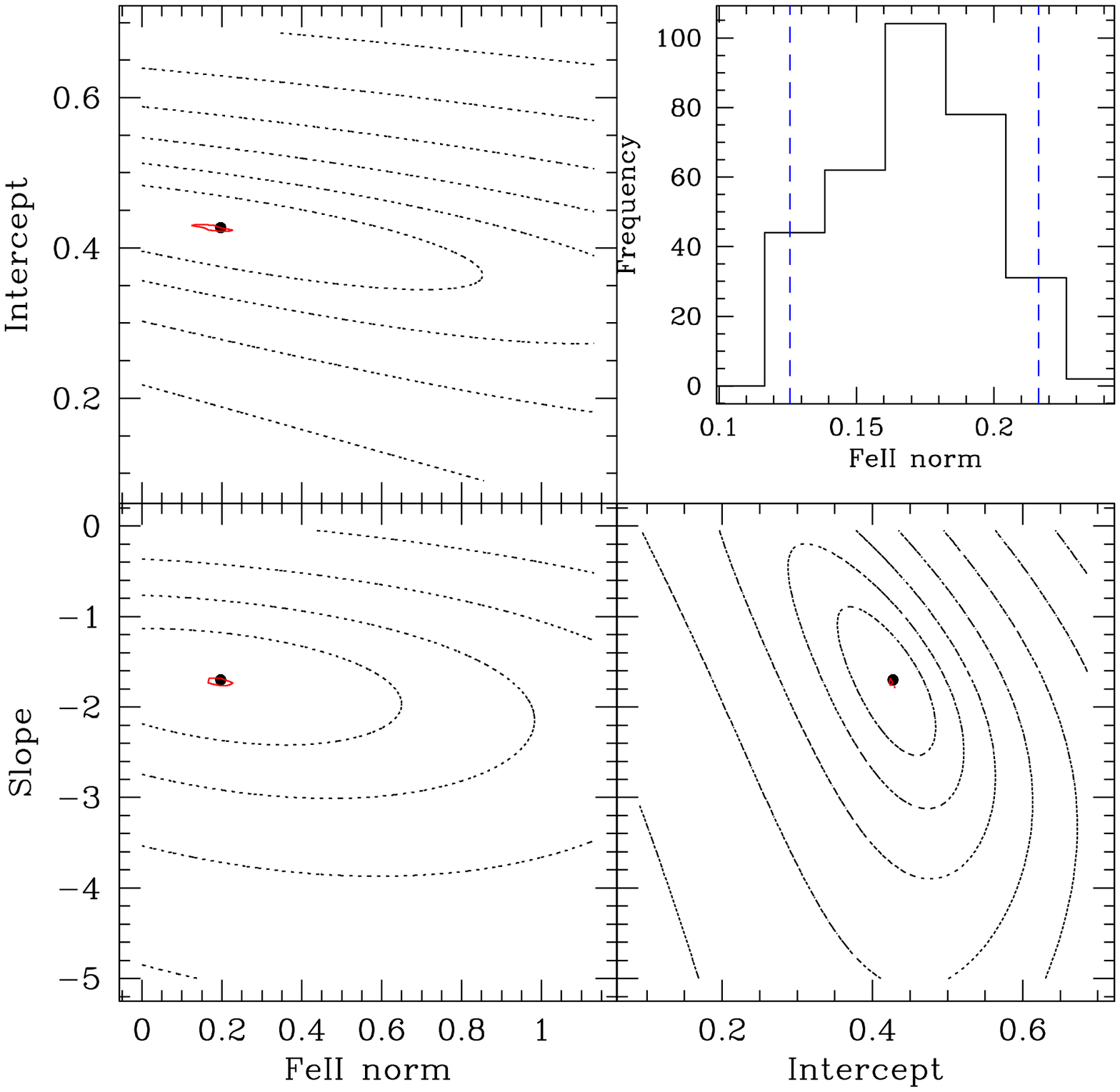}}
\end{figure*}
\end{center}

\begin{center}
\begin{figure*}[h]
\centering
\resizebox{0.3\textwidth}{!}{\includegraphics{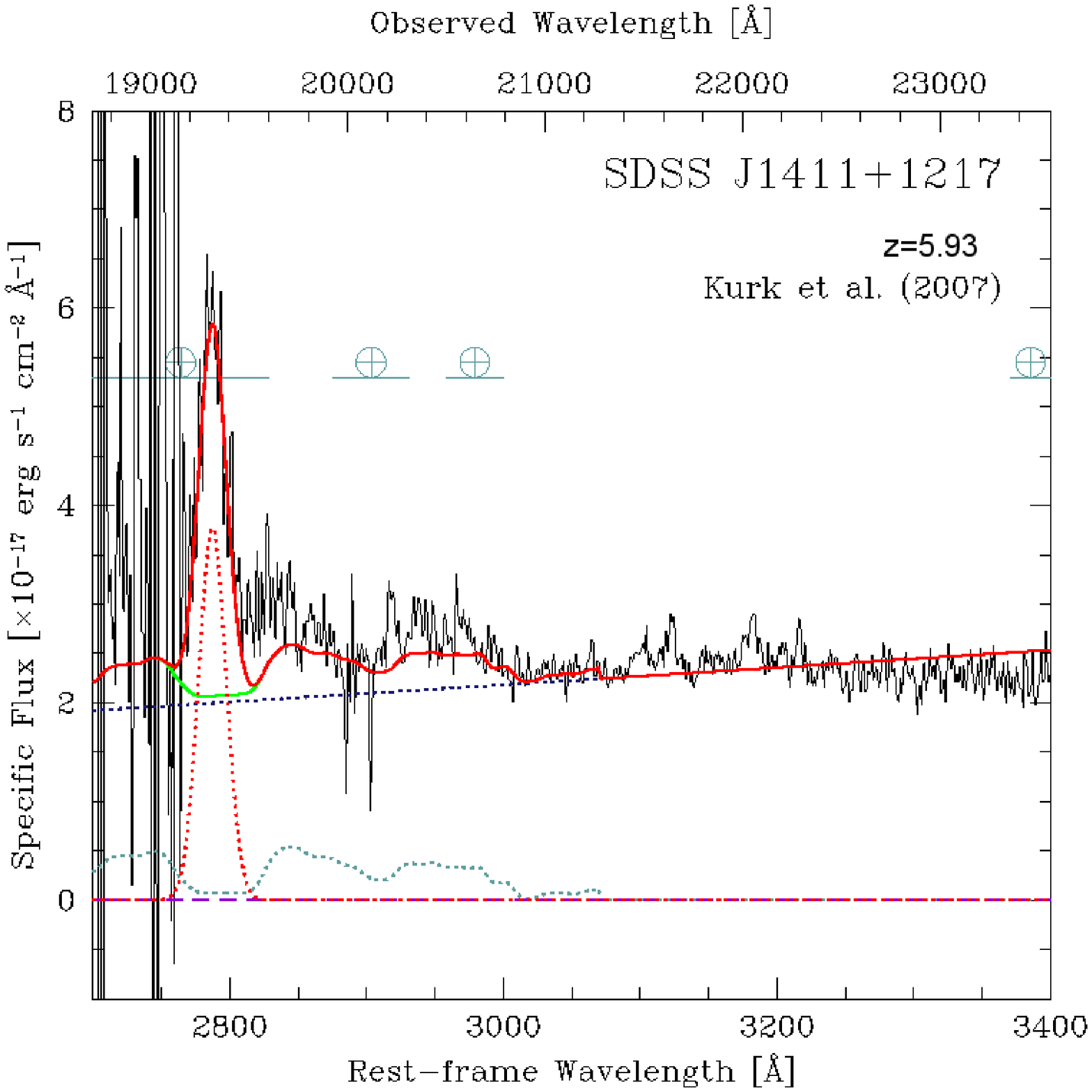}}
\resizebox{0.3\textwidth}{!}{\includegraphics{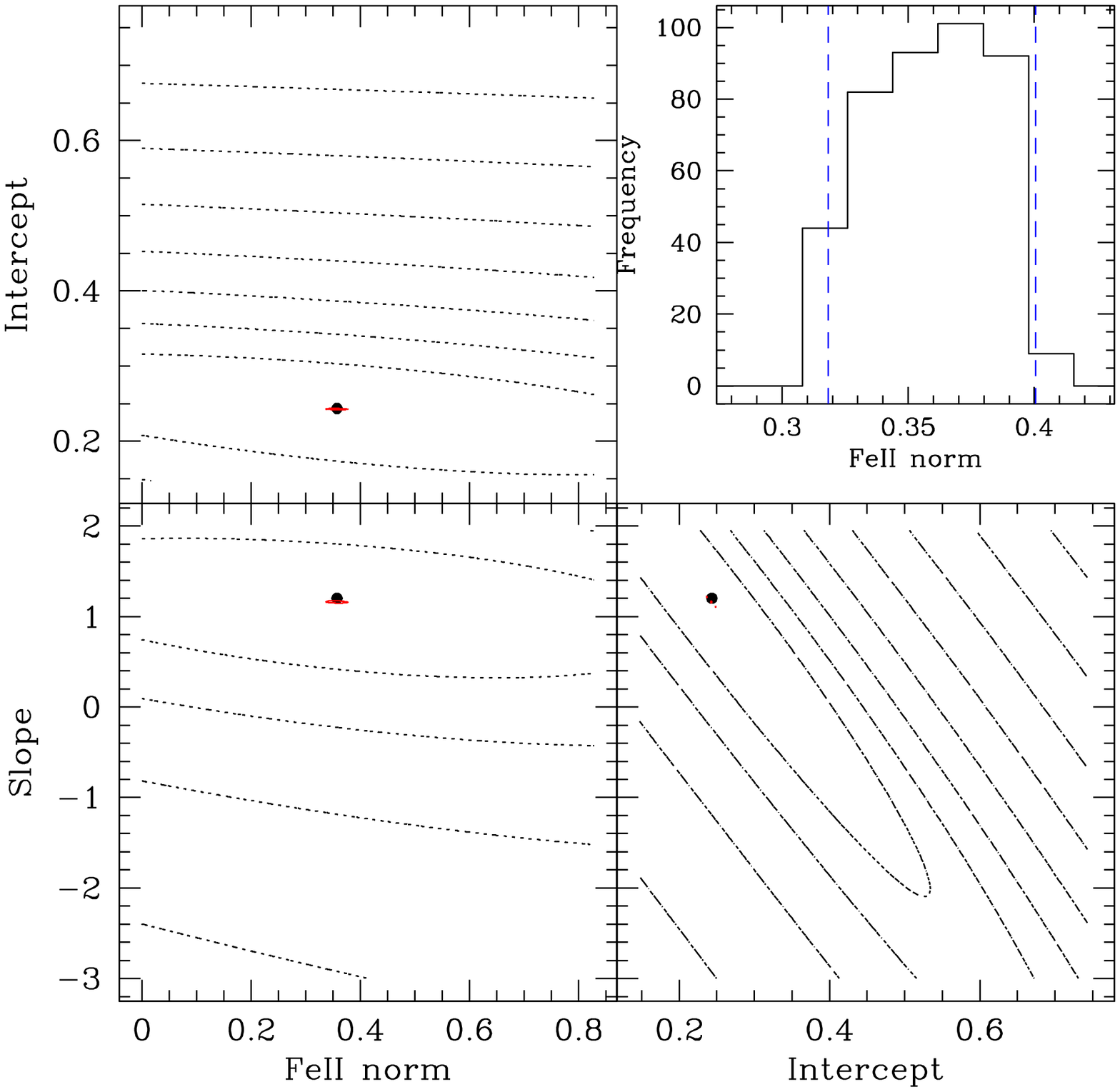}}
\end{figure*}
\end{center}

\begin{center}
\begin{figure*}[h]
\centering
\resizebox{0.3\textwidth}{!}{\includegraphics{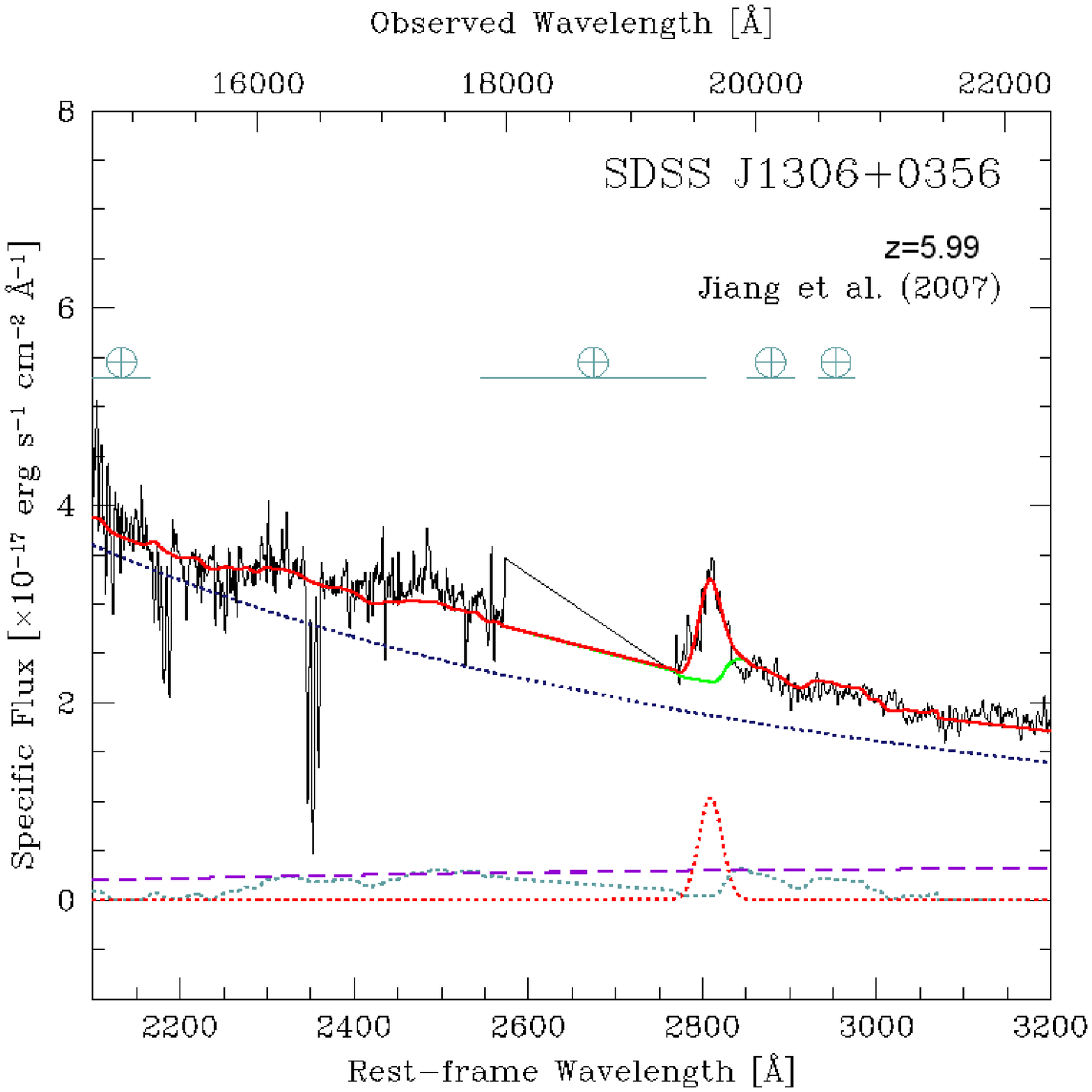}}
\resizebox{0.3\textwidth}{!}{\includegraphics{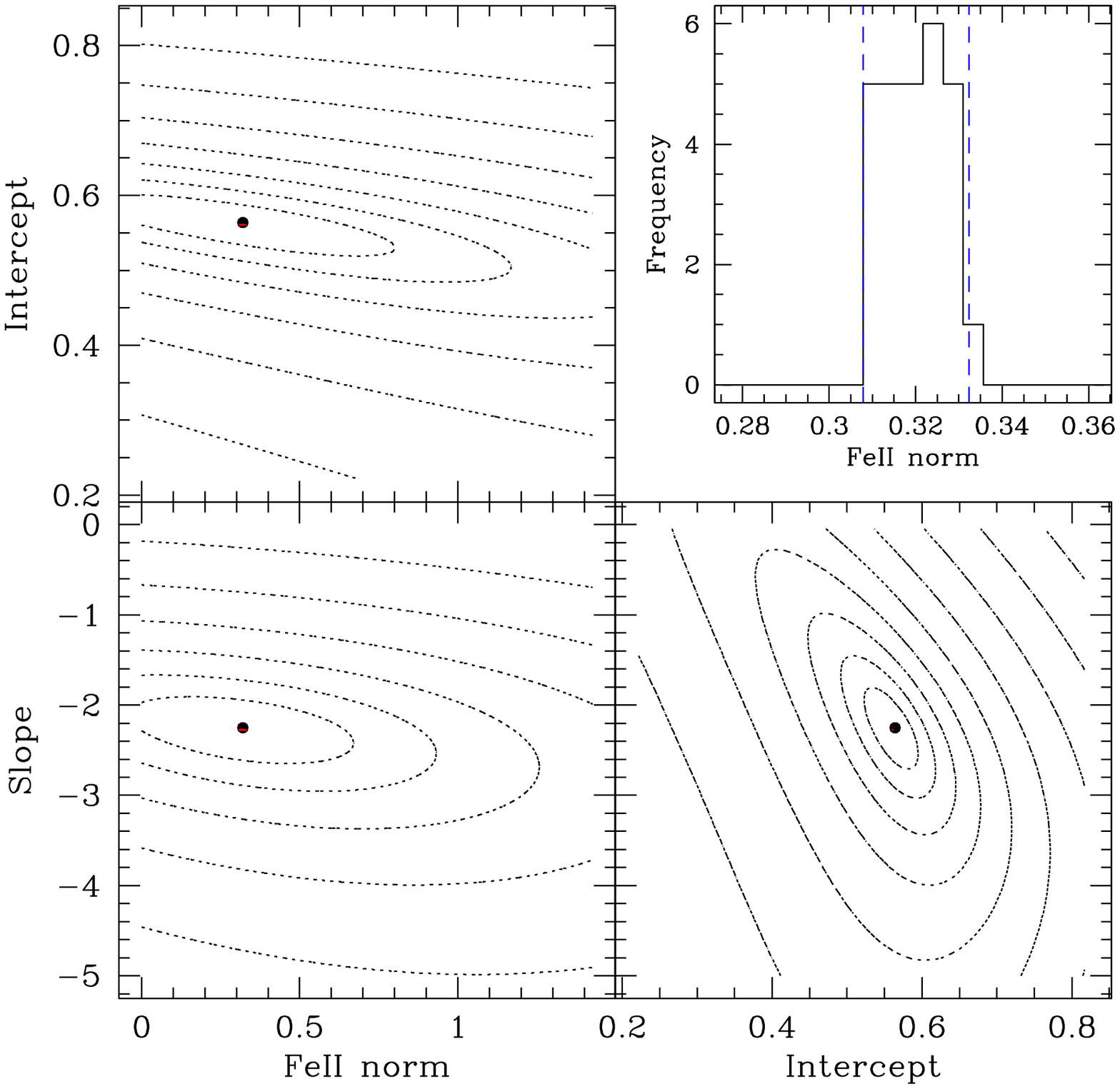}}
\end{figure*}
\end{center}

\begin{center}
\begin{figure*}[h]
\centering
\resizebox{0.3\textwidth}{!}{\includegraphics{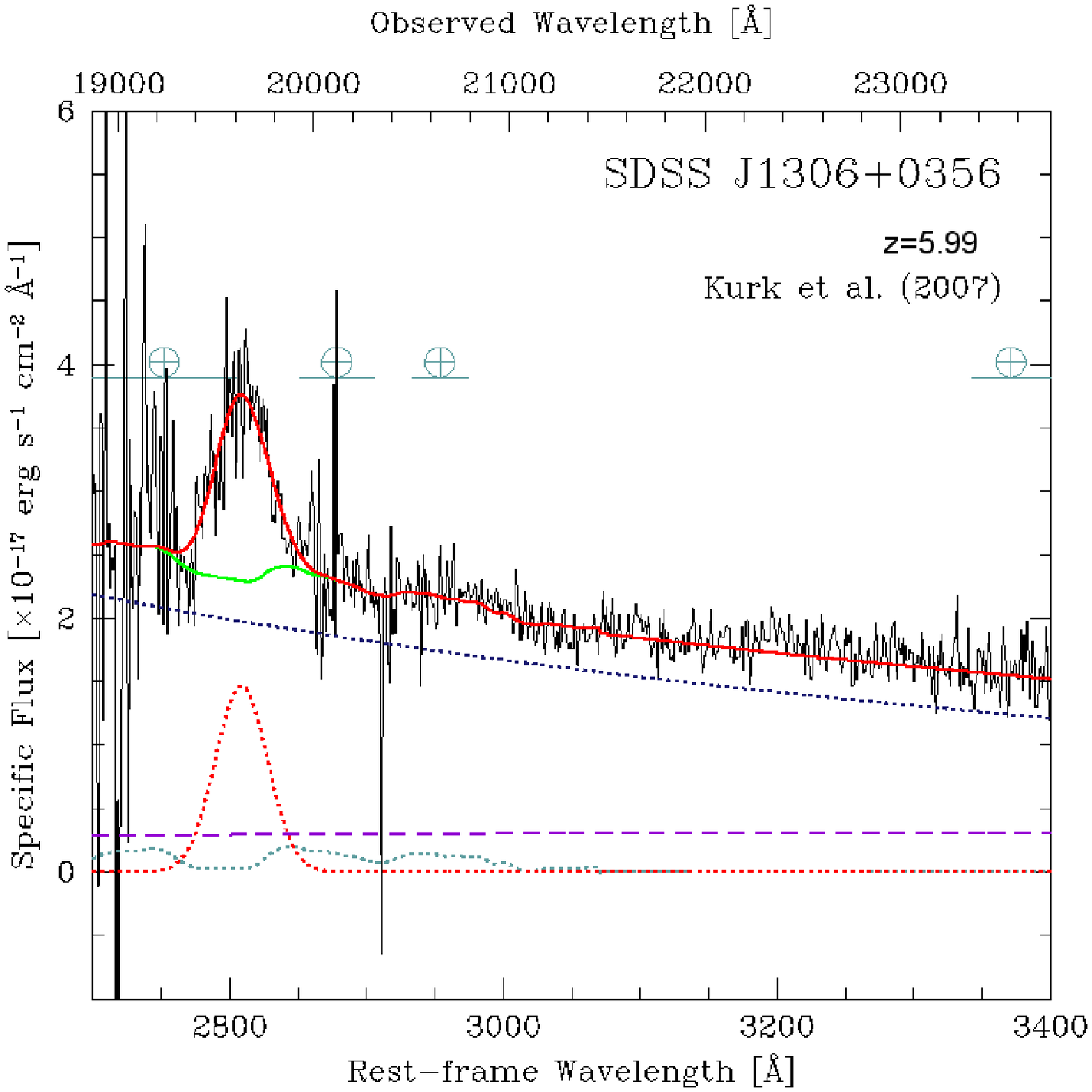}}
\resizebox{0.3\textwidth}{!}{\includegraphics{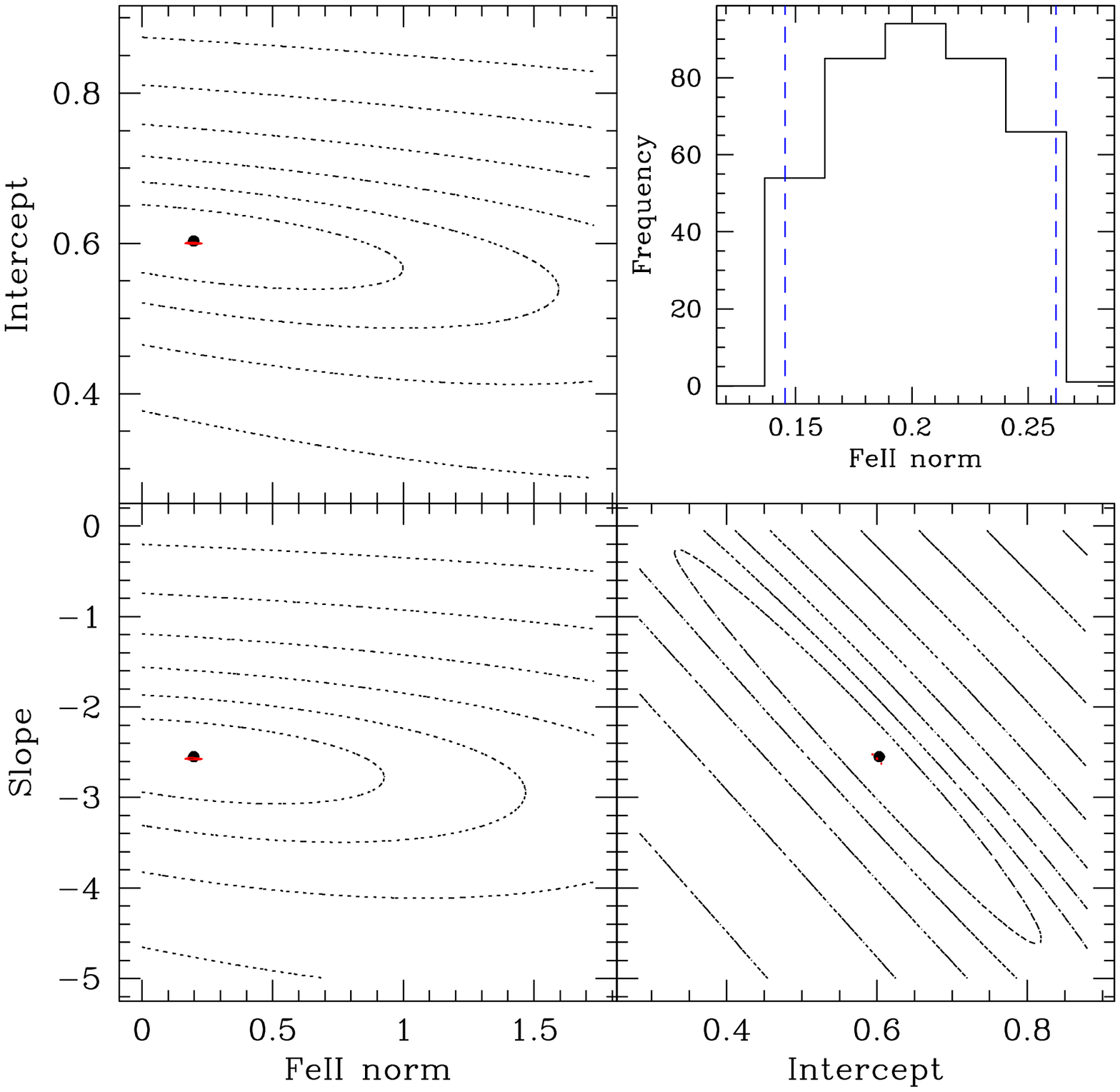}}
\end{figure*}
\end{center}

\begin{center}
\begin{figure*}[h]
\centering
\resizebox{0.3\textwidth}{!}{\includegraphics{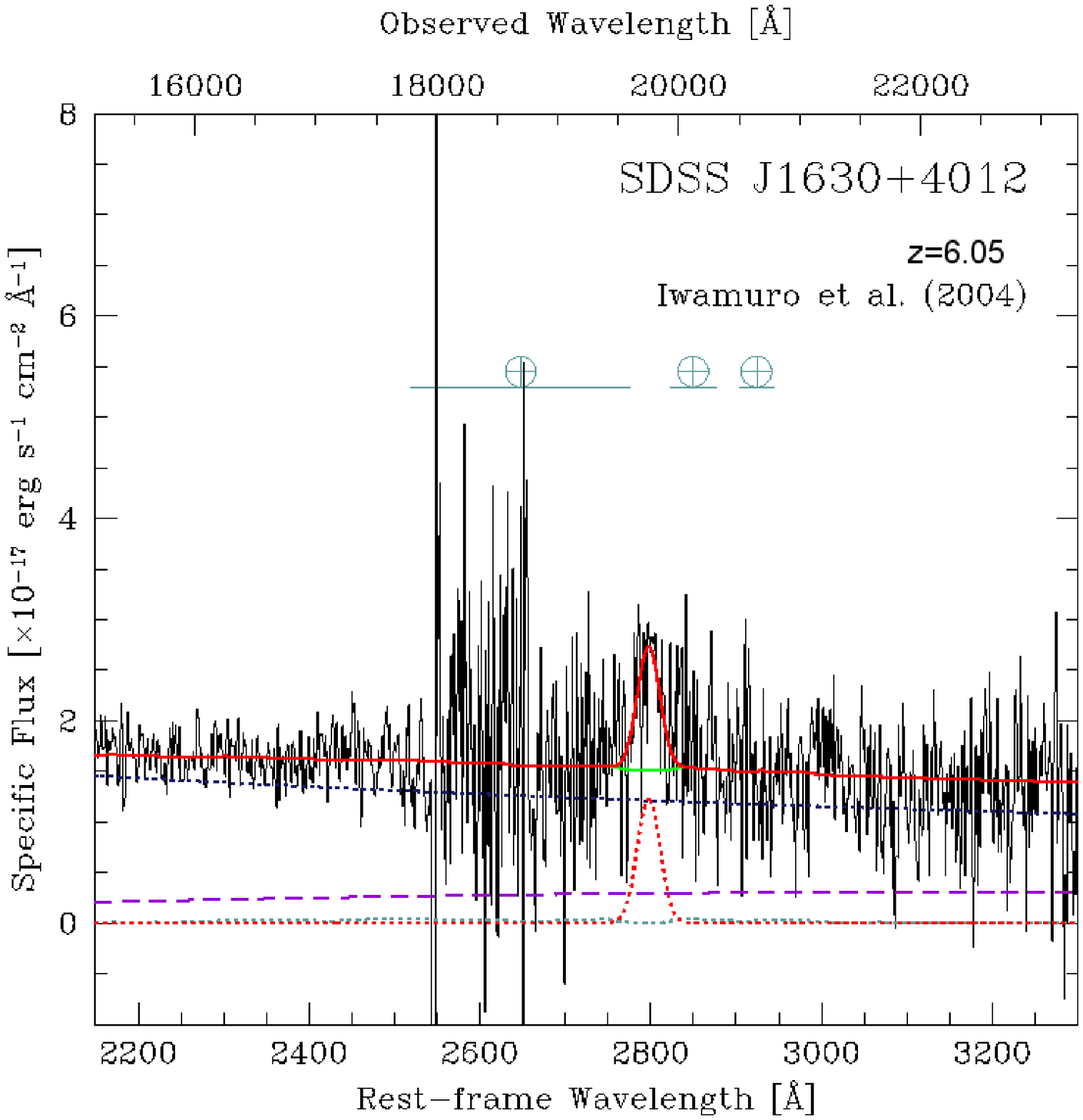}}
\resizebox{0.3\textwidth}{!}{\includegraphics{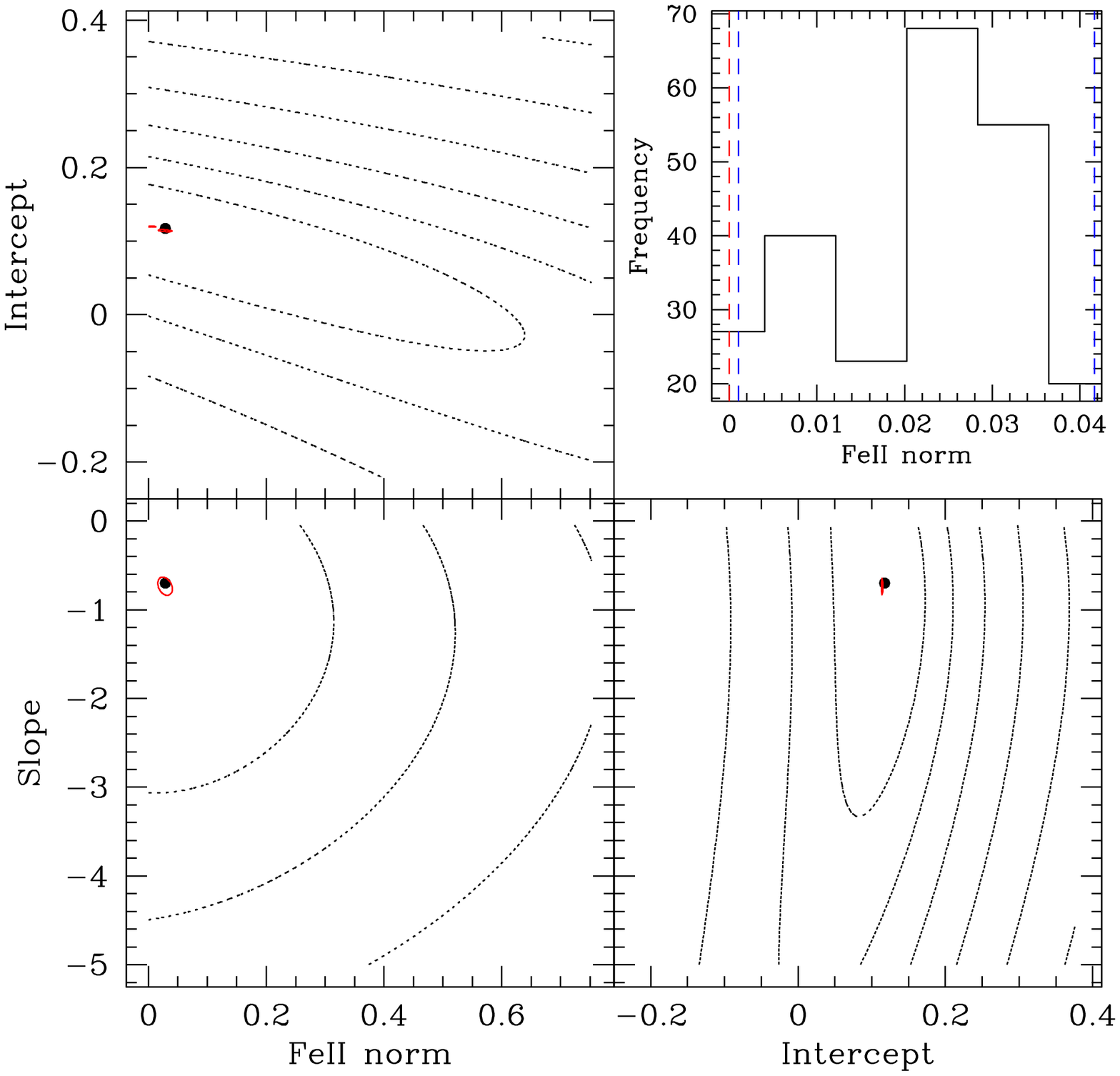}}
\end{figure*}
\end{center}

\begin{center}
\begin{figure*}[h]
\centering
\resizebox{0.3\textwidth}{!}{\includegraphics{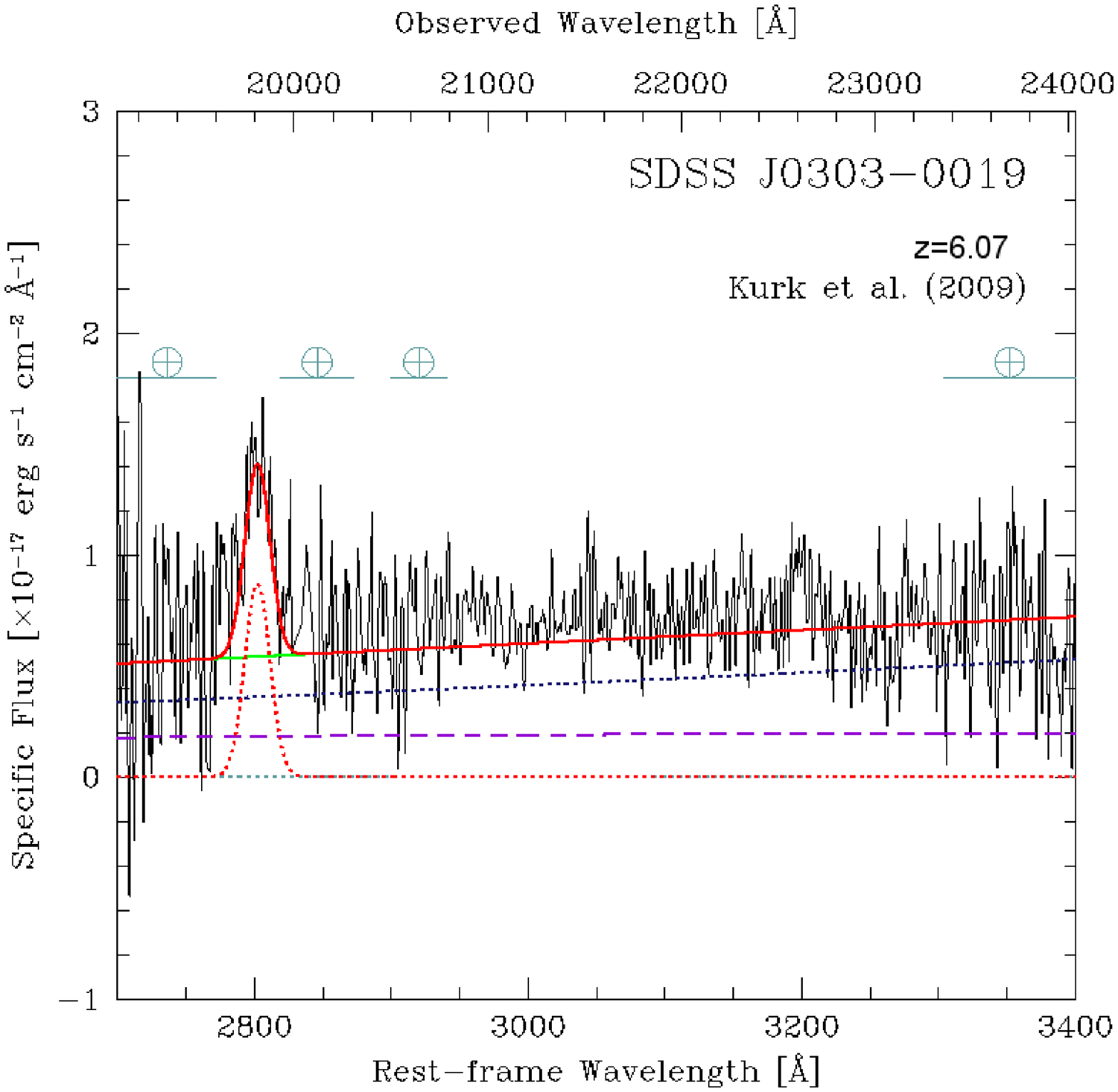}}
\resizebox{0.3\textwidth}{!}{\includegraphics{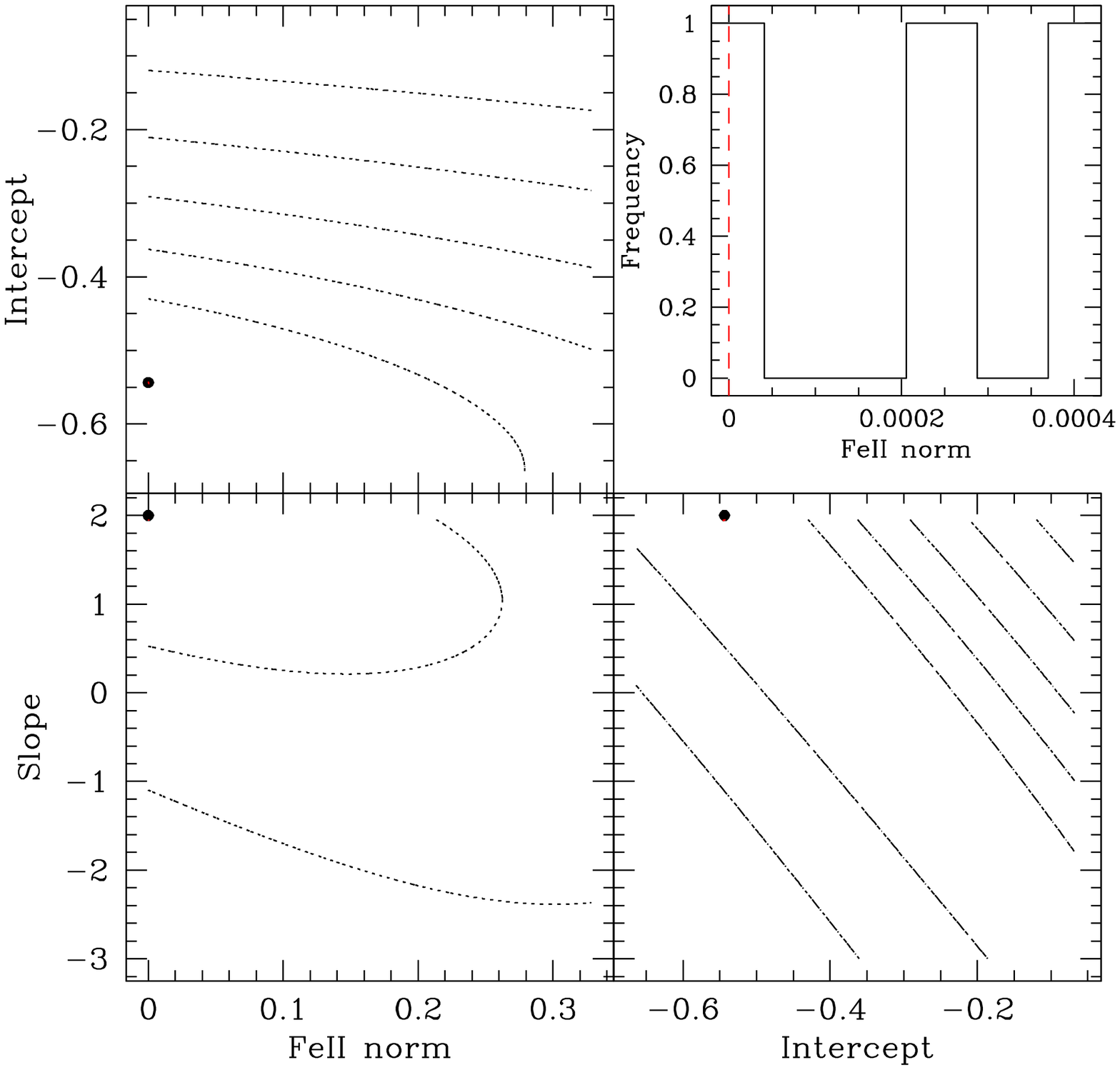}}
\end{figure*}
\end{center}

\begin{center}
\begin{figure*}[h]
\centering
\resizebox{0.3\textwidth}{!}{\includegraphics{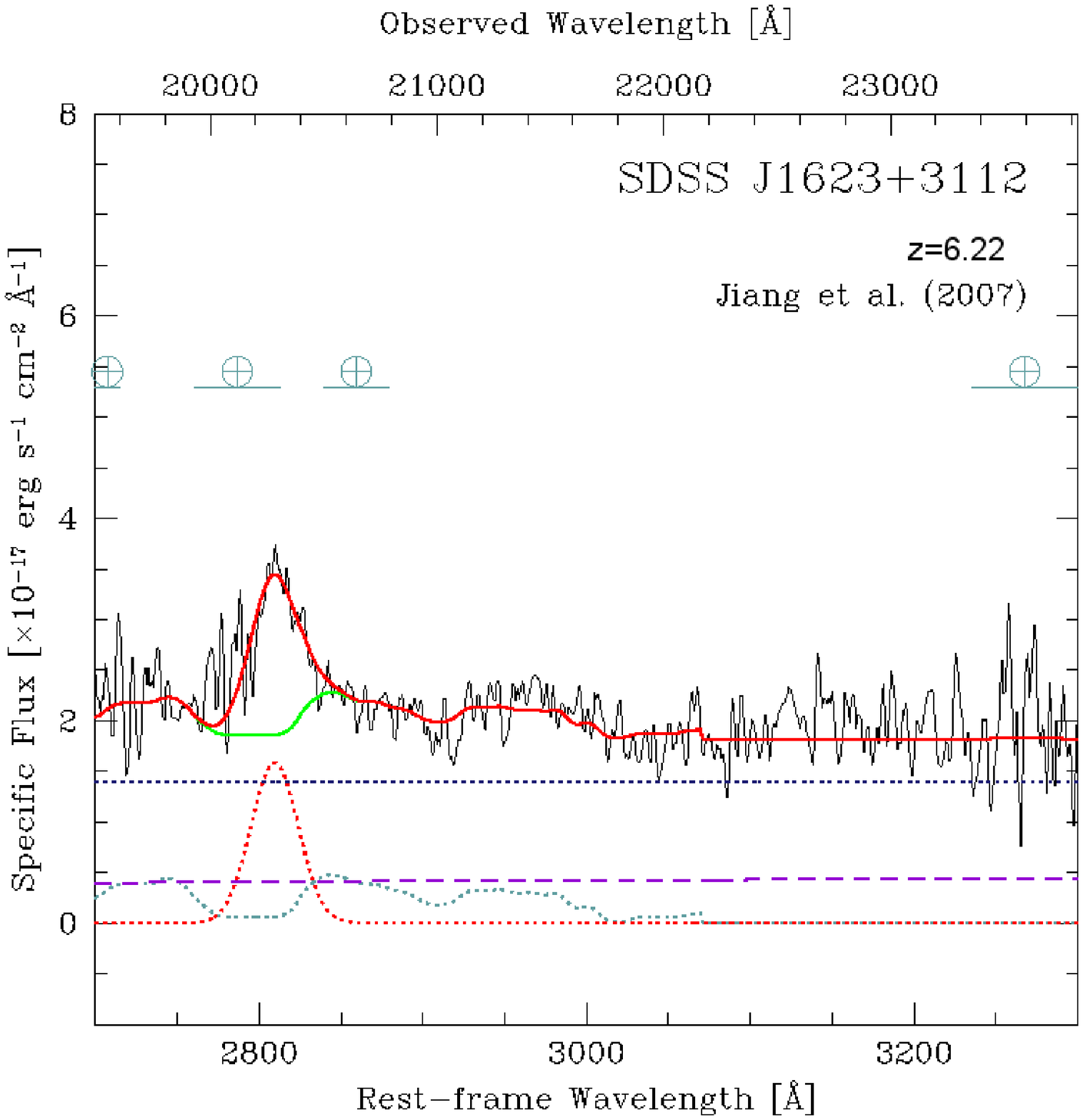}}
\resizebox{0.3\textwidth}{!}{\includegraphics{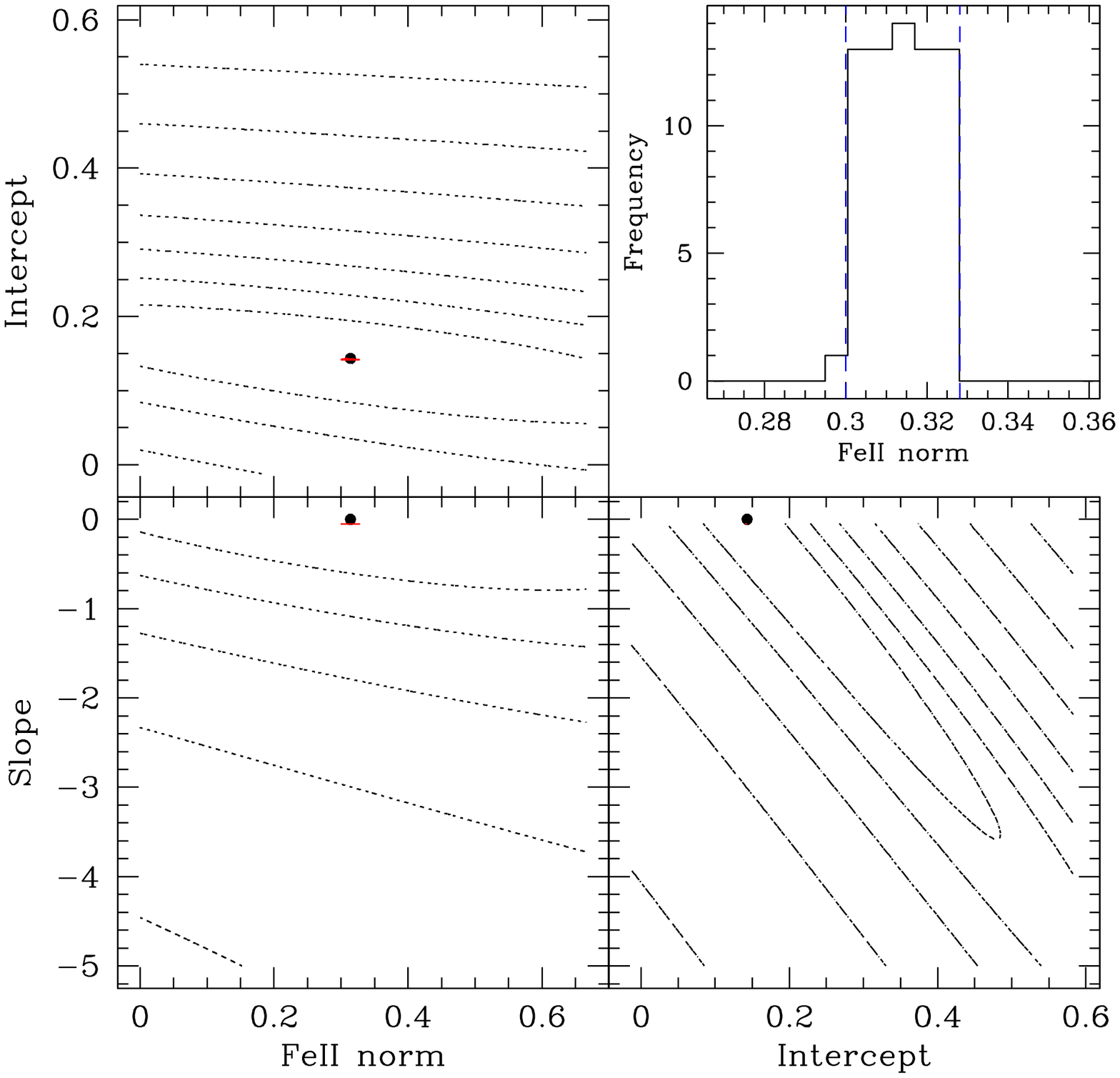}}
\end{figure*}
\end{center}

\begin{center}
\begin{figure*}[h]
\centering
\resizebox{0.3\textwidth}{!}{\includegraphics{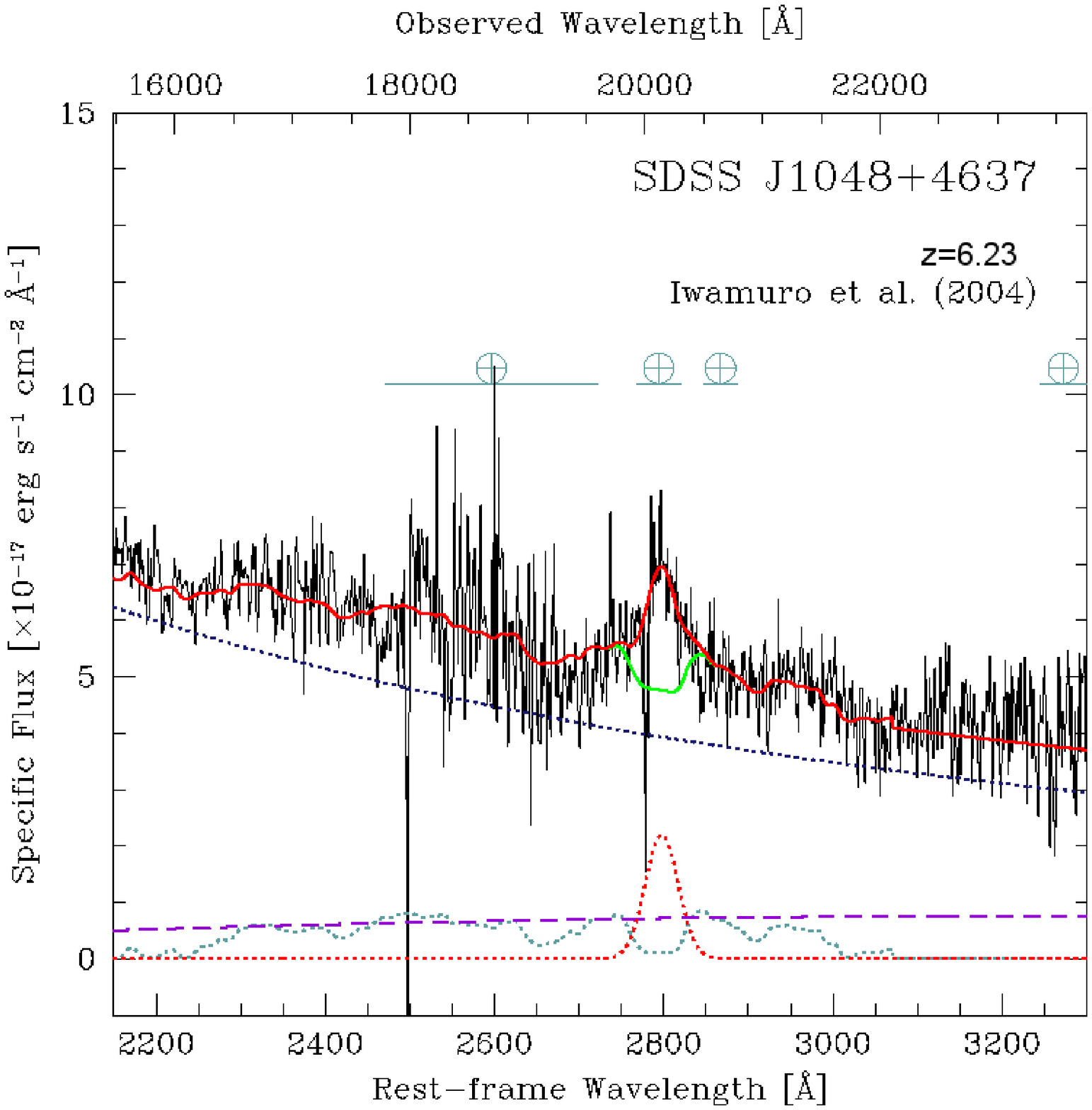}}
\resizebox{0.3\textwidth}{!}{\includegraphics{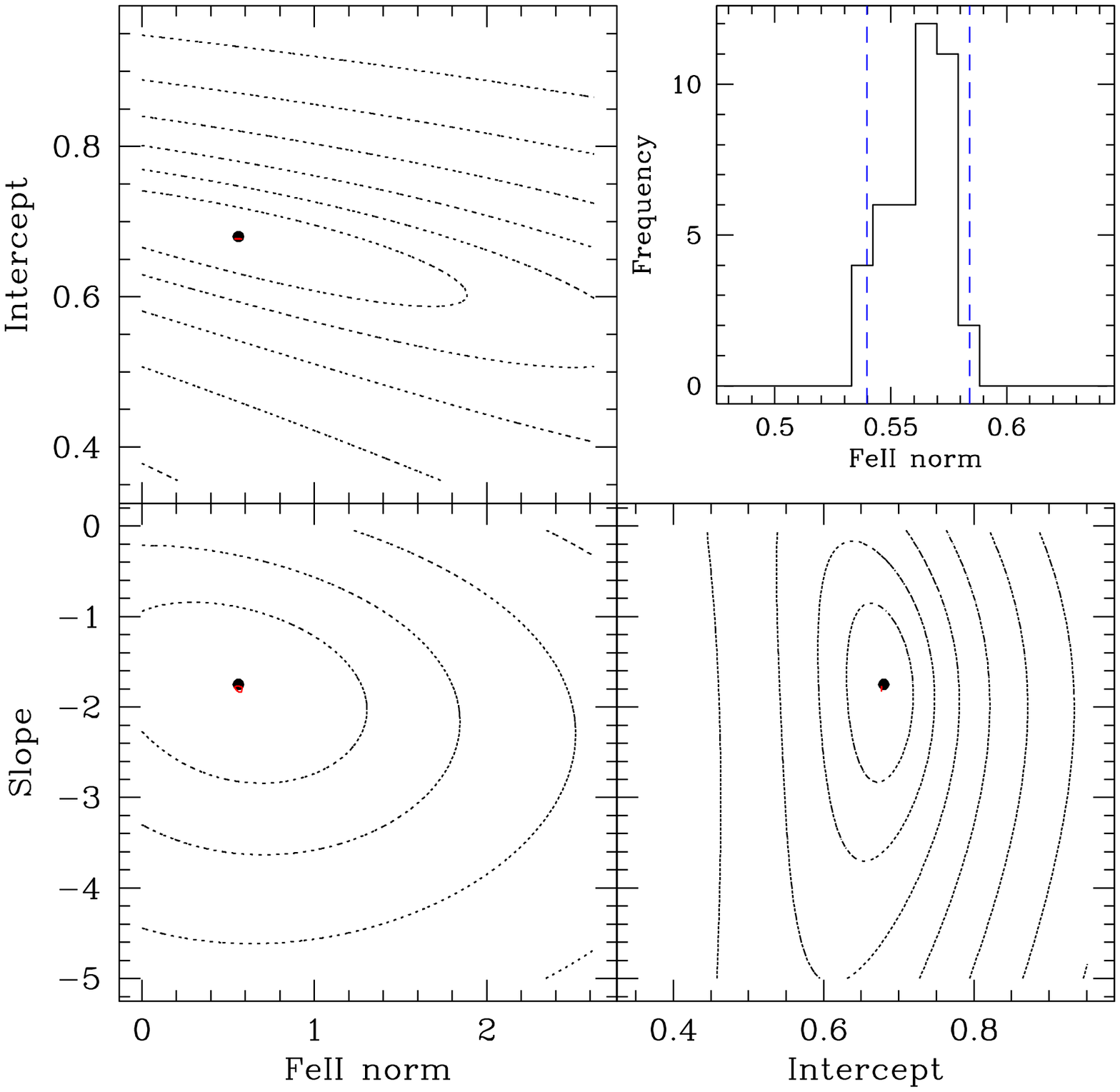}}
\end{figure*}
\end{center}

\begin{center}
\begin{figure*}[h]
\centering
\resizebox{0.3\textwidth}{!}{\includegraphics{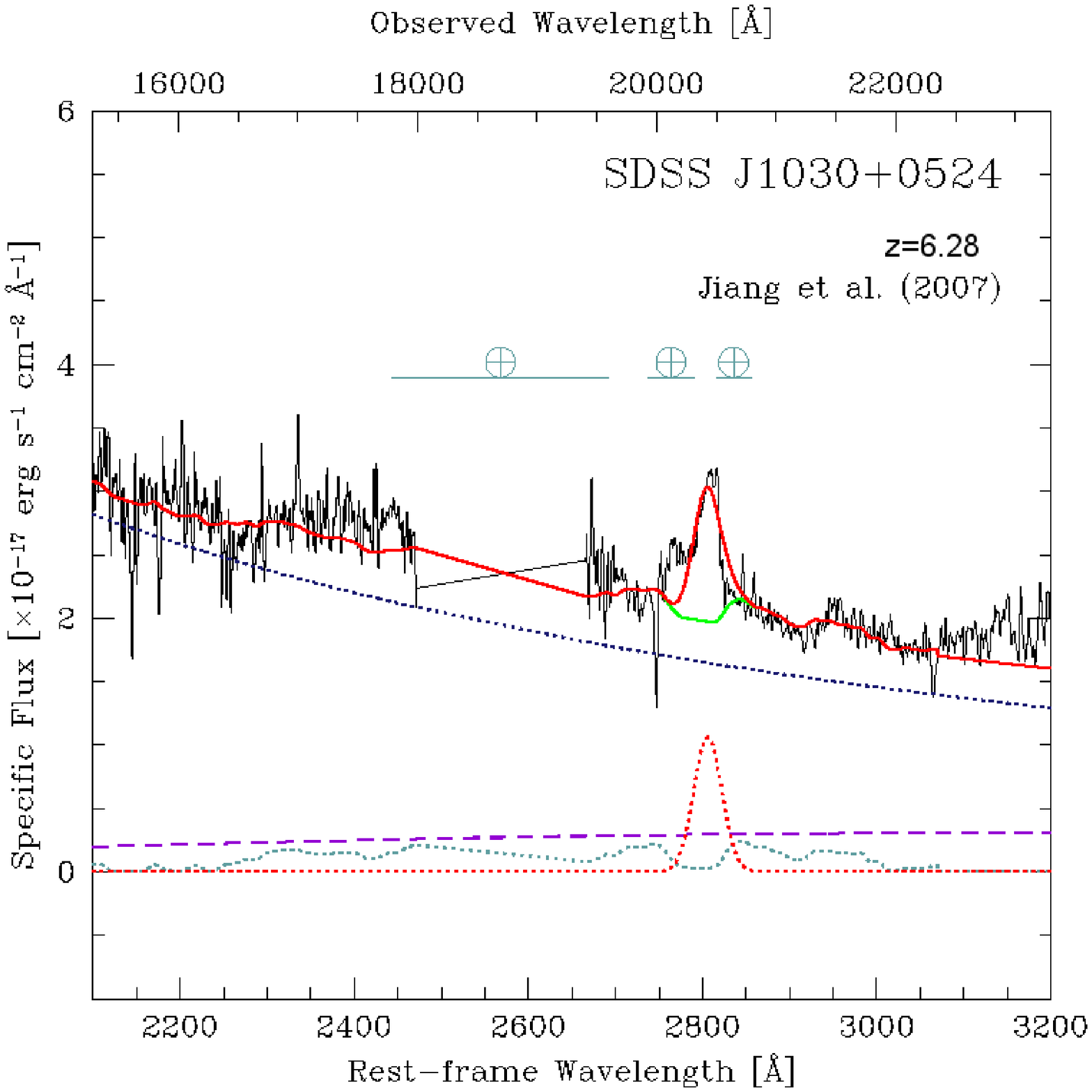}}
\resizebox{0.3\textwidth}{!}{\includegraphics{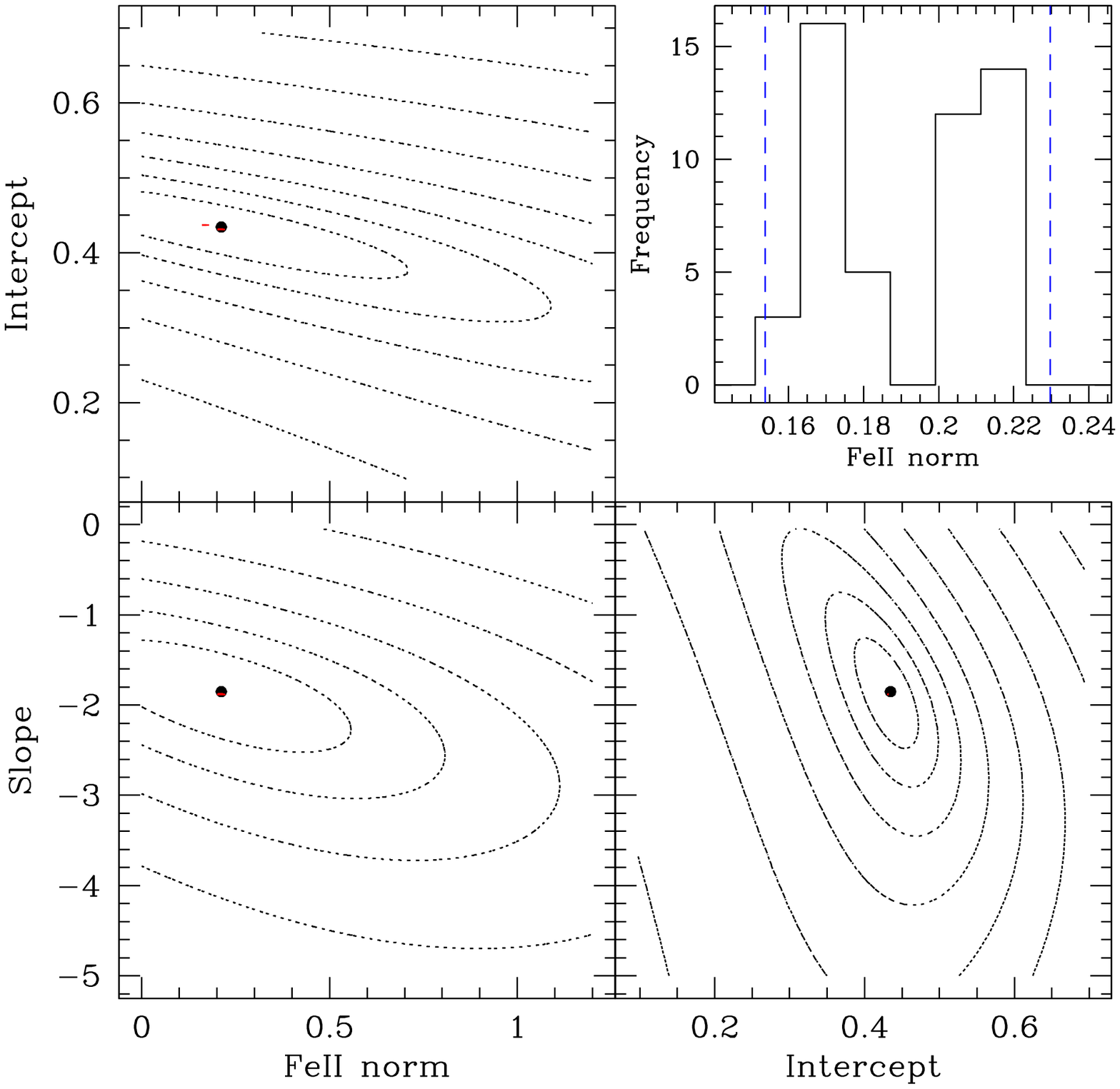}}
\end{figure*}
\end{center}

\begin{center}
\begin{figure*}[h]
\centering
\resizebox{0.3\textwidth}{!}{\includegraphics{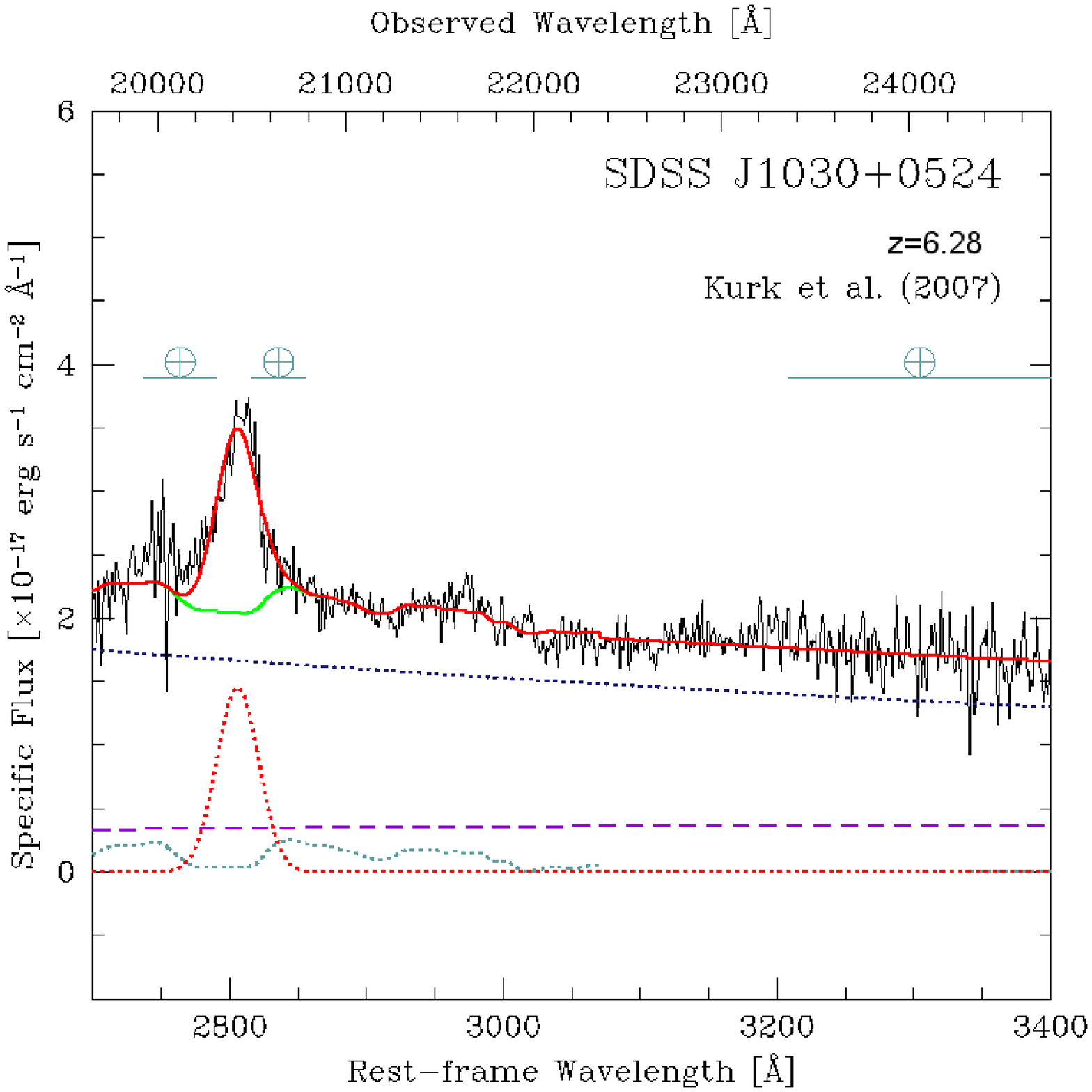}}
\resizebox{0.3\textwidth}{!}{\includegraphics{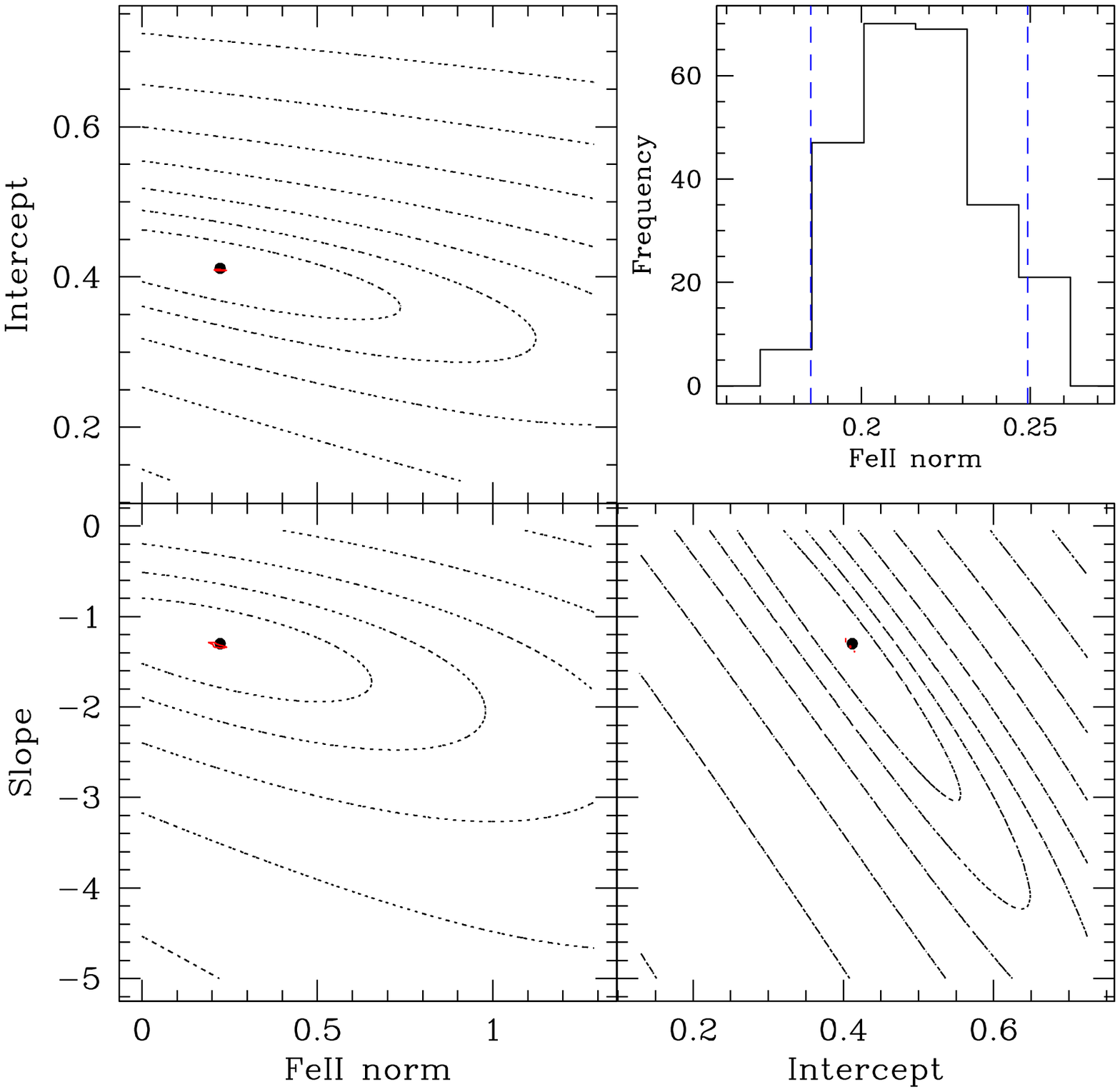}}
\end{figure*}
\end{center}

\begin{center}
\begin{figure*}[h]
\centering
\resizebox{0.3\textwidth}{!}{\includegraphics{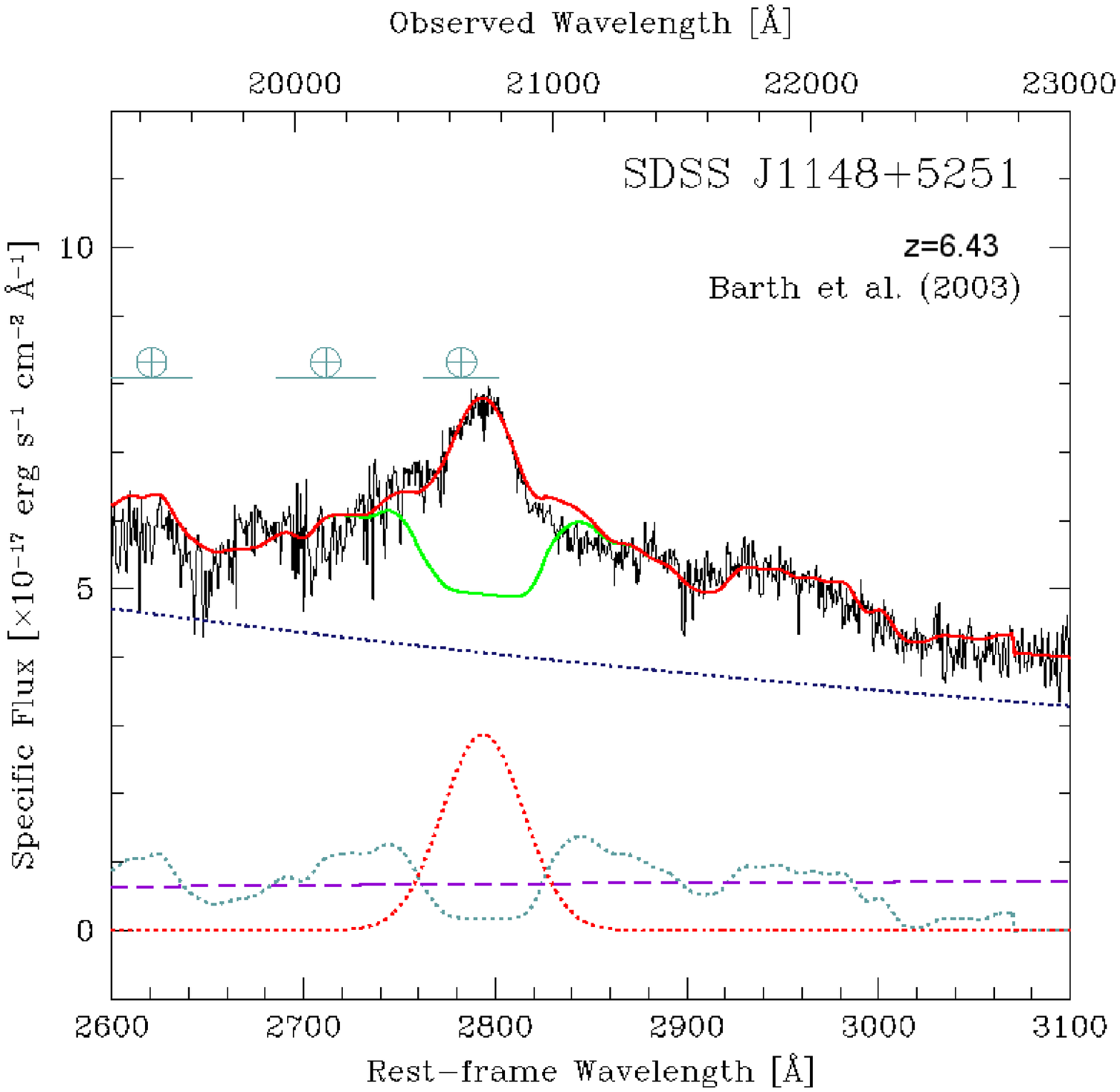}}
\resizebox{0.3\textwidth}{!}{\includegraphics{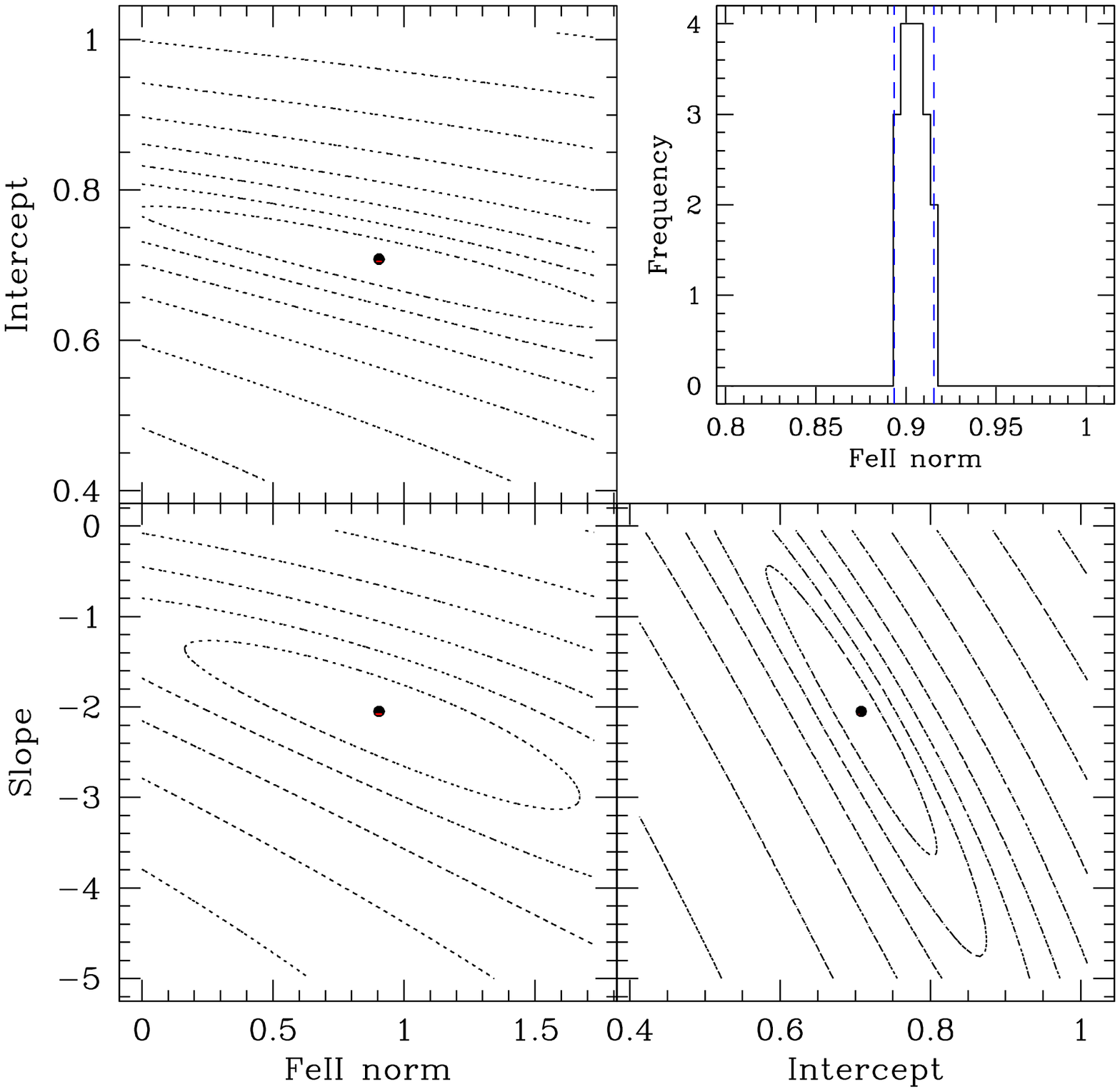}}
\end{figure*}
\end{center}

\end{document}